\documentclass[twocolumn,apj,numberedappendix]{openjournal}

\usepackage{graphicx}
\usepackage{amsmath,amssymb}
\usepackage{booktabs}
\usepackage{microtype}
\usepackage{bm}
\providecommand{\texorpdfstring}[2]{#1}

\shorttitle{HMF Constraints on Early Galaxies and AGN}
\shortauthors{Yogeswaren \& Schaefer}

\begin{document}

\makeatletter
\@firstsectionfalse
\makeatother

\onecolumngrid
\noindent{\footnotesize\textsc{Submitted to The Open Journal of Astrophysics}}\\[-1mm]
{\scriptsize Preprint typeset using \LaTeX\ style openjournal v. 09/06/15}\par
\vspace{0.35in}

\begin{center}
{\normalsize\bfseries\MakeUppercase{Halo Mass Function Constraints on Early Galaxy and AGN Populations in the JWST Era}\par}
\vspace{0.20in}
{\small\textsc{Elan Yogeswaren}\\
\footnotesize Terra Nova Sciences LLC, Houston, TX, 77054 USA\par}
\vspace{0.10in}
{\small\textsc{Robert K. Schaefer}\\
\footnotesize Johns Hopkins University Whiting School EP Applied Physics Program and the Applied Physics Laboratory, Laurel, MD, 20723 USA\par}
\vspace{0.20in}
{\itshape\footnotesize Preprint -- 2026\par}
\vspace{0.18in}
{\footnotesize ABSTRACT\par}
\end{center}

\begin{quote}
\small
The discovery by the \textit{James Webb Space Telescope} (JWST) of massive
galaxies and luminous active galactic nuclei (AGN) at high redshift provides
an increasingly sensitive test of the connection between cosmological
structure formation and the astrophysical processes that make dark-matter
halos observable. We develop a unified framework connecting the linear matter
power spectrum to the mass variance, peak height, halo mass function (HMF),
cumulative halo abundance, and observable galaxy and AGN populations. We use
the Sheth--Tormen (ST) HMF as the baseline and compare it with an illustrative
Bolshoi/Klypin-calibrated modification (KST) to quantify the sensitivity of
high-redshift predictions to the calibration of the HMF against cosmological N-body simulations. We apply the framework to three
complementary tests: the baryonic and stellar-mass reservoir available to
massive galaxies, the abundance of halos capable of hosting luminous AGN, and
the rest-frame ultraviolet luminosity function of high-redshift galaxies.
Because the most massive systems occupy the rare, high-peak tail of the halo
distribution, modest changes in inferred host mass, baryonic conversion
efficiency, black-hole occupation or duty cycle, and HMF calibration can
produce large changes in predicted abundances. The ST--KST comparison shows
that uncertainty in the high-redshift halo population can materially affect
the inferred host supply, particularly toward $z\sim10$, while leaving the
underlying halo-to-observable mappings unchanged. Comparison with
representative JWST massive-galaxy, AGN, and UV-luminosity-function
measurements demonstrates that matching the properties of an individual
source is not sufficient for population consistency: the corresponding host
halos must also exist in adequate numbers. Current observations do not by
themselves require a departure from standard $\Lambda$CDM, but they provide
increasingly stringent quantitative stress tests of both cosmological halo
abundances and the baryonic and black-hole physics connecting halos to
observable sources at cosmic dawn.
\end{quote}
\vspace{0.12in}
\twocolumngrid

\section{Introduction}
\label{sec:introduction}

The large-scale structure of the Universe is understood to have arisen from
the gravitational amplification of tiny primordial departures from exact
homogeneity. In the standard cosmological picture, nearly scale-invariant,
approximately Gaussian primordial perturbations evolved through the
radiation- and matter-dominated eras and ultimately seeded the nonlinear dark
matter halos that host galaxies and supermassive black holes. A successful
description of early galaxy and active galactic nucleus (AGN) formation must
therefore connect the primordial fluctuation field, its scale-dependent linear
evolution, the nonlinear collapse of overdense regions, and the observable
populations that occupy the resulting halos
\citep{Peebles1980,Peacock1999,Dodelson2020,Planck2018}.

This connection is conventionally constructed from the matter power spectrum.
Transfer functions encode the scale-dependent evolution of primordial
perturbations through radiation domination, matter--radiation equality and
recombination, while the linear growth factor describes their subsequent
redshift evolution \citep{Bardeen1986,EisensteinHu1998,Peacock1999,Dodelson2020}.
Smoothing the resulting matter spectrum on a mass scale gives the variance
$\sigma(M,z)$ and peak-height variable $\nu(M,z)$, which provide the statistical
inputs to analytic halo-abundance models. Spherical-collapse theory,
Press--Schechter theory, excursion-set methods and ellipsoidal-collapse
extensions then provide a bridge from the linearly evolved density field to
the abundance of gravitationally bound halos
\citep{GunnGott1972,PressSchechter1974,Bond1991,LaceyCole1993,ShethTormen1999,Tinker2008}.

The high-mass tail of this halo population is especially informative at early
times. If an observed galaxy or AGN requires a halo above some characteristic
mass, then the cumulative abundance of such halos limits the number of
possible hosts. Because increasingly massive halos correspond to increasingly
rare peaks at high redshift, observations of luminous galaxies and accreting
black holes during the first billion years provide a sensitive test of the
combined cosmological halo supply and the astrophysical processes that connect
halos to observable systems
\citep{KormendyHo2013,Volonteri2010,Inayoshi2020}.

This issue has become particularly important with the \textit{James Webb Space
Telescope} (JWST). Deep imaging and spectroscopy have revealed substantial
populations of luminous galaxies and candidate AGN at $z\gtrsim6$, including
objects whose inferred stellar masses, luminosities or black-hole properties
make them demanding tests of early structure and baryonic assembly
\citep{Labbe2023,BoylanKolchin2023,Atek2023a,Atek2023b,Harikane2023,
Kocevski2023,Maiolino2024,Napolitano2025}. At the population level, measurements
of high-redshift rest-frame ultraviolet luminosity functions provide an
additional constraint because a successful model must reproduce not only the
luminosity of individual objects but also their observed number densities
\citep{Donnan2023,Donnan2024,Franco2025}.

The central organizing principle of the present work is therefore a
number-density framework. For a chosen halo mass function (HMF) model $X$, we
write
\begin{equation}
\frac{dn_X}{dM_h}(M_h,z),
\label{eq:intro_hmf}
\end{equation}
for the differential comoving abundance of halos of mass $M_h$ at redshift
$z$. An observable population is then connected to this cosmological halo
supply through an occupation or selection function,
\begin{equation}
n_{{\rm obj},X}(z)
=
\int_0^\infty
\left\langle N_{\rm obj}\mid M_h,z\right\rangle
\frac{dn_X}{dM_h}(M_h,z)\,dM_h .
\label{eq:intro_general_number_density}
\end{equation}
Equation~\eqref{eq:intro_general_number_density} is deliberately general. The
HMF specifies how many halos cosmology provides, while
$\langle N_{\rm obj}\mid M_h,z\rangle$ specifies how those halos are mapped
into the observed population. This separation between \emph{halo supply} and
\emph{halo-to-observable physics} is the common structure underlying the
different tests developed below.

We apply this framework in three complementary ways. First, we construct a
\emph{baryonic-reservoir and massive-galaxy test}: the cosmic baryon fraction
limits the stellar mass that can be assembled in halos of a given mass, while
the cumulative HMF determines whether a sufficient number of such halos exists.
Second, we construct an \emph{AGN host-abundance test}: the available halo
population is combined with a minimum host mass, black-hole occupation
fraction and AGN duty cycle to constrain the number density of active systems.
Third, we construct a \emph{galaxy UV luminosity-function test}: a controlled
stellar halo-to-light mapping converts halo mass into rest-frame UV magnitude,
allowing the predicted population $\Phi(M_{\rm UV},z)$ to be compared directly
with high-redshift measurements. These tests probe related physics but should
not be conflated. In particular, the stellar UV mapping used for the galaxy
luminosity function is not an AGN luminosity model.

For the cosmological halo supply, the Sheth--Tormen (ST) mass function is used
as the principal analytic baseline. We additionally examine the
Bolshoi-calibrated Klypin modification of the ST abundance
\citep{Klypin2011}. Recent high-resolution multi-scale simulations provide
alternative HMF calibrations; for example, \citet{Benson2026} combine
cosmological-volume and zoom-in simulations to construct a flexible HMF
calibration spanning a broad range of halo masses and dark-matter power
spectra. Such developments illustrate that the simulation calibration of the
HMF remains an active area of research and provide complementary alternatives
to the Bolshoi/Klypin prescription considered here. This Klypin-corrected
Sheth--Tormen (KST) branch is not treated as a unique or definitive
high-redshift HMF. Rather, it provides a controlled example of how a
simulation-motivated change in halo abundance propagates through otherwise
identical baryonic and occupation assumptions.

An important feature of this organization is that it separates questions that
can otherwise appear degenerate. Increasing a stellar-conversion efficiency
can make a halo more luminous, but it does not create additional halos.
Changing an AGN duty cycle changes the fraction of eligible hosts observed in
an active phase, but it does not change the cosmological abundance of those
hosts. Conversely, changing the HMF alters the supply of halos without, by
itself, changing the luminosity assigned to an individual halo. The framework
therefore provides a direct way to identify whether an apparent high-redshift
discrepancy is primarily associated with halo abundance, baryonic conversion,
black-hole occupation and activity, or an observational mapping.

The purpose of this paper is to use the emerging JWST high-redshift
populations as a quantitative stress test of standard $\Lambda$CDM structure
formation and of the astrophysical assumptions that connect dark matter halos
to observable galaxies and AGN. Rather than interpreting any apparent
discrepancy as evidence for a failure of the standard cosmological model, we
use a transparent analytic and numerical framework to quantify how demanding
different high-redshift observations are under explicitly stated assumptions,
and to examine how those conclusions change when the adopted halo abundance
or halo-to-observable mapping is varied. This distinction is especially
important because stellar masses, star-formation efficiencies, black-hole
masses, Eddington ratios, occupation fractions, duty cycles, luminosity
scatter, survey completeness, and cosmic variance can all influence the
inferred level of tension.

\section{Cosmological and Halo-Model Inputs}
\label{sec:cosmological_halo_inputs}

The observational tests developed in this work require a common cosmological prediction for the abundance of dark-matter halos as a function of mass and redshift. 

The calculation begins with the linear matter power spectrum,
constructs the mass variance and peak-height parameter, and then uses these
quantities to evaluate the halo mass function (HMF).  The sequence may be
summarized as
\begin{equation}
\begin{aligned}
P_{\rm m}(k,0)
&\longrightarrow \sigma(M,0)
\longrightarrow \sigma(M,z)
\longrightarrow \nu(M,z)
\\
&\longrightarrow \frac{dn_X}{dM}
\longrightarrow n_X(>M,z),
\end{aligned}
\label{eq:sec2_chain}
\end{equation}
where $X$ denotes the adopted halo-abundance prescription.  The
Sheth--Tormen (ST) mass function provides the principal analytic baseline.
Press--Schechter (PS) is included as the spherical-collapse reference, and a
Bolshoi/Klypin-calibrated modification of ST is used to examine the sensitivity
of the high-redshift predictions to the adopted HMF calibration.

\subsection{Fiducial Cosmology}
\label{subsec:fiducial_cosmology}
The critical density $\rho_{crit}$ for a flat universe is defined as  
\begin{equation}
\rho_{{\rm crit},0} =
\frac{3H_0^2}{8\pi G},
\label{eq:critical_density}
\end{equation}
where $H_0$ is the present value of the Hubble constant, and $G$ is Newton's gravitational constant.

We adopt a Planck-like, spatially flat $\Lambda$CDM cosmology
\citep{Planck2018}.  The principal parameters are
\begin{equation}
\begin{aligned}
\Omega_{\rm m}&\simeq0.316, &
\Omega_{\rm b}&\simeq0.0493,\\
h&\simeq0.673, &
n_{\rm s}&\simeq0.965,\\
\sigma_8&=0.812. &&
\end{aligned}
\label{eq:fiducial_cosmology}
\end{equation}
where $\Omega_a=\rho_a/\rho_{\rm crit}$ denotes the density parameter
of component $a$, and
\begin{equation}
h \equiv
\frac{H_0}
{100\,{\rm km\,s^{-1}\,Mpc^{-1}}}.
\end{equation}
The parameter $\sigma_8$ is the rms linear matter fluctuation within
spheres of radius $8\,h^{-1}\,{\rm Mpc}$ at $z=0$.
Spatial flatness gives
\begin{equation}
\Omega_\Lambda=1-\Omega_{\rm m}\simeq0.684.
\label{eq:omega_lambda}
\end{equation}
The numerical calculations use the corresponding fiducial parameter values with their full internal precision.

The mean comoving matter density is
\begin{equation}
\bar{\rho}_{\rm m}
=
\Omega_{\rm m}\rho_{{\rm crit},0}.
\label{eq:mean_density}
\end{equation}
The corresponding cosmic baryon fraction,
\begin{equation}
f_b=\frac{\Omega_{\rm b}}{\Omega_{\rm m}}\simeq0.156,
\label{eq:sec2_baryon_fraction}
\end{equation}
sets the maximum baryonic reservoir associated with a halo and is an input for the massive-galaxy and luminosity applications developed below.

\subsection{Linear Matter Power Spectrum}
\label{subsec:matter_power}

The present-day linear matter power spectrum is written
\begin{equation}
P_{\rm m}(k,0)=A_{\rm norm}\,k^{n_{\rm s}}T^2(k),
\label{eq:matter_power_today}
\end{equation}
where $T(k)$ is the matter transfer function and $A_{\rm norm}$ fixes the
overall normalization.  The transfer function describes the scale-dependent
processing of primordial density perturbations through the radiation- and
matter-dominated eras.  The numerical calculation uses the Eisenstein--Hu
cold-dark-matter/baryon transfer-function prescription
\citep{Bardeen1986,EisensteinHu1998,Peacock1999,Dodelson2020,Mo2010}.

The amplitude is normalized directly to the adopted value of $\sigma_8$.
For a spherical top-hat filter of radius
\begin{equation}
R_8=8h^{-1}\,{\rm Mpc},
\label{eq:R8}
\end{equation}
the normalization condition is
\begin{equation}
\sigma_8^2
=
\frac{1}{2\pi^2}
\int_0^\infty
k^2P_{\rm m}(k,0)
W^2(kR_8)\,dk.
\label{eq:sigma8_normalization}
\end{equation}
Operationally, an unnormalized spectrum
\begin{equation}
P_{\rm unnorm}(k)=k^{n_{\rm s}}T^2(k)
\label{eq:unnormalized_power}
\end{equation}
is first used to calculate $\sigma_{8,\rm unnorm}$.  The normalization is then
\begin{equation}
A_{\rm norm}
=
\left(
\frac{\sigma_8}{\sigma_{8,\rm unnorm}}
\right)^2,
\label{eq:power_normalization}
\end{equation}
so that
\begin{equation}
P_{\rm m}(k,0)=A_{\rm norm}P_{\rm unnorm}(k).
\label{eq:normalized_power}
\end{equation}
This procedure fixes the power-spectrum amplitude once and ensures that the
spectrum used throughout the HMF calculation reproduces the specified
$\sigma_8$.

\subsection{Linear Growth Factor}
\label{subsec:linear_growth}

The redshift evolution of the linear matter spectrum is described by the
growth factor $D(z)$,
\begin{equation}
P_{\rm m}(k,z)
=
D^2(z)P_{\rm m}(k,0),
\qquad D(0)=1.
\label{eq:power_growth}
\end{equation}
For the spatially flat matter--$\Lambda$ background considered here, define
\begin{equation}
E(a)\equiv\frac{H(a)}{H_0}
=
\left[
\Omega_{{\rm m},0}a^{-3}
+
\Omega_{\Lambda,0}
\right]^{1/2}.
\label{eq:Ea}
\end{equation}
The growing-mode solution may then be expressed as
\begin{equation}
D_+(a)
=
\frac{5\Omega_{{\rm m},0}}{2}
E(a)
\int_0^a
\frac{da'}
{a'^3E^3(a')},
\label{eq:growth_unnormalized}
\end{equation}
with the normalized growth factor
\begin{equation}
D(a)
=
\frac{D_+(a)}{D_+(1)},
\qquad
a=(1+z)^{-1}.
\label{eq:growth_normalized}
\end{equation}
This form approaches the familiar matter-dominated behavior at high redshift
while including the suppression of late-time growth by the cosmological
constant \citep{Peebles1980,Peacock1999,Dodelson2020,Mo2010}.  The numerical
calculation evaluates Equation~\eqref{eq:growth_normalized} directly rather
than imposing the Einstein--de Sitter approximation $D=(1+z)^{-1}$.

\subsection{Mass Variance and Peak Height}
\label{subsec:growth_variance}

The linear density field is smoothed on a comoving scale $R$ using a spherical
top-hat window.  Its Fourier transform is
\begin{equation}
W(x)
=
\frac{3}{x^3}
\left(
\sin x-x\cos x
\right),
\label{eq:tophat_window}
\end{equation}
with $x=kR$.  The smoothing radius is related to the halo-mass coordinate by
\begin{equation}
M
=
\frac{4\pi}{3}
\bar{\rho}_{\rm m}R^3.
\label{eq:mass_radius}
\end{equation}

The variance of the smoothed linear density field as a function of mass and redshift is
\begin{equation}
\sigma^2(M,z)
=
\frac{1}{2\pi^2}
\int_0^\infty
k^2P_{\rm m}(k,z)
W^2(kR)\,dk.
\label{eq:mass_variance}
\end{equation}
Since the adopted linear growth is scale independent,
\begin{equation}
\sigma(M,z)=D(z)\sigma(M,0).
\label{eq:sigma_growth}
\end{equation}

For the collapse threshold we use the standard linearly extrapolated spherical-collapse value
\citep{GunnGott1972,PressSchechter1974}.
\begin{equation}
\delta_{\rm c}=1.686,
\label{eq:delta_c}
\end{equation}
  The peak-height parameter is
therefore
\begin{equation}
\nu(M,z)
=
\frac{\delta_{\rm c}}{\sigma(M,z)}
=
\frac{\delta_{\rm c}}
{D(z)\sigma(M,0)}.
\label{eq:peak_height}
\end{equation}
Increasing halo mass reduces $\sigma(M,z)$, while increasing redshift reduces
the linear growth factor.  Both effects increase $\nu$, so massive halos at
high redshift correspond to increasingly rare peaks in the primordial density
field.

\begin{figure*}
    \centering
    \includegraphics[width=\textwidth]{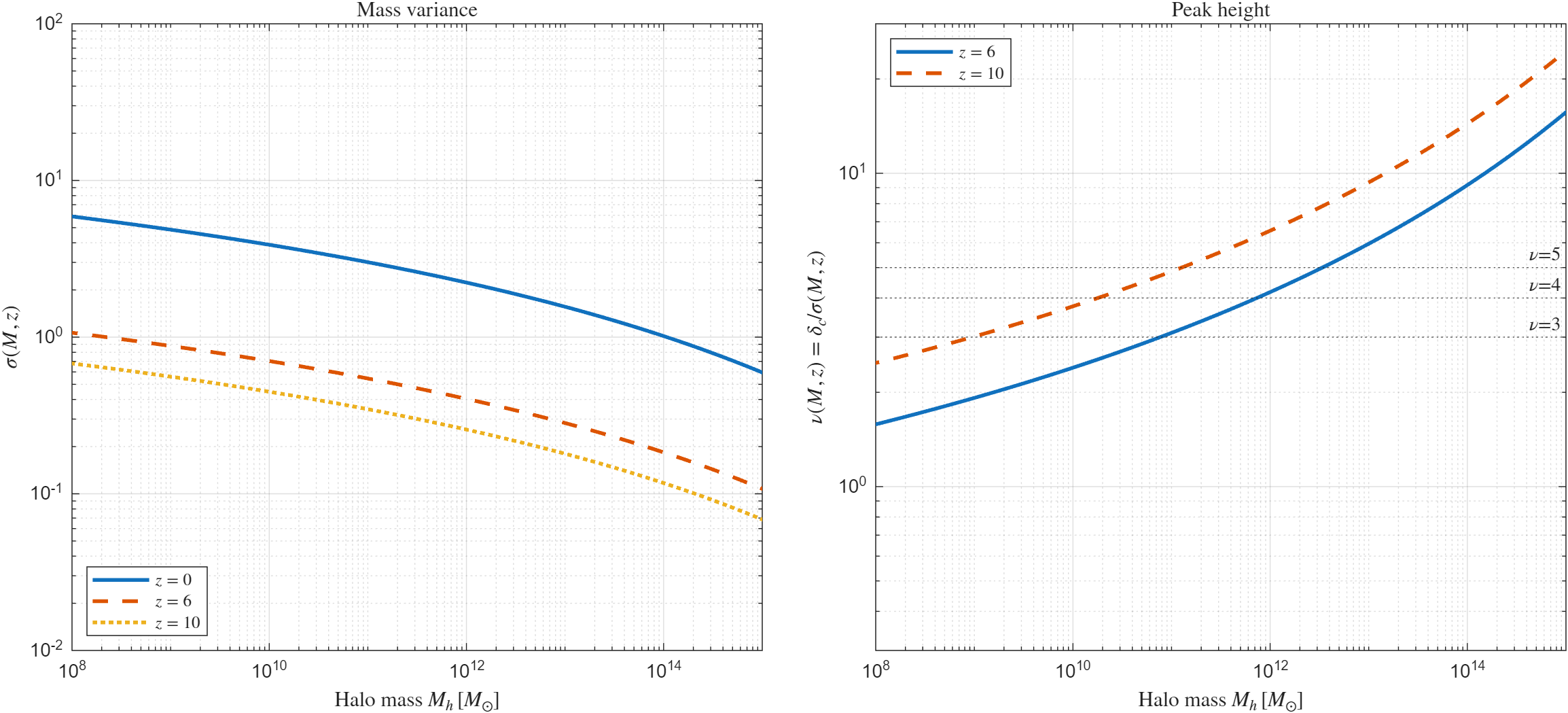}
    \caption{Mass variance $\sigma(M,z)$ and peak-height parameter
    $\nu(M,z)=\delta_{\rm c}/\sigma(M,z)$ at representative redshifts.
    Increasing mass and redshift move halos toward the rare, high-$\nu$ tail
    of the density field.}
    \label{fig:sigma_nu}
\end{figure*}

\subsection{Press--Schechter and Sheth--Tormen Halo Mass Functions}
\label{subsec:baseline_hmf}

The HMF gives the comoving number density of collapsed dark-matter halos as a
function of halo mass.  We use the multiplicity-function convention
\begin{equation}
\begin{aligned}
\frac{dn_X}{dM}(M,z)
={}&
\frac{\bar{\rho}_{\rm m}}{M^2}
f_X(\nu)
\\
&\times
\left|
\frac{d\ln\sigma^{-1}}
{d\ln M}
\right|.
\end{aligned}
\label{eq:hmf_master}
\end{equation}
The factor of $\nu$ is included explicitly in the multiplicity functions
defined below.  No additional factor of $\nu$ is therefore introduced through
a separate $d\nu/dM$ term.  The logarithmic form used for displaying the HMF is
\begin{equation}
\frac{dn_X}{d\log_{10}M}
=
(\ln10)M\frac{dn_X}{dM}.
\label{eq:hmf_logmass}
\end{equation}

For Gaussian initial conditions and spherical collapse, the
Press--Schechter multiplicity function is
\begin{equation}
f_{\rm PS}(\nu)
=
\sqrt{\frac{2}{\pi}}\,
\nu
\exp\left(-\frac{\nu^2}{2}\right)
\label{eq:ps_multiplicity}
\end{equation}
\citep{PressSchechter1974,Bond1991}.  Press--Schechter provides the basic
analytic connection between the Gaussian density field and the abundance of
collapsed objects.

The Sheth--Tormen prescription incorporates the effect of nonspherical,
approximately ellipsoidal collapse and provides the principal HMF baseline
used in this work \citep{ShethTormen1999,ShethMoTormen2001}.  Its multiplicity
function is
\begin{equation}
\begin{aligned}
f_{\rm ST}(\nu)
={}&
A_{\rm ST}
\sqrt{\frac{2a}{\pi}}
\left[
1+(a\nu^2)^{-p}
\right]
\\
&\times
\nu
\exp\left(
-\frac{a\nu^2}{2}
\right),
\end{aligned}
\label{eq:st_multiplicity}
\end{equation}
with
\begin{equation}
A_{\rm ST}=0.3222,
\qquad
a=0.707,
\qquad
p=0.3.
\label{eq:st_parameters}
\end{equation}
Relative to the spherical Press--Schechter result, ST predicts a different
abundance in the rare high-mass tail that is particularly relevant to
high-redshift galaxy and AGN applications.  More generally, HMF calibrations
depend on simulation methodology, halo finder, mass definition, cosmology, and
redshift range \citep{Tinker2008}.

The cumulative number density of halos above a threshold $M_{\min}$ is
\begin{equation}
n_X(>M_{\min},z)
=
\int_{M_{\min}}^\infty
\frac{dn_X}{dM'}(M',z)\,dM',
\label{eq:cumulative_number}
\end{equation}
and the corresponding cumulative collapsed mass density is
\begin{equation}
\begin{aligned}
\rho_{h,X}(>M_{\min},z)
={}&
\int_{M_{\min}}^\infty
M'
\frac{dn_X}{dM'}(M',z)
\,dM'.
\end{aligned}
\label{eq:sec3_cumulative_mass}
\end{equation}
These two integral quantities play different roles in the applications below:
$n_X$ measures the number of potential hosts, whereas $\rho_{h,X}$ measures
the total collapsed matter reservoir contained in those hosts.

\begin{figure*}
    \centering
    \includegraphics[width=\columnwidth]{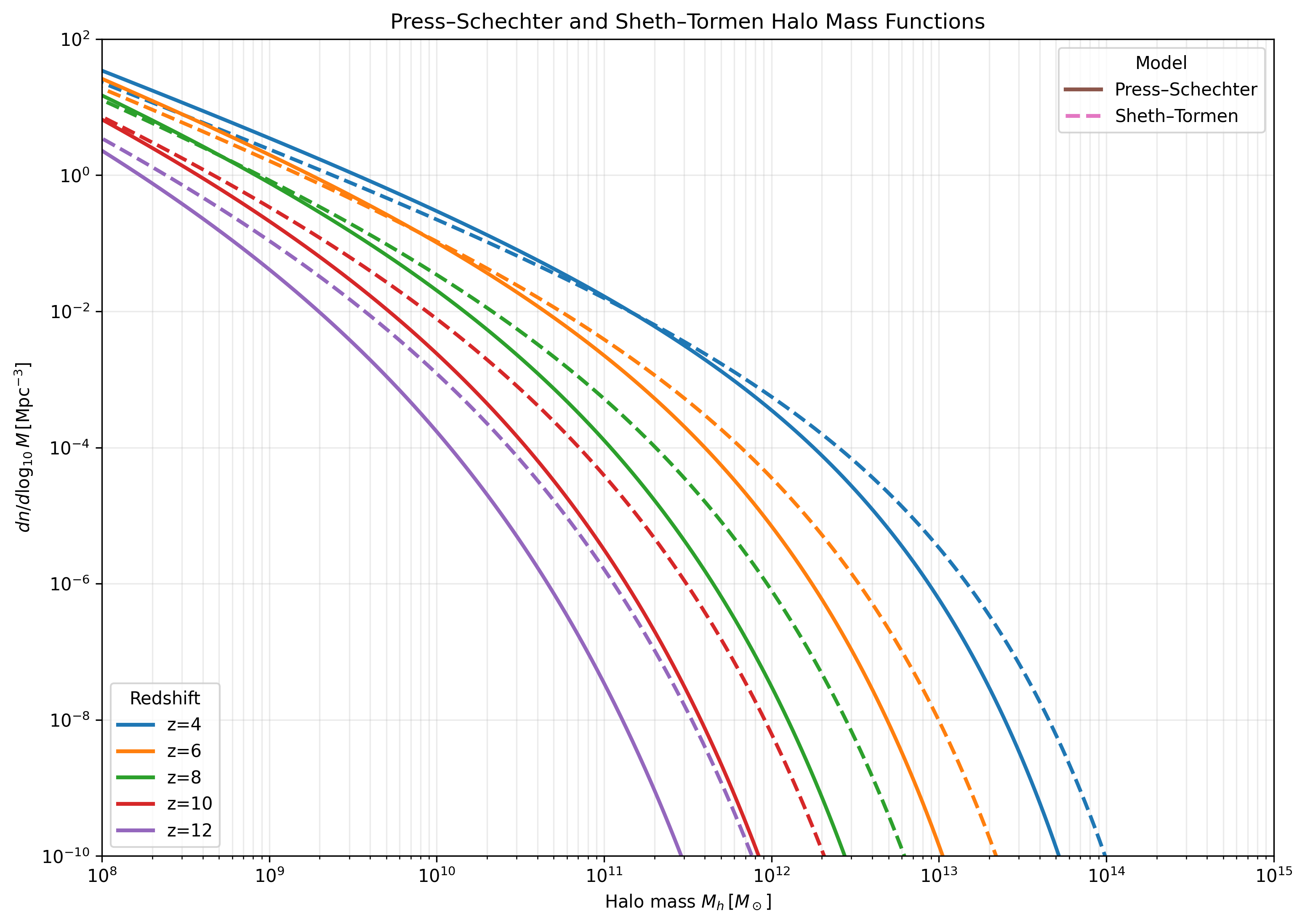}
    \caption{Press--Schechter and Sheth--Tormen halo mass functions at
    representative redshifts, calculated from the same normalized linear
    matter spectrum.  The comparison illustrates the sensitivity of the
    rare-halo population to the adopted collapse prescription.}
    \label{fig:ps_st_hmf}
\end{figure*}

\subsection{Bolshoi/Klypin-Calibrated Abundance Modification}
\label{subsec:kst}

Analytic HMF prescriptions provide a useful theoretical description of halo
collapse, but numerical simulations show that the abundance of halos can
depart systematically from a single universal analytic form.  Such
differences become especially important when the objects of interest occupy
the exponentially suppressed high-mass tail.  To quantify the sensitivity of
the present calculations to HMF calibration, we consider the
Bolshoi/Klypin high-redshift abundance modification \citep{Klypin2011}.

The redshift-dependent correction factor is
\begin{equation}
F_{\rm K}(z)
=
\frac{[5.501D(z)]^4}
{1+[5.500D(z)]^4},
\label{eq:klypin_factor}
\end{equation}
where $D(z)$ is the normalized growth factor defined in
Equation~\eqref{eq:growth_normalized}.  We define the
Klypin-corrected Sheth--Tormen branch, denoted KST, by
\begin{equation}
\frac{dn_{\rm KST}}{dM}(M,z) = 
F_{\rm K}(z) \frac{dn_{\rm ST}}{dM}(M,z).
\label{eq:kst_hmf}
\end{equation}

Since $F_{\rm K}(z)$ is independent of mass in this implementation, the
correction factors out of cumulative integrals:
\begin{equation}
n_{\rm KST}(>M,z)
=
F_{\rm K}(z)
n_{\rm ST}(>M,z),
\label{eq:kst_cumulative_number}
\end{equation}
and
\begin{equation}
\rho_{h,\rm KST}(>M,z)
=
F_{\rm K}(z)
\rho_{h,\rm ST}(>M,z).
\label{eq:kst_cumulative_mass}
\end{equation}
The ST and KST calculations therefore share the same background cosmology,
linear power spectrum, growth factor, mass variance, collapse threshold,
peak-height parameter, and halo-to-observable mappings.  Their difference
isolates the effect of the adopted abundance calibration.

For reference, the correction factors most directly relevant to the
high-redshift comparisons are
\begin{equation}
\begin{aligned}
F_{\rm K}(6)&\simeq0.496,\\
F_{\rm K}(8)&\simeq0.265,\\
F_{\rm K}(10)&\simeq0.139.
\end{aligned}
\label{eq:klypin_reference_values}
\end{equation}
These correspond to progressively stronger reductions of the predicted halo
supply toward cosmic dawn.  The KST branch is used as a controlled
simulation-motivated sensitivity test rather than as a unique or definitive
replacement for the ST mass function.

\subsubsection{Halo-mass definition and calibration range}
\label{subsubsec:kst_scope}

A simulation-calibrated HMF must be interpreted together with its halo finder
and halo-mass definition.  Friends-of-friends masses, virial masses, and
spherical-overdensity masses defined relative to either the critical density
or the mean matter density are not generally interchangeable.  This issue is
particularly important in the rare high-mass tail, where modest shifts in halo
mass can produce large changes in cumulative abundance.

In the analytic calculation used here, the halo-mass coordinate is associated
with the spherical smoothing scale through
\begin{equation}
M_h
=
\frac{4\pi}{3}
\bar{\rho}_{\rm m}R^3.
\label{eq:halo_mass_coordinate}
\end{equation}
The same mass coordinate and numerical mass grid are used in the ST and KST
branches.  Equation~\eqref{eq:kst_hmf} should therefore be understood as a
controlled abundance modification within this common calculation, rather than
as an exact transformation between different simulation halo-mass
definitions.  Precision comparisons with simulation catalogs or
observationally inferred host masses require all quantities to be placed on a
consistent mass definition.

The adopted Bolshoi/Klypin calibration is applied over
\begin{equation}
0\leq z\leq10.
\label{eq:kst_calibration_range}
\end{equation}
Equation~\eqref{eq:klypin_factor} can be evaluated mathematically at larger
redshift, but values beyond this interval are extrapolations rather than
simulation-calibrated predictions.  High-redshift observational points beyond
$z=10$ can therefore provide useful context without being interpreted as
calibrated KST constraints.

\subsection{Connection to the Observational Tests}
\label{subsec:sec2_bridge}

The quantities defined above provide the common cosmological halo supply for
the three observational applications.  Their relationship can be written
schematically as
\begin{equation}
\begin{aligned}
P_{\rm m}(k,z)
&\longrightarrow
\sigma(M,z)
\longrightarrow
\nu(M,z)
\\
&\longrightarrow
\frac{dn_X}{dM}
\longrightarrow
\left\{
n_X(>M,z),
\rho_{h,X}(>M,z)
\right\}.
\end{aligned}
\label{eq:sec2_summary_chain}
\end{equation}

The massive-galaxy test combines these quantities with the cosmic baryon
fraction and a stellar-conversion efficiency to determine whether the
available halo population can support the inferred stellar masses and number
densities.  The AGN test combines the cumulative host supply with black-hole
occupation, activity, and host-mass assumptions.  The UV luminosity-function
test instead maps individual halo masses into stellar UV luminosities and
propagates the differential HMF into an observable luminosity distribution.

This construction separates the cosmological and astrophysical components of
each comparison.  The HMF determines the number and mass reservoir of
potential hosts.  Baryonic conversion, black-hole occupation and duty cycle,
and halo-to-light relations determine how that halo population is converted
into observable galaxies and AGN.  The following sections use this common
framework to test each observable population separately.

\section{Baryonic-Reservoir and Massive-Galaxy Test}
\label{sec:baryonic_reservoir_test}

The first observational application of the halo-abundance framework is the
massive-galaxy test.  The central question is whether the dark-matter halo
population at a given redshift provides both (i) a sufficiently large baryonic
reservoir to support the inferred stellar masses and (ii) a sufficient number
of such halos to reproduce the observed population.  These are related but
distinct requirements.  The first is a mass-budget constraint set by the
cosmic baryon fraction and the efficiency of converting baryons into stars;
the second is an abundance constraint set by the halo mass function (HMF).

This distinction is especially important at high redshift.  Massive halos
correspond to increasingly rare peaks in the primordial density field, so a
moderate increase in the halo mass required to host an observed galaxy can
produce a large decrease in its predicted abundance.  The massive-galaxy test
therefore provides a direct way to connect the inferred properties of early
JWST galaxies to the cosmological supply of collapsed halos, while keeping the
baryonic assumptions explicit.

\subsection{From Halo Mass to the Available Baryonic Reservoir}
\label{subsec:baryonic_reservoir}

For the cosmological parameters adopted in this work, the cosmic baryon
fraction is
\begin{equation}
f_b=\frac{\Omega_b}{\Omega_m}\simeq 0.156 .
\label{eq:sec3_baryon_fraction}
\end{equation}
A halo of mass $M_h$ therefore contains, at most, a cosmic-mean baryonic
reservoir
\begin{equation}
M_b=f_bM_h .
\label{eq:sec3_baryonic_mass}
\end{equation}
If an effective fraction $\epsilon_\star$ of this reservoir is converted into
stars, the stellar mass satisfies
\begin{equation}
M_\star\leq \epsilon_\star f_bM_h,
\qquad 0<\epsilon_\star\leq 1 .
\label{eq:sec3_stellar_mass_bound}
\end{equation}
Consequently, an observed galaxy with inferred stellar mass $M_\star$ requires
a minimum host-halo mass
\begin{equation}
M_{h,\min}=
\frac{M_\star}{\epsilon_\star f_b}.
\label{eq:sec3_minimum_halo_mass}
\end{equation}

The limiting case $\epsilon_\star=1$ corresponds to conversion of the entire
cosmic-mean baryonic allotment into stars.  It is therefore an absolute
baryonic ceiling rather than a realistic star-formation efficiency.  For
$\epsilon_\star<1$, a larger halo mass is required to support the same
$M_\star$, moving the putative host farther into the rare high-mass tail of the
HMF.  This simple relation is useful because it separates the finite baryonic
mass budget from the subsequent question of whether halos of the required mass
exist in sufficient numbers \citep{Behroozi2013,Moster2013,BoylanKolchin2023}.

For an HMF prescription $X\in\{\mathrm{ST},\mathrm{KST}\}$, the cumulative
halo abundance above the required threshold is
\begin{equation}
n_X(>M_{h,\min},z)
=
\int_{M_{h,\min}}^\infty
\frac{dn_X}{dM_h'}(M_h',z)\,dM_h' .
\label{eq:sec3_halo_supply}
\end{equation}
The corresponding massive-galaxy number-density test is therefore
\begin{equation}
n_{\rm gal,obs}(>M_\star,z)
\quad \hbox{versus}\quad
n_X\!\left(
>\frac{M_\star}{\epsilon_\star f_b},z
\right).
\label{eq:sec3_galaxy_number_test}
\end{equation}
An inferred stellar mass alone is thus not a cosmological abundance
constraint: it becomes one only after specifying the stellar-to-halo mapping
and comparing the resulting halo requirement with the observed galaxy number
density.

\subsection{Differential Halo Supply}
\label{subsec:sec3_differential_hmf}

\begin{figure*}
    \centering
    \includegraphics[width=\textwidth]{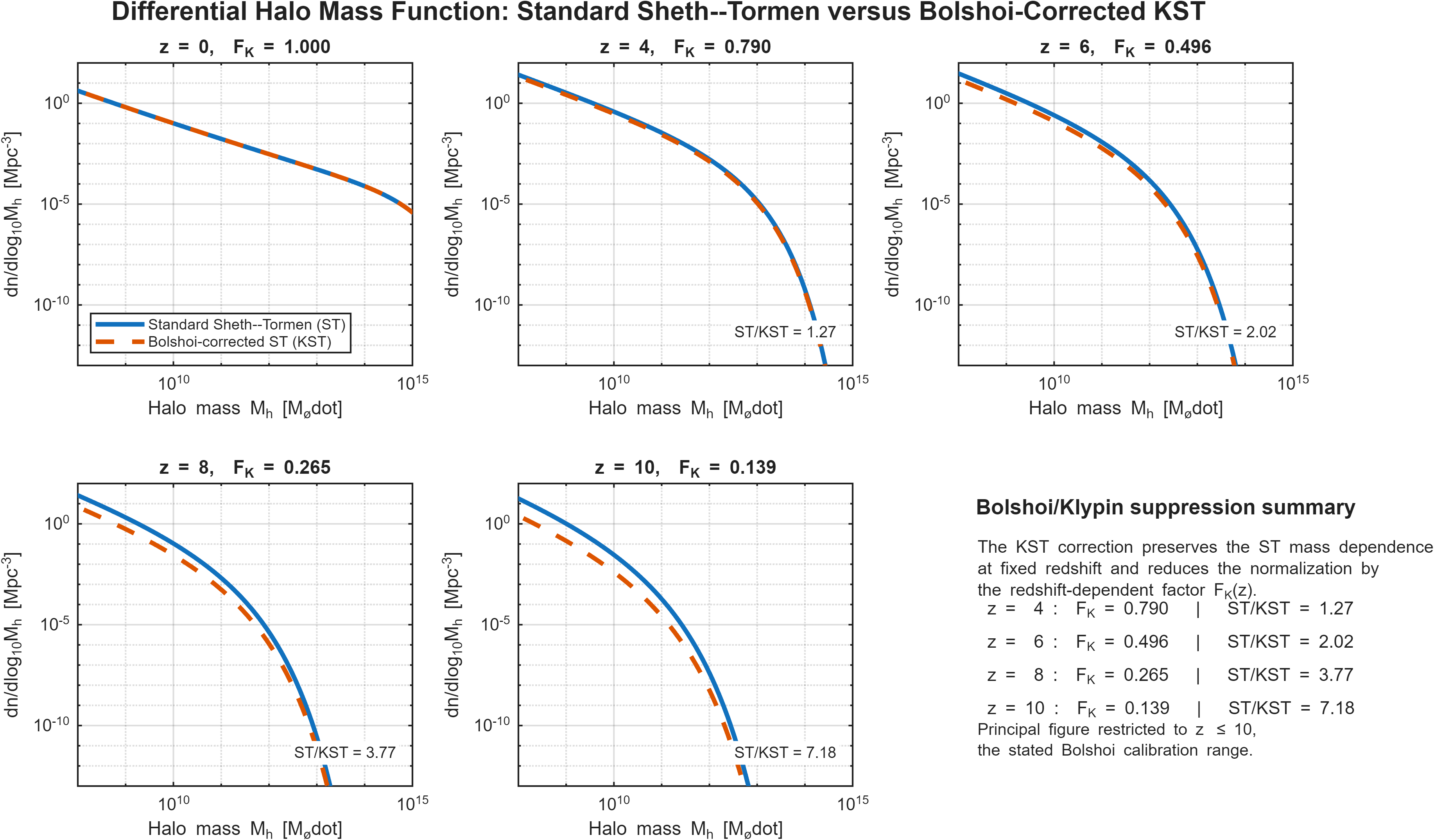}
    \caption{Differential halo mass function comparing the standard
    Sheth--Tormen (ST) prediction with the Bolshoi/Klypin-corrected ST (KST)
    prescription at $z=0$, 4, 6, 8, and 10.  At fixed redshift the KST
    correction rescales the ST abundance by $F_{\rm K}(z)$ while preserving
    the mass dependence of the implemented ST curve.  The separation grows
    toward high redshift as the available population of massive halos becomes
    increasingly sensitive to the adopted HMF calibration.  The calibrated
    KST comparison is restricted to $z\leq10$.}
    \label{fig:st_kst_differential}
\end{figure*}

Figure~\ref{fig:st_kst_differential} compares the differential halo abundance
predicted by the standard Sheth--Tormen (ST) mass function with the
Bolshoi/Klypin-corrected ST (KST) prescription at representative redshifts.
The ST calculation provides the principal analytic baseline.  The KST branch
uses the redshift-dependent correction $F_{\rm K}(z)$ defined in
Section~\ref{subsec:kst}, so that
\begin{equation}
\frac{dn_{\rm KST}}{dM_h}
=
F_{\rm K}(z)
\frac{dn_{\rm ST}}{dM_h}.
\label{eq:sec3_kst_differential}
\end{equation}
Because the adopted correction is independent of halo mass at fixed redshift,
it changes the abundance normalization while preserving the mass dependence of
the implemented ST curve.  The underlying matter power spectrum, growth
factor, mass variance, and peak-height parameter are identical in the two
branches.

The comparison illustrates the increasing sensitivity of the available halo
population to HMF calibration toward cosmic dawn.  At low redshift the ST and
KST predictions are close, whereas their separation becomes progressively
larger at high redshift.  For the massive-galaxy test this means that the same
stellar-mass requirement and the same baryonic conversion efficiency can
correspond to substantially different predicted host abundances depending on
the adopted HMF calibration.

\subsection{Boylan--Kolchin/Labb\'e Number-Density Stress Test}
\label{subsec:boylan_labbe_test}

The cumulative abundance of halos above a mass threshold $M_h$ is
\begin{equation}
n_X(>M_h,z)
=
\int_{M_h}^{\infty}
\frac{dn_X}{dM'}(M',z)\,dM',
\qquad X\in\{\mathrm{ST},\mathrm{KST}\}.
\label{eq:sec3_cumulative_abundance}
\end{equation}
Combining this quantity with
Equation~\eqref{eq:sec3_minimum_halo_mass} maps an inferred stellar mass into
the abundance of halos capable of hosting such a galaxy.

\begin{figure*}
    \centering
    \includegraphics[width=\textwidth]{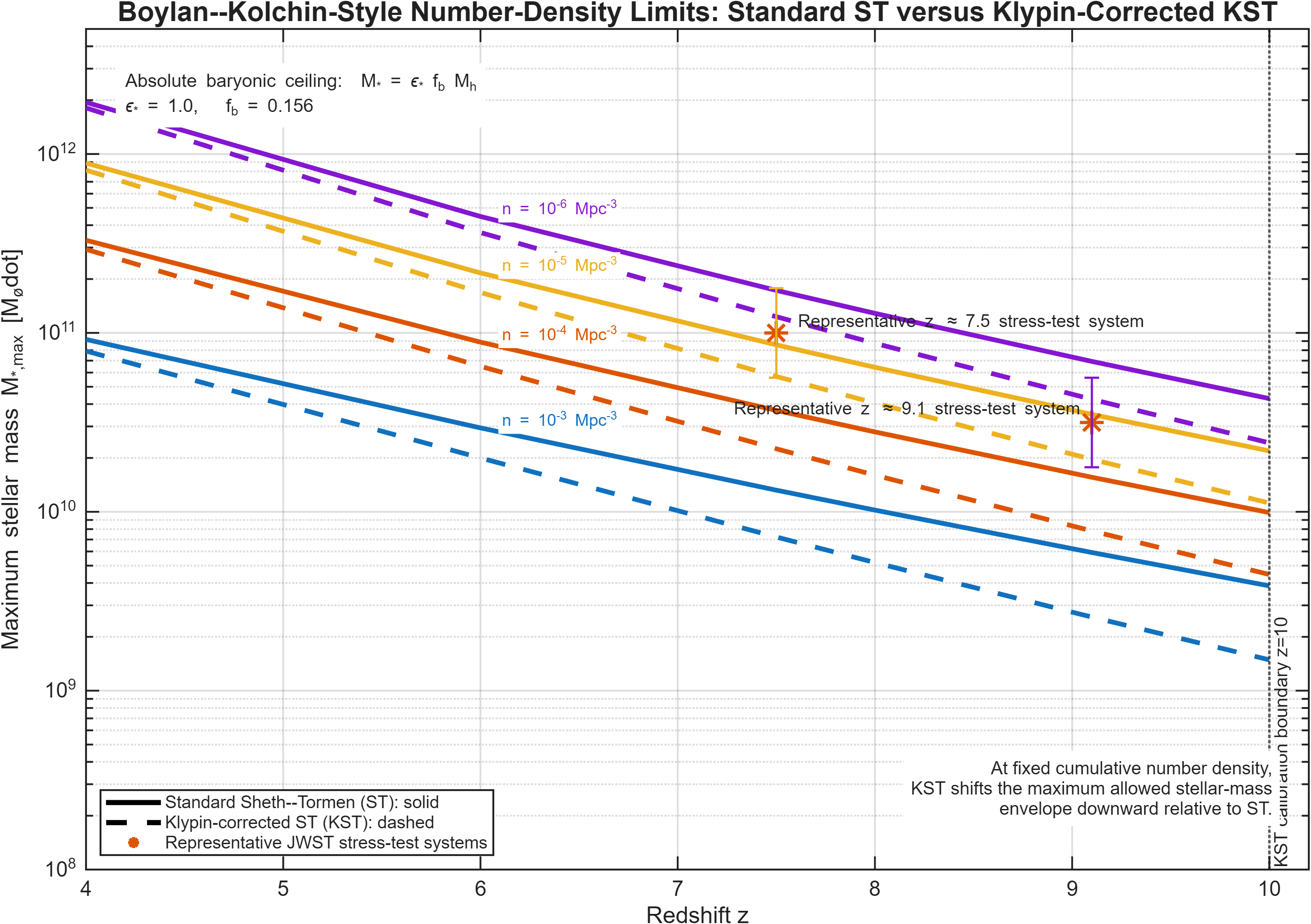}
    \caption{Boylan--Kolchin-style number-density stress test comparing the
    standard ST and KST halo reservoirs.  Curves show the maximum stellar mass
    supportable at fixed cumulative halo number density under
    $M_\star=\epsilon_\star f_bM_h$, with $\epsilon_\star=1$ representing the
    absolute baryon-conversion ceiling.  Representative systems near
    $z\simeq7.5$ and $z\simeq9.1$ are shown with their observational
    uncertainty ranges.  The lower KST halo abundance makes a fixed
    stellar-mass/number-density combination more demanding.  The $z=10$
    marker denotes the upper boundary of the adopted KST calibration.}
    \label{fig:boylan_kolchin_number_density}
\end{figure*}

Figure~\ref{fig:boylan_kolchin_number_density} presents this comparison in the
redshift--stellar-mass plane.  At a fixed cumulative number density, the HMF
sets the corresponding halo-mass threshold, and the baryonic relation converts
that threshold into the maximum stellar mass supportable for an assumed
$\epsilon_\star$.  Since the KST abundance is lower than the ST abundance at a
fixed halo mass and redshift, a given cumulative number density is reached at a
lower halo-mass threshold in KST.  The maximum stellar mass supportable at that
number density is therefore also reduced.

Representative Boylan--Kolchin/Labb\'e stress-test systems near
$z\simeq7.5$ and $z\simeq9.1$ illustrate the region in which the required host
halos are already rare \citep{Labbe2023,BoylanKolchin2023}.  The
$\epsilon_\star=1$ curves provide the most permissive possible baryonic
interpretation.  Any smaller conversion efficiency increases the required
host-halo mass and makes the abundance constraint correspondingly stronger.

The comparison should not be interpreted as a model-independent exclusion of
$\Lambda$CDM.  Its significance depends jointly on the inferred stellar mass,
redshift, survey volume, completeness, cosmic variance, lensing where
applicable, and the assumed stellar-conversion efficiency.  Because the
high-mass HMF is steep, comparatively modest changes in the inferred host mass
can produce large changes in the predicted abundance.  The calculation is
therefore most appropriately interpreted as a stress test of the combined
halo-supply and baryonic mapping.

\subsection{Peak Height and the Rarity of Massive Hosts}
\label{subsec:peak_height_baryons}

A complementary description of the massive-galaxy problem is provided by the
peak-height parameter
\begin{equation}
\nu(M_h,z)=\frac{\delta_c}{\sigma(M_h,z)} ,
\label{eq:sec3_peak_height}
\end{equation}
where $\delta_c\simeq1.686$.  At fixed halo mass, $\sigma(M_h,z)$ decreases
toward high redshift and $\nu$ therefore increases.  Massive halos at cosmic
dawn consequently correspond to increasingly rare fluctuations in the
primordial density field.

\begin{figure*}
    \centering
    \includegraphics[width=\columnwidth]{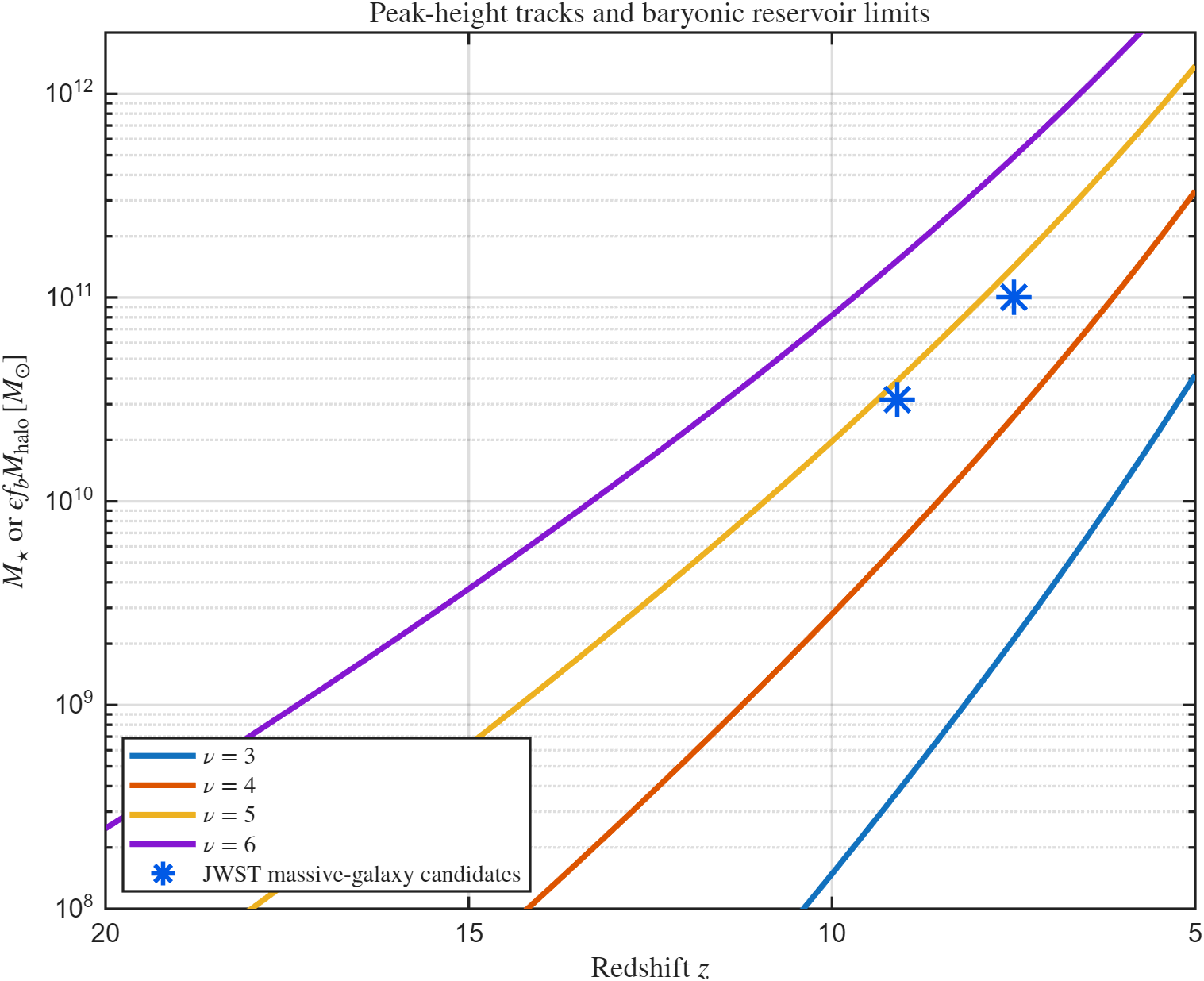}
    \caption{Peak-height tracks and baryonic-reservoir limits.  Curves of
    fixed $\nu$ identify halo masses corresponding to increasingly rare
    primordial density peaks.  At high redshift, massive galaxy hosts occupy
    progressively rarer high-$\nu$ regions.  The baryonic relation
    $M_b=f_bM_h$ converts the halo-mass scale into the corresponding maximum
    cosmic-mean baryonic reservoir.}
    \label{fig:peak_height_baryons}
\end{figure*}

Figure~\ref{fig:peak_height_baryons} shows fixed-$\nu$ tracks together with
the baryonic relation $M_b=f_bM_h$.  The diagram provides a direct physical
view of how the stellar masses associated with the most massive early galaxies
map onto increasingly rare host halos.  The peak-height relation is determined
by the linear density field, whereas the HMF calibration determines the
abundance assigned to the resulting halo population.  The two quantities
therefore provide complementary information: $\nu$ characterizes the rarity of
the underlying fluctuation, while $dn/dM_h$ determines the number of collapsed
halos available to host the observed systems.

\subsection{Cumulative Matter and Stellar-Mass-Density Constraints}
\label{subsec:cumulative_stellar_density}

The number-density test considers whether enough individual halos exist above
a specified mass threshold.  A complementary integral constraint follows from
the total collapsed matter contained in those halos.  The cumulative halo-mass
density is

\begin{equation}
\begin{aligned}
\rho_{h,X}(>M_h,z)
&=
\int_{M_h}^{\infty}
M'\,
\frac{dn_X}{dM'}(M',z)\,
dM',
\\[-1mm]
&\hspace{18mm}
X\in\{\mathrm{ST},\mathrm{KST}\}.
\end{aligned}
\label{eq:cumulative_mass}
\end{equation}

For the adopted mass-independent KST prescription,
\begin{equation}
\rho_{h,\mathrm{KST}}(>M_h,z)
=
F_{\rm K}(z)\,
\rho_{h,\mathrm{ST}}(>M_h,z).
\label{eq:sec3_kst_mass_density}
\end{equation}
The maximum stellar-mass density associated with galaxies above a stellar-mass
threshold $M_\star$ is then
\begin{equation}
\begin{aligned}
\rho_{\star,\max,X}(>M_\star,z)
={}&
\epsilon_\star f_b\,
\rho_{h,X}
\left(
>\frac{M_\star}{\epsilon_\star f_b},
z
\right),
\\[-1mm]
& X\in\{\mathrm{ST},\mathrm{KST}\}.
\end{aligned}
\label{eq:sec3_stellar_density_ceiling}
\end{equation}

\begin{figure*}
    \centering
    \includegraphics[width=\textwidth]{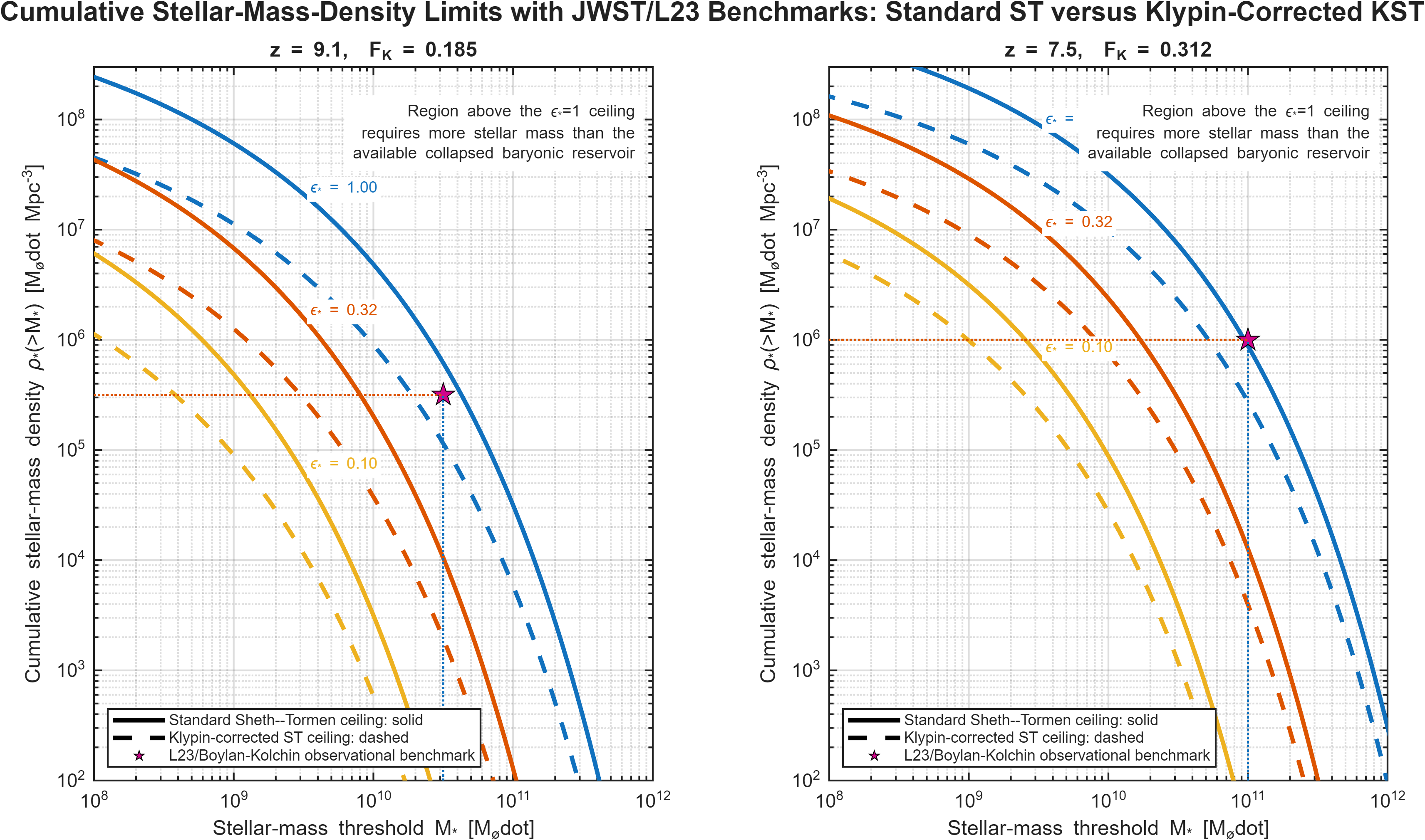}
    \caption{Cumulative matter-density and maximum stellar-mass-density
    constraints comparing standard ST and KST predictions at representative
    high redshifts.  The halo-mass reservoir is converted into
    stellar-mass-density ceilings using the same baryonic mapping for both HMF
    branches.  The $\epsilon_\star=1$ sequence gives the absolute
    baryon-conversion ceiling, while lower efficiencies require more massive
    and rarer host halos.  The Boylan--Kolchin/Labb\'e benchmark illustrates
    the observational massive-galaxy stress test.}
    \label{fig:cumulative_stellar_density}
\end{figure*}

Figure~\ref{fig:cumulative_stellar_density} applies this integral constraint at
representative high redshifts.  The halo-mass reservoir is converted into a
maximum stellar-mass density using the same baryonic prescription for both HMF
branches.  The $\epsilon_\star=1$ sequence again represents an absolute
baryonic ceiling: an observed stellar-mass density above this limit cannot be
accommodated under the simultaneous assumptions entering the calculation.  A
measurement below but close to the ceiling is less decisive, because modest
changes in stellar mass, redshift, completeness, survey volume, or the
stellar-to-halo mapping can shift the comparison appreciably.

The Boylan--Kolchin/Labb\'e benchmark provides a direct observational
comparison with the calculated mass reservoir.  The KST abundance reduction
propagates into the integrated halo-mass density before any baryonic conversion
efficiency is applied.  Consequently, its stellar-mass-density ceiling lies
below the corresponding ST prediction.  This result emphasizes that an
assessment of very massive high-redshift galaxies depends not only on the
available cosmic baryon fraction but also on the abundance calibration of the
rare host halos in which those baryons reside.

\subsection{Interpretation of the Massive-Galaxy Test}
\label{subsec:massive_galaxy_interpretation}

The massive-galaxy calculation can be summarized by the sequence
\begin{equation}
\begin{aligned}
M_\star
&\longrightarrow
M_{h,\min}
=\frac{M_\star}{\epsilon_\star f_b}
\\
&\longrightarrow
\left\{
n_X(>M_{h,\min},z),
\,
\rho_{h,X}(>M_{h,\min},z)
\right\}
\\
&\longrightarrow
\text{massive-galaxy consistency test}.
\end{aligned}
\label{eq:sec3_logic}
\end{equation}
The first step is controlled by the baryonic mapping; the second is controlled
by the cosmological halo population.  Treating these ingredients separately
makes clear why neither a large inferred stellar mass nor a low predicted halo
abundance is sufficient by itself to establish a cosmological inconsistency.

The calculations show that massive high-redshift galaxies are particularly
sensitive probes because their required hosts can lie on the exponentially
suppressed rare-halo tail.  In this regime the inferred degree of tension
depends jointly on the stellar-to-halo conversion efficiency and the adopted
HMF calibration.  Standard ST provides a comparatively larger high-redshift
halo reservoir, whereas the illustrative KST calibration lowers both the
cumulative number of suitable halos and the integrated collapsed mass
available to support their stellar populations.

These results are therefore best interpreted as quantitative stress tests of
the combined cosmological and baryonic model rather than as binary
falsification criteria.  A robust claim of inconsistency would require the
discrepancy to persist across physically plausible uncertainties in stellar
population modeling, stellar mass, redshift, lensing magnification,
completeness, survey volume, cosmic variance, halo-mass definition, and
baryonic conversion efficiency \citep{BoylanKolchin2023}.  Within the
controlled comparison developed here, the background cosmology and baryonic
mapping are held fixed between ST and KST, so their difference isolates the
sensitivity of the massive-galaxy constraint to the adopted halo-abundance
calibration.

The next application considers active galactic nuclei, for which the same
cosmological halo supply must be supplemented by black-hole occupation, duty
cycle, accretion physics, and the relation between black-hole properties and
host-halo mass.

\section{AGN Host-Abundance Test}
\label{sec:agn_abundance_test}

The second application of the halo-abundance framework concerns active
galactic nuclei (AGN) at high redshift.  The cosmological part of the problem
is set by the number of dark-matter halos available to host the observed
population.  The astrophysical part is more complex than for the
massive-galaxy test because an eligible halo need not contain a black hole of
the required mass, and a black hole that is present need not be observed in an
active accretion phase.  The predicted AGN abundance therefore depends on the
minimum host-halo mass, black-hole occupation fraction, AGN duty cycle, and the
mapping between AGN luminosity, black-hole mass, and halo mass.

The purpose of the calculation is to keep these ingredients explicit.  The HMF
determines the cosmological supply of potential hosts, while the occupation and
activity factors determine what fraction of those hosts can appear in a
selected AGN sample.  This provides a direct way to identify combinations of
host mass and AGN astrophysics that become increasingly demanding at early
cosmic times.

\subsection{AGN Halo Occupation}
\label{subsec:agn_occupation}

For a halo of mass $M_h$ at redshift $z$, the mean number of observable AGN
may be written
\begin{equation}
\begin{aligned}
\left\langle N_{\rm AGN}\mid M_h,z\right\rangle
={}&
f_{\rm occ}(M_h,z)\,
f_{\rm duty}(M_h,z)
\\
&\times
\left\langle
N_{\rm BH,eligible}\mid M_h,z
\right\rangle ,
\end{aligned}
\label{eq:agn_occupation_general}
\end{equation}
where $f_{\rm occ}$ is the fraction of eligible halos containing a black hole
capable of participating in the selected population, $f_{\rm duty}$ is the
fraction of those systems observed in the relevant active state, and
$\langle N_{\rm BH,eligible}\mid M_h,z\rangle$ is the mean number of black
holes satisfying the adopted eligibility conditions.

For the rare high-redshift systems considered here, a transparent limiting
model is the one-central-black-hole approximation,
\begin{equation}
\left\langle
N_{\rm BH,eligible}\mid M_h,z
\right\rangle
\simeq
\Theta(M_h-M_{\min}),
\label{eq:agn_central_bh}
\end{equation}
where $\Theta$ is the Heaviside step function and $M_{\min}$ is the minimum
host-halo mass associated with the selected AGN population.  This
approximation treats the abundance test in terms of one relevant central black
hole per eligible halo.

The corresponding AGN number density for HMF prescription $X$ is
\begin{equation}
\begin{aligned}
n_{{\rm AGN},X}(z)
={}&
\int_{M_{\min}}^\infty
f_{\rm occ}(M_h,z)
f_{\rm duty}(M_h,z)
\\
&\times
\frac{dn_X}{dM_h}(M_h,z)\,dM_h ,
\end{aligned}
\label{eq:agn_abundance_general}
\end{equation}
with $X\in\{\mathrm{ST},\mathrm{KST}\}$.  If $f_{\rm occ}$ and
$f_{\rm duty}$ are approximated as constants above $M_{\min}$, this becomes
\begin{equation}
n_{{\rm AGN},X}(z)
=
f_{\rm occ}f_{\rm duty}\,
n_X(>M_{\min},z).
\label{eq:agn_abundance_simple}
\end{equation}

Equation~\eqref{eq:agn_abundance_simple} separates the cosmological and
astrophysical factors particularly clearly.  The cumulative HMF specifies the
number of potential hosts, while $f_{\rm occ}f_{\rm duty}$ specifies the
fraction that appear in the selected active population.  Since both factors
are bounded above by unity,
\begin{equation}
0\leq f_{\rm occ}\leq1,
\qquad
0\leq f_{\rm duty}\leq1,
\label{eq:occ_duty_bounds}
\end{equation}
the cumulative halo abundance provides an absolute host-supply ceiling for a
specified $M_{\min}$.

The occupation fraction and duty cycle represent different physics.
Black-hole occupation depends on seeding, retention, mergers, and subsequent
black-hole growth, whereas the duty cycle describes the episodic nature of
accretion and depends on gas supply, angular-momentum transport, feedback,
environment, and the luminosity threshold defining the sample
\citep{Volonteri2010,Inayoshi2020}.  They should therefore not be interpreted
as interchangeable parameters.

\subsection{AGN Duty Cycle and the Host-Supply Ceiling}
\label{subsec:agn_duty_cycle}

In a simple time-domain interpretation, the duty cycle may be written
\begin{equation}
f_{\rm duty}
=
\frac{t_{\rm active}}{t_{\rm available}},
\label{eq:duty_cycle}
\end{equation}
where $t_{\rm active}$ is the time spent above the adopted activity or
luminosity threshold and $t_{\rm available}$ is the relevant comparison
interval.  The precise interpretation is sample dependent: the available
interval may correspond to a redshift-bin duration, a black-hole growth
interval, or another physically motivated timescale.

The limiting choice
\begin{equation}
f_{\rm duty}=1
\label{eq:duty_unity}
\end{equation}
assumes that every occupied and otherwise eligible system is observed in the
selected active state.  For a fixed host population this maximizes the
predicted AGN abundance and therefore provides the least restrictive
halo-supply bound.  If an observed abundance is difficult to accommodate even
under $f_{\rm occ}=f_{\rm duty}=1$, reducing either factor cannot alleviate
that particular host-supply constraint.

Conversely, the duty cycle cannot be interpreted independently of
$M_{\min}$.  The cumulative halo abundance can change extremely rapidly with
host mass in the rare-halo regime.  A moderate reduction in the inferred
minimum host mass can therefore increase the available halo population by a
large factor, while a smaller duty cycle reduces the fraction of that
population observed as AGN.  The two effects must be considered jointly.

For the ST and KST branches, identical occupation, duty-cycle, and host-mass
assumptions give
\begin{equation}
n_{{\rm AGN},{\rm KST}}(z)
=
F_{\rm K}(z)\,
n_{{\rm AGN},{\rm ST}}(z).
\label{eq:agn_kst_scaling}
\end{equation}
Thus the HMF calibration changes the cosmological supply of eligible hosts but
does not prescribe a different AGN lifetime, black-hole seeding efficiency, or
accretion history.

\subsection{AGN Halo-Supply Constraints}
\label{subsec:agn_halo_supply}

\begin{figure*}
    \centering
    \includegraphics[width=\textwidth]{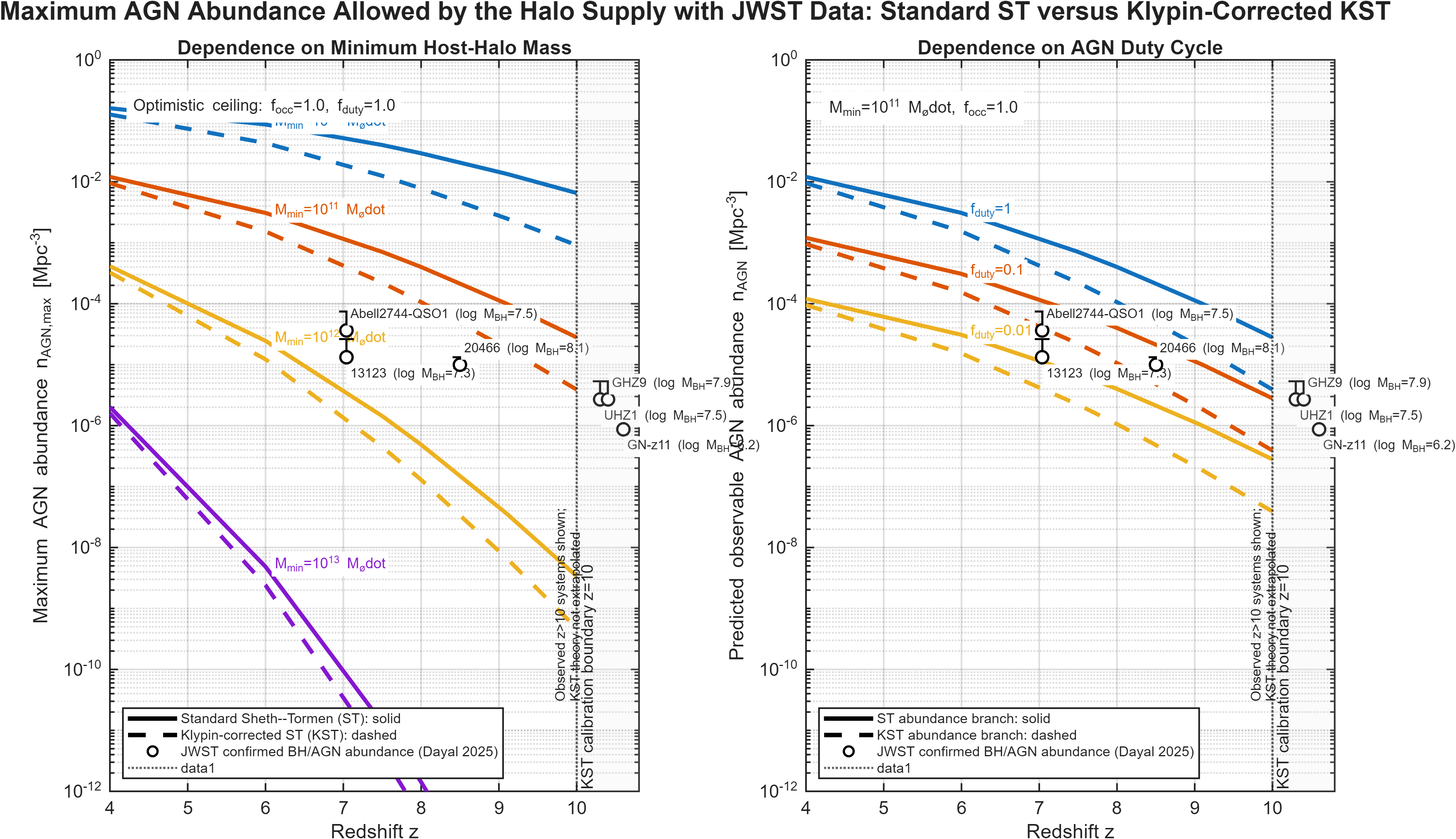}
\caption{
Maximum AGN abundance permitted by the available halo population as a
function of redshift. The curves show the cumulative abundance of halos
above the adopted minimum host-halo mass for the ST and KST halo mass
functions, thereby providing an upper limit on the abundance of AGN that
can be hosted by the corresponding halo population. The theoretical
halo-supply limits are compared with the high-redshift AGN abundance
constraints of \citet{Dayal2025}, based on the emerging JWST population
of faint and obscured AGN. The comparison illustrates whether the
observed AGN abundance can be accommodated by the available host-halo
population before introducing additional reductions associated with
AGN occupation fractions or duty cycles.
}
    \label{fig:agn_halo_supply}
\end{figure*}

Figure~\ref{fig:agn_halo_supply} shows the maximum AGN abundance permitted by
the halo population as a function of redshift, compared directly with the
high-redshift AGN abundance data of \citet{Dayal2025} plotted in the figure.  The left panel illustrates the
strong dependence on minimum host-halo mass under the maximally permissive
$f_{\rm occ}=f_{\rm duty}=1$ assumption.  If the selected AGN can occupy halos
with $M_h\sim10^{11}\,M_\odot$, the potential host population is much larger
than if halos of $10^{12}$--$10^{13}\,M_\odot$ are required.  The latter
rapidly enter the rare-halo tail toward high redshift.

The right panel shows the additional suppression associated with the duty
cycle for a representative host threshold.  For an observed AGN number
density $n_{\rm AGN,obs}$, Equation~\eqref{eq:agn_abundance_simple} implies the
necessary condition
\begin{equation}
f_{\rm duty}f_{\rm occ}
\geq
\frac{n_{\rm AGN,obs}}
{n_X(>M_{\min},z)}.
\label{eq:agn_required_fraction}
\end{equation}
Since $f_{\rm duty}f_{\rm occ}\leq1$, a specified host threshold cannot supply
the observed population within HMF branch $X$ if
\begin{equation}
n_{\rm AGN,obs}
>
n_X(>M_{\min},z).
\label{eq:agn_impossible_condition}
\end{equation}

Representative JWST high-redshift black-hole and AGN abundance estimates are
shown as observational benchmarks.  These points should not be interpreted as
direct measurements of host-halo mass.  Their conversion into a halo
constraint depends on black-hole masses, bolometric corrections, Eddington
ratios, AGN classification, selection functions, survey volume, completeness,
and the black-hole--halo relation.  The comparison instead asks whether a
specified set of host and activity assumptions provides a sufficiently large
cosmological reservoir of candidate halos.

\subsection{From AGN Luminosity to Black-Hole Mass}
\label{subsec:agn_luminosity_bh}

The halo abundance cannot be compared with a luminosity-selected AGN sample
until the observed luminosity is related to a characteristic black-hole and
host-halo mass.  For a black hole of mass $M_{\rm BH}$, the Eddington
luminosity is
\begin{equation}
L_{\rm Edd}
=
\frac{4\pi G M_{\rm BH}m_pc}{\sigma_{\rm T}},
\label{eq:eddington_luminosity}
\end{equation}
where $m_p$ is the proton mass and $\sigma_{\rm T}$ is the Thomson cross
section.  Numerically,
\begin{equation}
L_{\rm Edd}
\simeq
1.26\times10^{38}
\left(
\frac{M_{\rm BH}}{M_\odot}
\right)
{\rm erg\,s^{-1}}.
\label{eq:eddington_numeric}
\end{equation}

Defining the Eddington ratio
\begin{equation}
\lambda_{\rm Edd}
=
\frac{L_{\rm bol}}{L_{\rm Edd}},
\label{eq:eddington_ratio}
\end{equation}
the bolometric luminosity is
\begin{equation}
L_{\rm bol}
=
\lambda_{\rm Edd}L_{\rm Edd}.
\label{eq:bolometric_luminosity}
\end{equation}
A luminosity threshold $L_{\min}$ therefore corresponds to a minimum
black-hole mass
\begin{equation}
M_{{\rm BH},\min}
=
\frac{L_{\min}\sigma_{\rm T}}
{4\pi Gm_pc\,\lambda_{\rm Edd}}.
\label{eq:minimum_bh_mass}
\end{equation}

This relation illustrates an important degeneracy.  At fixed luminosity, a
smaller Eddington ratio requires a more massive black hole, whereas a source
radiating closer to the Eddington limit can produce the same luminosity with a
smaller $M_{\rm BH}$.  A luminosity-selected sample therefore does not define a
unique black-hole mass unless the Eddington-ratio distribution is specified.

\subsection{From Black-Hole Mass to Host-Halo Mass}
\label{subsec:bh_halo_mapping}

A second relation is required to map black-hole mass into host-halo mass.  For
a transparent phenomenological parameterization, define
\begin{equation}
\epsilon_{\rm BH}
=
\frac{M_{\rm BH}}{M_h},
\label{eq:bh_halo_ratio}
\end{equation}
so that
\begin{equation}
M_{\rm BH}
=
\epsilon_{\rm BH}M_h.
\label{eq:bh_halo_mapping}
\end{equation}
Here $\epsilon_{\rm BH}$ is an effective black-hole-to-halo mass ratio.  It is
not a radiative efficiency and is not assumed to be a universal constant.
Observed correlations between black-hole mass and host properties provide
physical motivation for such a mapping, but their form, scatter, and
redshift evolution remain important uncertainties
\citep{KormendyHo2013,Volonteri2010,Inayoshi2020}.

Combining Equations~\eqref{eq:minimum_bh_mass} and
\eqref{eq:bh_halo_mapping} gives the characteristic minimum host mass
\begin{equation}
M_{\min}
=
\frac{M_{{\rm BH},\min}}
{\epsilon_{\rm BH}},
\label{eq:minimum_halo_from_bh}
\end{equation}
or, explicitly,
\begin{equation}
\begin{aligned}
M_{\min}
={}&
\frac{L_{\min}\sigma_{\rm T}}
{4\pi Gm_pc}
\\
&\times
\frac{1}
{\lambda_{\rm Edd}\epsilon_{\rm BH}}.
\end{aligned}
\label{eq:minimum_halo_from_luminosity}
\end{equation}

Equation~\eqref{eq:minimum_halo_from_luminosity} makes explicit that the
required host mass varies inversely with both the assumed Eddington ratio
and the effective black-hole-to-halo mass ratio.

\begin{figure*}
    \centering
    \includegraphics[width=\columnwidth]{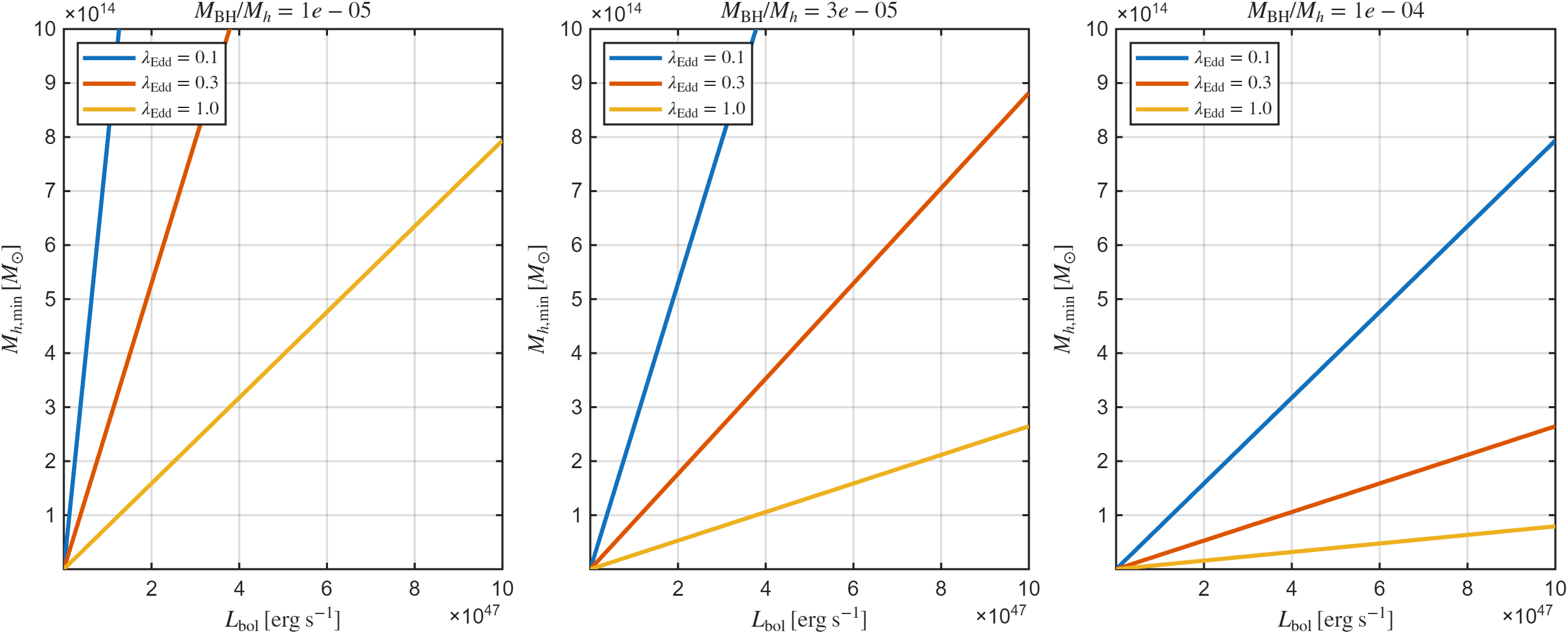}
    \caption{Mapping between AGN luminosity, black-hole mass, and
    characteristic host-halo mass.  The Eddington ratio
    $\lambda_{\rm Edd}$ converts luminosity into black-hole mass, while the
    effective ratio $\epsilon_{\rm BH}=M_{\rm BH}/M_h$ maps black-hole mass
    into a host requirement.  Different combinations of these parameters can
    therefore correspond to the same observed AGN luminosity.}
    \label{fig:agn_host_mapping}
\end{figure*}

Figure~\ref{fig:agn_host_mapping} illustrates this mapping.  A luminous AGN
can be associated with a larger black hole, a larger Eddington ratio, a larger
black-hole-to-halo mass ratio, or some combination of these quantities.
Consequently, luminosity alone does not uniquely determine $M_{\min}$.  Once a
mapping is specified, however, the cumulative HMF supplies a direct abundance
test of the required hosts.

\subsection{Clustering as an Independent Host-Mass Test}
\label{subsec:agn_clustering}

AGN clustering provides a complementary way to test the host masses inferred
from abundance arguments.  Massive halos correspond to rarer peaks and
generally have larger large-scale bias.  An AGN population assigned to very
massive halos should therefore exhibit stronger clustering than one occupying
more common, lower-mass halos \citep{Bardeen1986,ShethMoTormen2001}.

For HMF prescription $X$, the effective AGN bias can be written
\begin{equation}
\begin{aligned}
b_{{\rm AGN},X}(z)
=
\frac{
\displaystyle
\int_0^\infty
b_h(M_h,z)
\left\langle N_{\rm AGN}\mid M_h,z\right\rangle
\frac{dn_X}{dM_h}\,dM_h
}{
\displaystyle
\int_0^\infty
\left\langle N_{\rm AGN}\mid M_h,z\right\rangle
\frac{dn_X}{dM_h}\,dM_h
},
\end{aligned}
\label{eq:agn_effective_bias}
\end{equation}
where $b_h(M_h,z)$ is the linear halo bias.  For the threshold model with
mass-independent $f_{\rm occ}$ and $f_{\rm duty}$ above $M_{\min}$, these
constant factors cancel:
\begin{equation}
\begin{aligned}
b_{{\rm AGN},X}(z)
=
\frac{
\displaystyle
\int_{M_{\min}}^\infty
b_h(M_h,z)
\frac{dn_X}{dM_h}\,dM_h
}{
\displaystyle
\int_{M_{\min}}^\infty
\frac{dn_X}{dM_h}\,dM_h
}.
\end{aligned}
\label{eq:agn_bias_threshold}
\end{equation}

The abundance and clustering observables consequently constrain different
aspects of the AGN--halo connection.  The abundance measures how many eligible
hosts exist, whereas the bias measures the characteristic rarity of those
hosts.  Increasing $M_{\min}$ lowers the available host abundance while
generally increasing the predicted bias.  A clustering measurement can
therefore test whether the characteristic halo masses required by an abundance
model are plausible.

For the particular KST prescription used here, the factor $F_{\rm K}(z)$ is
independent of halo mass at fixed redshift.  If the same halo-bias relation and
the same occupation model are applied to both HMF branches, this common factor
cancels between the numerator and denominator of
Equation~\eqref{eq:agn_effective_bias}.  Under these assumptions,
\begin{equation}
b_{{\rm AGN},{\rm KST}}(z)
=
b_{{\rm AGN},{\rm ST}}(z).
\label{eq:agn_bias_kst}
\end{equation}
Thus the KST modification changes the predicted AGN abundance but does not by
itself generate a different normalized clustering bias.  A difference in
clustering would require a mass-dependent HMF modification, a different
halo-bias calibration, or a change in the inferred occupation or host-mass
distribution.

\subsection{Interpretation of the AGN Abundance Test}
\label{subsec:agn_interpretation}

The AGN calculation can be summarized by two coupled mappings.  The first
connects an observed AGN property to a characteristic host requirement,
\begin{equation}
\begin{aligned}
L_{\rm bol}
&\longrightarrow
M_{\rm BH}
\longrightarrow
M_{\min},
\end{aligned}
\label{eq:agn_property_chain}
\end{equation}
where the conversion depends principally on $\lambda_{\rm Edd}$ and
$\epsilon_{\rm BH}$.  The second maps the host requirement into an observable
population,
\begin{equation}
\begin{aligned}
M_{\min}
&\longrightarrow
n_X(>M_{\min},z)
\\
&\longrightarrow
f_{\rm occ}f_{\rm duty}\,
n_X(>M_{\min},z)
\\
&\longrightarrow
n_{{\rm AGN},X}(z).
\end{aligned}
\label{eq:agn_abundance_chain}
\end{equation}

These relations make clear why an observed AGN abundance does not uniquely
determine a host-halo mass.  A low number density may reflect rare massive
hosts, a small black-hole occupation fraction, a short duty cycle, or a
combination of these effects.  Conversely, explaining a comparatively large
AGN abundance with very massive hosts can require occupation and duty-cycle
factors approaching their physical upper limits.

The high-redshift comparison is therefore a conditional halo-supply test.  A
deficit of predicted AGN relative to an observed abundance can arise from the
adopted HMF, the inferred minimum host mass, black-hole occupation, duty cycle,
the black-hole--halo relation, the Eddington-ratio distribution, or
observational selection.  The HMF constrains the cosmological reservoir, but
does not by itself provide a complete AGN luminosity function.

The quantitative KST abundance comparison is restricted to the calibration
interval $z\leq10$ defined in Section~\ref{subsubsec:kst_scope}.  Sources at
higher redshift remain valuable for testing the luminosity--black-hole--halo
mapping, but KST abundance estimates there are extrapolative.  Within the
calibrated interval, comparing ST and KST with identical occupation and
activity assumptions isolates the sensitivity of the AGN host-supply
constraint to the adopted halo-abundance calibration.

A more complete forward model would replace the single host threshold and
constant occupation factors with distributions of black-hole mass, Eddington
ratio, obscuration, variability, host-halo mass, and survey selection.  The
present framework provides the analytic baseline for such an extension by
showing explicitly where each of these ingredients enters the abundance
calculation.

The next section considers the third observational application: the
high-redshift galaxy UV luminosity function.  In that calculation the HMF is
mapped into luminosity space through stellar and baryonic physics rather than
through the AGN occupation and duty-cycle factors defined here.

\section{High-Redshift UV Luminosity-Function Test}
\label{sec:uv_luminosity_test}

The third observational application connects the cosmological halo population
to the rest-frame ultraviolet (UV) luminosities measured for galaxies at
cosmic dawn.  The halo mass function determines the abundance of collapsed
dark-matter structures, whereas observations measure the radiation emitted by
the baryonic component.  A comparison between the two therefore requires two
separate ingredients: a mapping from halo mass to UV luminosity and a
calculation of the abundance of halos that populate each luminosity interval.

For a halo of mass $M_h$, the first ingredient may be written schematically as
\begin{equation}
M_h
\longrightarrow
M_{\rm UV}(M_h,z),
\label{eq:sec5_halo_to_uv}
\end{equation}
whereas the second is supplied by the differential HMF,
\begin{equation}
\frac{dn_X}{dM_h}(M_h,z),
\qquad
X\in\{\mathrm{ST},\mathrm{KST}\}.
\label{eq:sec5_hmf_input}
\end{equation}
Combining these quantities produces the theoretical UV luminosity function
$\Phi_X(M_{\rm UV},z)$ that can be compared directly with observed galaxy
number densities.

The distinction between luminosity and abundance is central to the
interpretation.  A baryonic model may allow a halo of a particular mass to
produce a luminous galaxy, but population consistency additionally requires
that halos of the required mass exist in sufficient numbers.  Conversely, an
adequate halo supply does not guarantee agreement with observations if the
adopted baryonic prescription produces galaxies that are systematically too
faint.  The calculation below therefore keeps the halo-to-light physics and
the cosmological abundance calculation explicit.

The observational motivation follows the rapidly developing JWST view of
cosmic dawn, including massive galaxies, luminous UV sources, and candidate
AGN at high redshift
\citep{Labbe2023,BoylanKolchin2023,Harikane2023,Atek2023a,Atek2023b,
Maiolino2024,Franco2025,Donnan2023,Donnan2024}.  The model is intended as a
controlled first-order description of the connection between halo supply,
baryonic conversion efficiency, and UV luminosity.

\subsection{Baryonic Conversion and Stellar Mass}
\label{subsec:sec5_baryonic_conversion}

For a halo of mass $M_h$, the baryonic mass associated with the cosmic mean
baryon fraction is
\begin{equation}
M_b=f_bM_h,
\label{eq:sec5_baryonic_mass}
\end{equation}
where
\begin{equation}
f_b=\frac{\Omega_b}{\Omega_m}\simeq0.156.
\label{eq:sec5_fb}
\end{equation}
The stellar mass is parameterized as
\begin{equation}
M_\star
=
\epsilon_\star f_bM_h,
\label{eq:sec5_stellar_mass}
\end{equation}
where $\epsilon_\star$ is an effective integrated baryon-to-star conversion
efficiency.  The numerical calculations consider
\begin{equation}
\epsilon_\star
=
0.03,\;0.10,\;0.20,
\label{eq:sec5_efficiencies}
\end{equation}
representing illustrative low, intermediate, and relatively high conversion
efficiencies \citep{Behroozi2013,Moster2013}.

These values are not intended to define a universal efficiency law.  In
general, $\epsilon_\star$ may depend on halo mass, redshift, assembly history,
feedback, and environment.  Here it serves as a transparent parameter for
examining how uncertainty in baryonic conversion propagates into the
halo-to-UV mapping.

\subsection{Star-Formation Rate and Rest-Frame UV Luminosity}
\label{subsec:sec5_sfr_uv}

The stellar mass is converted into a characteristic star-formation rate using
the timescale
\begin{equation}
t_\star(z)
=
0.1H^{-1}(z).
\label{eq:sec5_starformation_timescale}
\end{equation}
The corresponding star-formation rate is
\begin{equation}
{\rm SFR}(M_h,z)
=
\frac{\epsilon_\star f_bM_h}
{t_\star(z)}.
\label{eq:sec5_sfr}
\end{equation}
This prescription captures the approximate shortening of characteristic
dynamical timescales toward high redshift without attempting to describe
individual star-formation histories.

The rest-frame UV spectral luminosity is taken to be proportional to the
star-formation rate,
\begin{equation}
L_{\nu,{\rm UV}}
=
\kappa_{\rm UV}\,{\rm SFR},
\label{eq:sec5_uv_luminosity}
\end{equation}
with
\begin{equation}
\kappa_{\rm UV}
=
8\times10^{27}
\frac{{\rm erg\,s^{-1}\,Hz^{-1}}}
{M_\odot\,{\rm yr^{-1}}}.
\label{eq:sec5_kappa_uv}
\end{equation}
The UV conversion factor is an astrophysical calibration and may depend on the
stellar initial mass function, metallicity, stellar population age, and
star-formation history.  It should therefore be regarded as part of the
halo-to-light model rather than as a cosmological prediction
\citep{Bouwens2021,FinkelsteinBagley2022,Napolitano2025}.

The corresponding AB absolute magnitude is
\begin{equation}
\begin{aligned}
M_{\rm UV}
=
51.60
-2.5\log_{10}
\left(
\frac{L_{\nu,{\rm UV}}}
{{\rm erg\,s^{-1}\,Hz^{-1}}}
\right).
\end{aligned}
\label{eq:sec5_ab_magnitude}
\end{equation}
Equations~\eqref{eq:sec5_sfr}--\eqref{eq:sec5_ab_magnitude} define the
deterministic mapping
\begin{equation}
M_{\rm UV}
=
M_{\rm UV}(M_h,z;\epsilon_\star).
\label{eq:sec5_deterministic_mapping}
\end{equation}

At fixed redshift, increasing either $M_h$ or $\epsilon_\star$ increases the
UV luminosity and produces a more negative $M_{\rm UV}$.  At fixed halo mass
and efficiency, redshift enters through $H(z)$ in the adopted
star-formation timescale.  The mapping itself is independent of the HMF
prescription: ST and KST assign the same luminosity to a halo of specified
$M_h$, $z$, and $\epsilon_\star$.  Their distinction arises only when the
abundance of such halos is evaluated.

\subsection{JWST Redshift--UV Luminosity Plane}
\label{subsec:sec5_jwst_synthesis}

\begin{figure*}
    \centering
    \includegraphics[width=\textwidth]{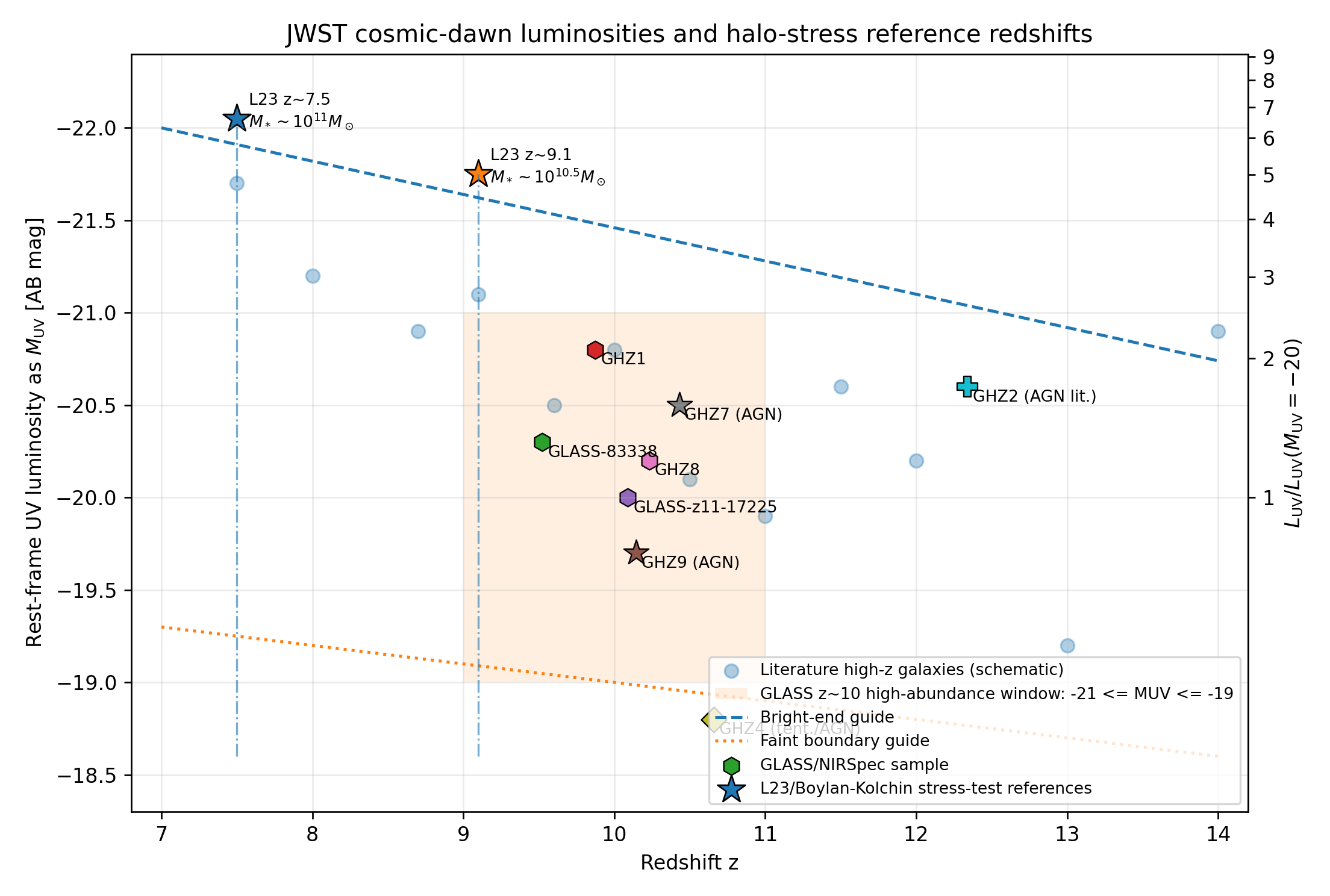}
    \caption{Synthesis of cosmic-dawn rest-frame UV luminosities in the
    redshift--$M_{\rm UV}$ plane.  Representative high-redshift JWST galaxies
    and candidate AGN illustrate the luminosity and redshift regime relevant
    to the halo-to-light calculation.  Brighter sources correspond to more
    negative $M_{\rm UV}$
    \citep{Labbe2023,BoylanKolchin2023,Harikane2023,Atek2023a,Atek2023b,
    Maiolino2024}.}
    \label{fig:jwst_uv_synthesis}
\end{figure*}

Figure~\ref{fig:jwst_uv_synthesis} places representative high-redshift JWST
galaxies and candidate AGN in the redshift--$M_{\rm UV}$ plane.  Brighter
sources have more negative UV absolute magnitudes.  The observed population
spans a regime in which luminous systems are already present at early cosmic
times, motivating a direct connection between their measured UV luminosities
and the halo masses capable of supporting them
\citep{Labbe2023,BoylanKolchin2023,Harikane2023,Atek2023a,Atek2023b,
Maiolino2024}.

The observational plane raises two related questions.  First, what
combinations of halo mass and baryonic conversion efficiency can produce the
measured luminosities?  Second, how abundant are halos satisfying those
requirements at the corresponding redshifts?  The first question is addressed
by the halo-to-UV tracks below; the second requires transformation of the HMF
into a UV luminosity function.

\subsection{Halo-to-UV Luminosity Tracks}
\label{subsec:sec5_uv_tracks}

\begin{figure*}
    \centering
    \includegraphics[width=\textwidth]{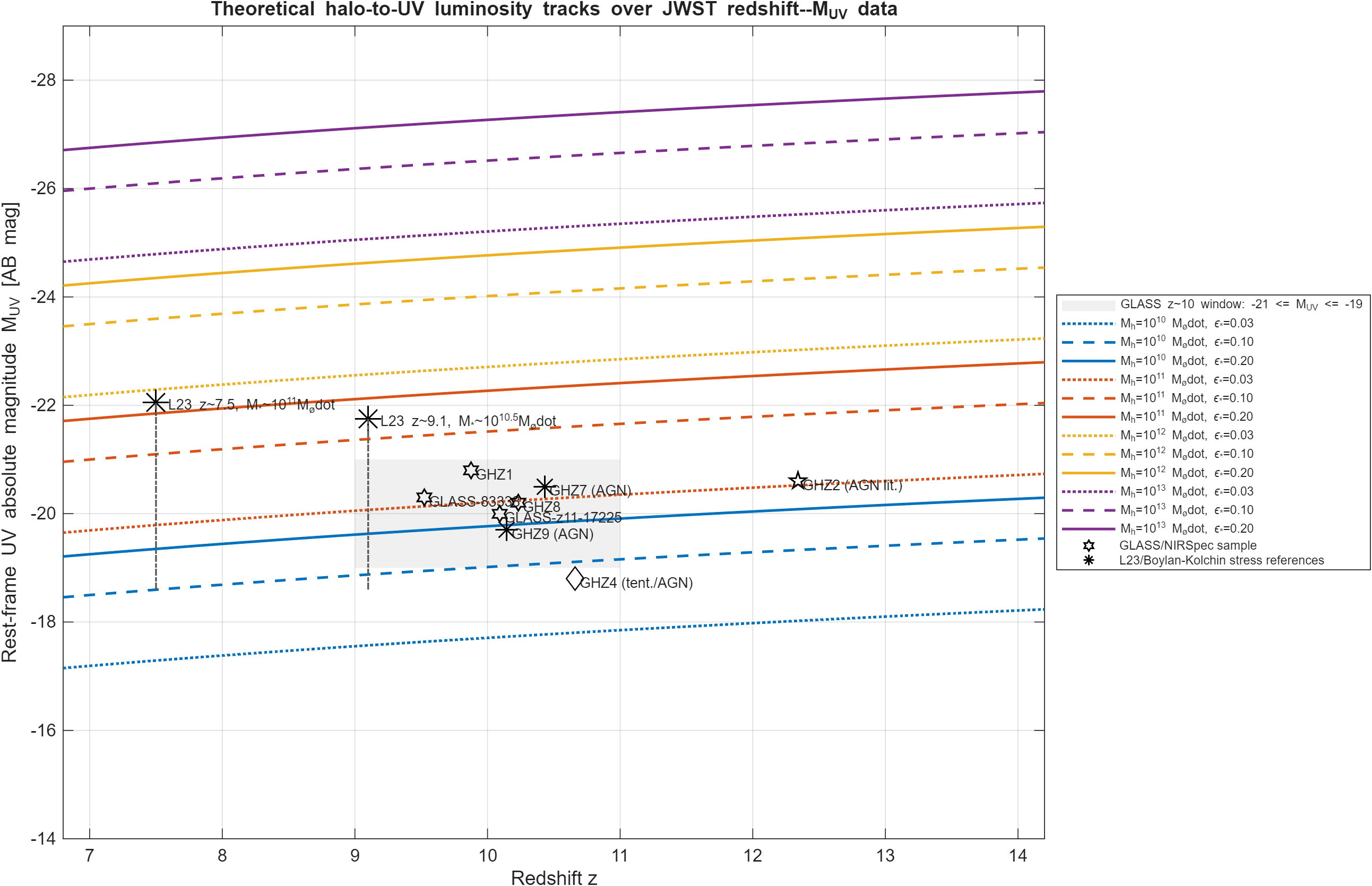}
\caption{
Halo-to-UV luminosity tracks in the redshift--$M_{\rm UV}$ plane for
$M_h=10^{10},\,10^{11},\,10^{12}$, and $10^{13}\,M_\odot$ and
$\epsilon_\star=0.03,\,0.10$, and $0.20$.
For each halo mass, a common colour is used, while dotted, dashed, and
solid curves denote $\epsilon_\star=0.03$, $0.10$, and $0.20$,
respectively.
The calculation uses $M_b=f_bM_h$,
${\rm SFR}=\epsilon_\star f_bM_h/t_\star$,
$t_\star=0.1H^{-1}(z)$, and
$L_{\nu,{\rm UV}}=\kappa_{\rm UV}{\rm SFR}$.
The theoretical tracks are compared with the high-redshift
GLASS/NIRSpec sources of \citet{Napolitano2025} and with the massive
galaxy candidates reported by \citet{Labbe2023} (L23), whose halo-abundance
and baryonic-reservoir implications were examined by
\citet{BoylanKolchin2023}.
Increasing halo mass or stellar-conversion efficiency produces a
brighter rest-frame UV magnitude, allowing the observed JWST sources
to be placed directly in the context of the halo-to-light mapping.
}
    \label{fig:halo_uv_tracks}
\end{figure*}

Figure~\ref{fig:halo_uv_tracks} evaluates
Equation~(88) for representative halo masses
\begin{equation}
M_h = 10^{10},\,10^{11},\,10^{12},\,10^{13}\,M_\odot
\label{eq:representative_halo_masses}
\end{equation}
and the efficiencies in Equation~(82). The resulting halo-to-UV
luminosity tracks are compared with the spectroscopically confirmed
high-redshift GLASS/NIRSpec sources of \citet{Napolitano2025} and with
the massive-galaxy benchmarks reported by \citet{Labbe2023} (L23),
whose implications for the available halo and baryonic reservoirs were
examined by \citet{BoylanKolchin2023}. These comparisons place the
simple halo-to-light mapping in direct observational context across the
JWST redshift range.  Increasing
halo mass produces a brighter UV source, while increasing
$\epsilon_\star$ allows a lower-mass halo to reach a comparable luminosity.

A measured $M_{\rm UV}$ therefore does not uniquely determine $M_h$.  A
luminous source may correspond to a comparatively massive halo with moderate
conversion efficiency, a less massive halo undergoing unusually efficient or
bursty star formation, or a UV continuum containing an AGN contribution.
Stellar population age, metallicity, dust attenuation, and star-formation
history introduce further uncertainty in the inversion from luminosity to
halo mass.

The tracks thus define the conditional luminosity associated with a halo
under the adopted baryonic model.  The probability of observing such a system
depends additionally on the abundance of halos at the corresponding mass and
redshift.

\subsection{From the Halo Mass Function to the UV Luminosity Function}
\label{subsec:sec5_uvlf}

The population-level prediction follows by combining the deterministic
halo-to-light relation with the HMF.  For a monotonic mapping, conservation of
number gives
\begin{equation}
\begin{aligned}
\Phi_X(M_{\rm UV},z)\,dM_{\rm UV}
={}&
\frac{dn_X}{dM_h}(M_h,z)
\left|dM_h\right|.
\end{aligned}
\label{eq:sec5_number_conservation}
\end{equation}
Hence,
\begin{equation}
\begin{aligned}
\Phi_X(M_{\rm UV},z)
={}&
\left.
\frac{dn_X}{dM_h}(M_h,z)
\left|
\frac{dM_h}{dM_{\rm UV}}
\right|
\right|_{M_h=M_h(M_{\rm UV},z)},
\end{aligned}
\label{eq:sec5_uvlf_general}
\end{equation}
where $\Phi_X$ has units of comoving number density per magnitude.

For the adopted linear luminosity scaling,
\begin{equation}
L_{\nu,{\rm UV}}\propto M_h,
\label{eq:sec5_luv_mass}
\end{equation}
and the magnitude relation can be written
\begin{equation}
M_{\rm UV}
=
C(z,\epsilon_\star)
-
2.5\log_{10}M_h,
\label{eq:sec5_muv_logmh}
\end{equation}
where $C(z,\epsilon_\star)$ contains the baryon fraction, conversion
efficiency, star-formation timescale, UV calibration, and unit normalization.
It follows that
\begin{equation}
\frac{dM_{\rm UV}}{dM_h}
=
-\frac{2.5}{\ln10}\frac{1}{M_h},
\label{eq:sec5_dmuv_dmh}
\end{equation}
and therefore
\begin{equation}
\left|
\frac{dM_h}{dM_{\rm UV}}
\right|
=
\frac{\ln10}{2.5}M_h.
\label{eq:sec5_dmh_dmuv}
\end{equation}

The deterministic luminosity function can consequently be written in the
compact form
\begin{equation}
\begin{aligned}
\Phi_X(M_{\rm UV},z)
={}&
\left.
\frac{\ln10}{2.5}\,
M_h
\frac{dn_X}{dM_h}(M_h,z)
\right|_{M_h=M_h(M_{\rm UV},z)} .
\end{aligned}
\label{eq:sec5_uvlf_compact}
\end{equation}

Since the same halo-to-light mapping is used for ST and KST and the adopted
KST correction is mass independent at fixed redshift,
\begin{equation}
\Phi_{\rm KST}(M_{\rm UV},z)
=
F_{\rm K}(z)
\Phi_{\rm ST}(M_{\rm UV},z).
\label{eq:sec5_uvlf_kst}
\end{equation}
This relation separates two physically different model dependencies.
Changing $\epsilon_\star$ changes the halo mass associated with a given
$M_{\rm UV}$ and therefore shifts the mapping through luminosity space.  The
KST factor instead changes the number of halos available at that mass.  A
change in star-formation efficiency and a change in HMF calibration are
therefore not equivalent effects.

\subsection{Comparison with Observed UV Luminosity Functions}
\label{subsec:sec5_uvlf_observations}

\begin{figure*}
    \centering
    \includegraphics[width=\textwidth]{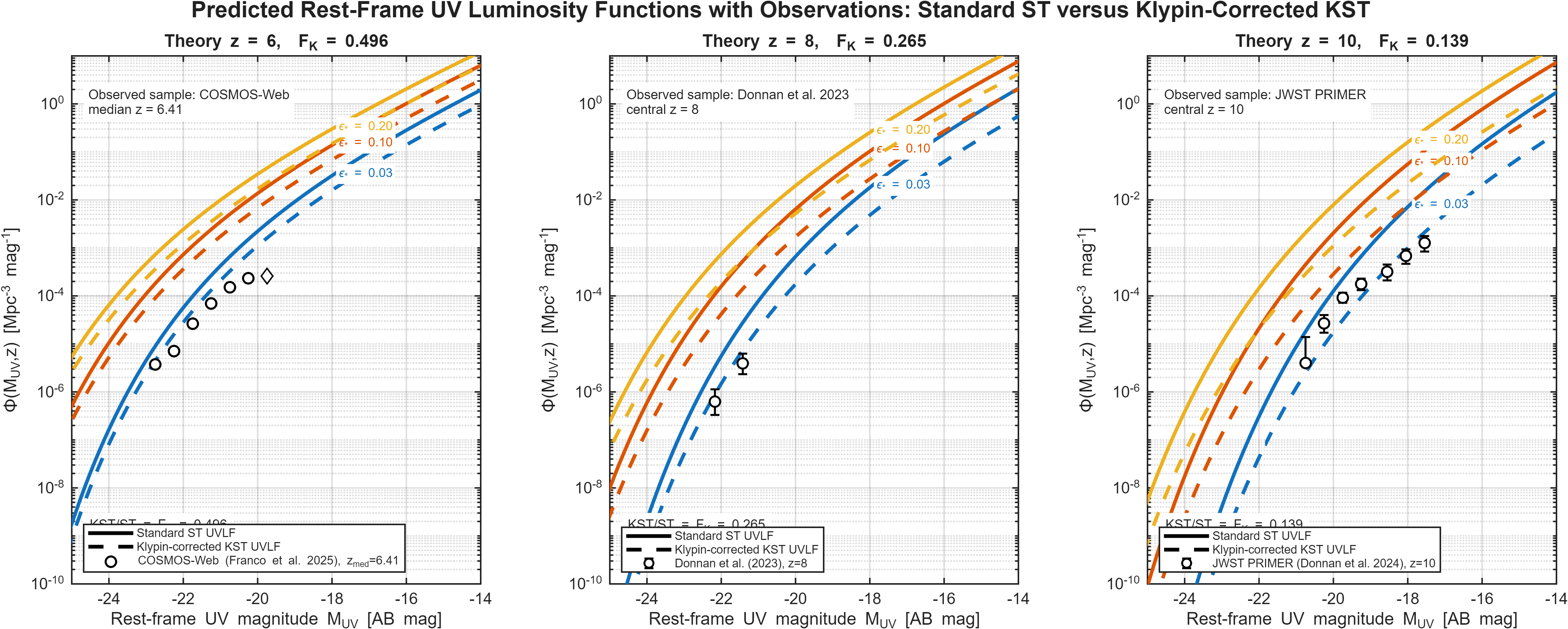}
    \caption{Predicted rest-frame UV luminosity functions compared with
    representative measurements at $z\simeq6$--10.  Solid curves show the
    standard Sheth--Tormen predictions and dashed curves show the corresponding
    KST predictions for $\epsilon_\star=0.03$, 0.10, and 0.20.  The $z=6$
    calculation is compared with COSMOS-Web measurements at
    $z_{\rm med}\simeq6.41$ \citep{Franco2025}; the $z=8$ calculation is
    compared with \citet{Donnan2023}; and the $z=10$ calculation is compared
    with \citet{Donnan2024}.  The increasing ST--KST separation toward high
    redshift reflects the growing sensitivity of the predicted UV-source
    abundance to the adopted HMF calibration.}
    \label{fig:uvlf_st_kst}
\end{figure*}

Figure~\ref{fig:uvlf_st_kst} compares the predicted rest-frame UV luminosity
functions at $z=6$, 8, and 10 with representative high-redshift measurements.
The model sequences use $\epsilon_\star=0.03$, 0.10, and 0.20 for both HMF
branches.

At the lower-redshift end, the COSMOS-Web determination of
\citet{Franco2025}, with median redshift $z_{\rm med}\simeq6.41$, provides a
nearby empirical benchmark for the $z=6$ theoretical calculation.  The
difference in redshift is relevant because the abundance of massive halos
evolves rapidly during this epoch.  At $z\simeq8$, the comparison uses the
measurements of \citet{Donnan2023}, while the $z\simeq10$ panel uses the JWST
determination of \citet{Donnan2024}.

The separation between the ST and KST predictions grows toward high redshift
according to the correction defined in Section~\ref{subsec:kst}.  Because the
halo-to-light relation is common to both branches, this separation directly
measures the effect of the different halo abundance assigned to the same
baryonic mapping.  The result demonstrates that HMF calibration can become an
important component of the predicted bright-galaxy abundance at cosmic dawn.

The dependence on $\epsilon_\star$ has a different origin.  Increasing
$\epsilon_\star$ makes a halo of fixed mass brighter and therefore allows a
specified observed $M_{\rm UV}$ to be associated with a lower halo mass.
Since lower-mass halos are more numerous, this can substantially increase the
predicted luminosity-function amplitude at a fixed magnitude.  The
high-redshift UV luminosity function therefore tests the combination of
baryonic conversion and cosmological halo supply rather than either ingredient
in isolation.

\subsection{Scatter in the Halo-to-Light Relation}
\label{subsec:sec5_scatter}

The deterministic calculation assigns one UV magnitude to each halo mass at a
fixed redshift.  Real galaxies exhibit a distribution of luminosities at fixed
$M_h$ because of variations in star-formation history, gas accretion,
metallicity, dust attenuation, feedback, merger history, halo assembly, and
possible AGN contamination.  A more general formulation introduces a
conditional probability distribution
\begin{equation}
P(M_{\rm UV}\mid M_h,z),
\label{eq:sec5_conditional_probability}
\end{equation}
normalized according to
\begin{equation}
\int_{-\infty}^{+\infty}
P(M_{\rm UV}\mid M_h,z)\,dM_{\rm UV}
=
1.
\label{eq:sec5_probability_normalization}
\end{equation}
The luminosity function then becomes
\begin{equation}
\begin{aligned}
\Phi_X(M_{\rm UV},z)
={}&
\int_0^\infty
\frac{dn_X}{dM_h}(M_h,z)
\\
&\times
P(M_{\rm UV}\mid M_h,z)\,dM_h .
\end{aligned}
\label{eq:sec5_uvlf_scatter}
\end{equation}
The deterministic model corresponds to the limiting case
\begin{equation}
\begin{aligned}
P(M_{\rm UV}\mid M_h,z)
\rightarrow
\delta_{\rm D}
\left[
M_{\rm UV}
-
M_{\rm UV}(M_h,z)
\right].
\end{aligned}
\label{eq:sec5_delta_mapping}
\end{equation}

Scatter can be particularly important at the bright end because the HMF
declines steeply with increasing mass.  Upward luminosity fluctuations from
the more numerous lower-mass halo population can therefore increase the
number of bright sources relative to a one-to-one mapping.  A full
statistical comparison should consequently include scatter together with
redshift uncertainties, completeness, cosmic variance, dust attenuation,
bursty star formation, and survey selection.

\subsection{Interpretation of the UV Luminosity-Function Test}
\label{subsec:sec5_uvlf_interpretation}

The complete observational mapping may be summarized as
\begin{equation}
\begin{aligned}
{\rm JWST}\;(z,M_{\rm UV})
&\longrightarrow
M_h(M_{\rm UV},z;\epsilon_\star)
\\
&\longrightarrow
\frac{dn_X}{dM_h}
\longrightarrow
\Phi_X(M_{\rm UV},z).
\end{aligned}
\label{eq:sec5_observational_chain}
\end{equation}
The first step depends on baryonic and stellar-population physics; the second
depends on the cosmological halo population.  This separation explains why
matching the luminosity of an individual source is necessary but not
sufficient for population consistency.

The UV luminosity-function comparison extends the halo-supply analysis from
individual massive or luminous objects to the population level.  The
measurements near $z\simeq6.4$, 8, and 10
\citep{Franco2025,Donnan2023,Donnan2024} probe a regime in which both the
halo-to-light conversion and the abundance of rare halos can materially affect
the prediction.  Within the deterministic model, the ST--KST difference is
entirely an abundance effect, while variation of $\epsilon_\star$ changes the
mapping between halo mass and observed magnitude.

The comparison is not a formal goodness-of-fit analysis.  The theoretical
curves use a deliberately simple halo-to-light relation, whereas the
observational measurements are affected by selection functions, photometric
and spectroscopic redshift uncertainties, completeness, survey volume, cosmic
variance, and possible AGN contributions to the UV continuum.  On the
theoretical side, stellar-population modeling, metallicity, dust attenuation,
feedback, bursty star formation, and scatter in the halo--luminosity relation
can all modify the predicted luminosity function.

The appropriate interpretation is therefore a quantitative test of the
combined cosmological and astrophysical framework.  A discrepancy between a
single deterministic curve and the measured luminosity function does not by
itself identify its origin as cosmological.  Conversely, adjusting the
halo-to-light mapping cannot create additional dark-matter halos; the
cosmological supply remains constrained by the HMF.  The most informative
comparison is consequently one in which uncertainties in both the baryonic
mapping and the halo abundance are propagated together.

The quantitative KST luminosity-function comparison is restricted to the
calibration interval $z\leq10$ specified in
Section~\ref{subsubsec:kst_scope}.  Halo-to-light relations and observational
redshift--luminosity diagrams may be considered at higher redshift, but a KST
abundance prediction beyond this boundary is an extrapolation rather than a
simulation-calibrated result.

Together with the massive-galaxy and AGN abundance tests, the UV luminosity
function completes the three observational applications of the common halo
framework.  The resulting comparison separates the cosmological abundance of
potential hosts from the baryonic and black-hole processes that determine
which of those hosts become observable at cosmic dawn.

\section{Discussion and Cosmological Implications}
\label{sec:discussion}

The high-redshift galaxy and AGN populations observed by JWST provide a
sensitive test of the connection between cosmological structure formation and
the baryonic processes that make dark-matter halos observable.  The three
applications considered in this work---massive galaxies, AGN host abundance,
and the rest-frame UV luminosity function---probe different observables, but
they share the same underlying question: does the cosmological halo population
provide a sufficiently large reservoir of hosts once the mapping from halo
mass to the observed population is specified?

This distinction separates the problem into two coupled components.  The
cosmological component determines the abundance and mass distribution of
halos.  The astrophysical component determines how stellar mass, black-hole
activity, and UV luminosity are assigned to those halos.  At low redshift,
moderate changes in either component may have relatively limited effects on
predicted number densities.  At cosmic dawn, however, massive systems occupy
the rare, high-$\nu$ tail of the halo distribution, where the abundance
changes rapidly with both halo mass and redshift.  Uncertainties in the
observable-to-halo mapping and in the HMF calibration can therefore be
strongly amplified.

The standard Sheth--Tormen (ST) HMF provides the principal analytic baseline
throughout the analysis.  The Bolshoi/Klypin-corrected ST (KST) prescription
provides an illustrative simulation-motivated test of the sensitivity to
high-redshift HMF calibration.  Since the KST factor in the adopted
implementation is independent of halo mass at fixed redshift, it changes the
available halo abundance without changing the common power spectrum, mass
variance, peak-height calculation, or the baryonic and luminosity mappings.
This separation makes it possible to identify which conclusions arise from
the cosmological halo supply and which arise from the astrophysical mapping.

\subsection{Rare Halos and the High-Redshift Host Supply}
\label{subsec:discussion_rare_halos}

The mass variance and peak-height calculations show that increasingly massive
halos at high redshift correspond to increasingly rare fluctuations in the
primordial density field.  This is the fundamental reason that the earliest
massive galaxies and luminous AGN can provide stringent tests of structure
formation.  Once a source is associated with a minimum host mass $M_{\min}$,
the relevant cosmological quantity is the cumulative abundance
\begin{equation}
n_X(>M_{\min},z),
\qquad
X\in\{\mathrm{ST},\mathrm{KST}\}.
\label{eq:discussion_host_supply}
\end{equation}

The steepness of the rare-halo tail has an important consequence: a modest
change in the inferred host mass can produce a much larger change in the
predicted abundance.  The observational significance of an apparently extreme
source therefore cannot be assessed from its luminosity or stellar mass
alone.  It depends on how those quantities map into halo mass and on the HMF
used to calculate the abundance of such halos.

Within the calibrated redshift interval, the KST prescription lowers the
high-redshift halo supply relative to the ST baseline.  The effect becomes
progressively more important toward cosmic dawn.  Because the correction has
already been defined quantitatively in Section~\ref{subsec:kst}, it is useful
here to emphasize its interpretation rather than repeat its numerical
definition: the KST branch represents a controlled change in halo abundance,
not a modification of the underlying linear density field or of the
halo-to-observable physics.

This distinction is especially relevant when considering the most extreme
objects.  A large stellar mass, black-hole mass, or UV luminosity may imply a
rare host under one astrophysical mapping but a substantially more common host
under another.  Conversely, for a fixed host requirement, different HMF
calibrations can predict materially different numbers of eligible systems.
High-redshift abundance tests therefore constrain the combination of these
ingredients rather than either one in isolation.

\subsection{Massive Galaxies as Baryonic-Reservoir Tests}
\label{subsec:discussion_massive_galaxies}

The massive-galaxy application is governed first by the finite baryonic
reservoir associated with a halo.  The relation
\begin{equation}
M_\star
\leq
\epsilon_\star f_bM_h
\label{eq:discussion_baryon_ceiling}
\end{equation}
provides a direct consistency condition between an inferred stellar mass and
its host.  The limiting value $\epsilon_\star=1$ represents an absolute
baryonic ceiling; realistic stellar-conversion efficiencies are expected to
be smaller.

The importance of this constraint increases at high redshift because the halo
mass required to support a large stellar mass may lie on the rapidly declining
high-mass tail.  Lowering $\epsilon_\star$ raises the required host mass,
thereby reducing the number of eligible halos.  The abundance constraint is
therefore not determined by the baryon fraction alone: it depends jointly on
the stellar-mass estimate, the adopted conversion efficiency, and the
cosmological abundance of the resulting host population
\citep{Labbe2023,BoylanKolchin2023}.

The cumulative mass-density test provides a complementary integral
constraint.  Even if an individual halo can in principle contain the required
stellar mass, the total collapsed matter contained in all eligible halos must
be sufficient to support the observed stellar-mass density.  The individual
host test and the integrated mass-reservoir test thus probe different aspects
of the same physical requirement.

These considerations also illustrate why the most massive JWST systems should
be treated as quantitative stress tests rather than as binary cosmological
falsification criteria.  Stellar masses at these redshifts depend on
stellar-population synthesis, star-formation history, dust, nebular emission,
redshift, and possible AGN contamination.  Lensing magnification, survey
volume, completeness, and cosmic variance can additionally affect the inferred
population abundance.  Because the halo abundance is so steep, relatively
modest changes in these quantities can produce appreciable shifts in the
inferred cosmological constraint.

\subsection{AGN Abundance and Early Black-Hole Growth}
\label{subsec:discussion_agn}

The AGN application introduces additional astrophysical degrees of freedom.
For a threshold host model, the observable abundance can be expressed as
\begin{equation}
n_{{\rm AGN},X}(z)
=
f_{\rm occ}f_{\rm duty}
n_X(>M_{\min},z),
\label{eq:discussion_agn_abundance}
\end{equation}
where $f_{\rm occ}$ is the black-hole occupation fraction and
$f_{\rm duty}$ is the active fraction.  The host threshold itself depends on
the mapping between AGN luminosity, black-hole mass, Eddington ratio, and
halo mass.

A pure abundance measurement therefore cannot uniquely determine all of these
quantities.  A low AGN number density may indicate rare massive hosts, a low
occupation fraction, a short duty cycle, or some combination of the three.
Similarly, assigning a luminous AGN to a very massive host can make the
available halo population extremely small, whereas a lower characteristic
host mass can substantially relax the halo-supply requirement.

The host-abundance calculation also has implications for early black-hole
growth.  Light stellar-remnant seeds require substantially more subsequent
growth than heavy-seed scenarios, while the limited time available at high
redshift places additional demands on the accretion history.  The halo
calculation does not distinguish among seed-formation mechanisms, but it
constrains the number of environments in which a successful growth history can
occur.  Depending on the inferred black-hole and host masses, viable scenarios
may require sustained near-Eddington or super-Eddington accretion, heavier
initial seeds, elevated black-hole-to-halo mass ratios, or lower characteristic
host masses \citep{Volonteri2010,Inayoshi2020}.

Clustering provides an important complementary observable because it probes
the characteristic host mass through halo bias.  More massive and rarer hosts
generally imply stronger large-scale clustering.  Abundance and clustering
therefore constrain different aspects of the AGN--halo connection: abundance
measures the available number of hosts, whereas clustering constrains their
typical rarity.  Under the mass-independent KST rescaling used here, the common
abundance factor cancels from a normalized effective-bias calculation when the
same bias relation and occupation model are used.  The KST prescription
therefore changes the predicted host abundance without, by itself, changing
the normalized AGN bias.

\subsection{UV Luminosity Functions as Population-Level Tests}
\label{subsec:discussion_uvlf}

The UV luminosity-function application moves the comparison from individual
extreme systems to the population level.  A deterministic halo-to-light model
assigns a UV magnitude to each halo mass through the baryonic conversion
efficiency, star-formation timescale, and UV luminosity calibration.  The HMF
then determines how many halos populate each luminosity interval.

This separation can be represented schematically as
\begin{equation}
\begin{aligned}
M_h
&\longrightarrow
M_{\rm UV}(M_h,z;\epsilon_\star),
\\
\frac{dn_X}{dM_h}
&\longrightarrow
\Phi_X(M_{\rm UV},z).
\end{aligned}
\label{eq:discussion_uv_separation}
\end{equation}
Changing $\epsilon_\star$ modifies the luminosity assigned to a halo and hence
the halo mass associated with a fixed observed magnitude.  Changing the HMF
calibration modifies the number of halos available at that mass.  These are
physically distinct effects and should not be absorbed into a single
efficiency parameter.

The comparison with high-redshift luminosity-function measurements near
$z\simeq6.4$, 8, and 10
\citep{Franco2025,Donnan2023,Donnan2024} illustrates why both effects become
important toward cosmic dawn.  Increasing the stellar-conversion efficiency
allows a given UV luminosity to be produced by a lower-mass and more abundant
halo.  The KST abundance correction acts in the opposite direction by reducing
the number of halos available to populate the luminosity function.  The
observed bright-end abundance therefore constrains the combination of
halo-to-light physics and cosmological halo supply.

A deterministic relation is useful for exposing these dependencies, but real
galaxies exhibit intrinsic scatter in luminosity at fixed halo mass.  The more
general prediction is
\begin{equation}
\begin{aligned}
\Phi_X(M_{\rm UV},z)
={}&
\int_0^\infty
\frac{dn_X}{dM_h}(M_h,z)
\\
&\times
P(M_{\rm UV}\mid M_h,z)\,dM_h .
\end{aligned}
\label{eq:discussion_uvlf_scatter}
\end{equation}
Scatter is especially important at the bright end because numerous lower-mass
halos can scatter toward higher luminosities.  A realistic forward model
should therefore distinguish intrinsic halo-to-light scatter from
observational uncertainties in flux, redshift, completeness, and inferred
number density.

\subsection{The Role of Halo-Mass-Function Calibration}
\label{subsec:discussion_hmf_calibration}

The comparison between ST and KST highlights a broader issue: the HMF itself
is a theoretical and numerical ingredient in high-redshift abundance tests.
Analytic collapse prescriptions are not exact descriptions of the halo
population, and simulation-calibrated mass functions depend on halo finder,
mass definition, cosmological parameters, numerical resolution, simulation
volume, and calibration redshift \citep{Tinker2008,Klypin2011}.

This issue becomes particularly important for rare objects.  A calibration
difference that appears moderate in the bulk of the halo population can
produce a substantial change in the inferred abundance of the most massive
systems.  Theoretical uncertainty in the HMF must therefore be considered
alongside observational and baryonic uncertainties when assessing early
galaxy and AGN populations.

The ST--KST comparison should not be interpreted as exhausting the range of
possible HMF calibrations.  Rather, it demonstrates quantitatively that the
choice of halo-abundance prescription can materially affect the inferred
high-redshift host reservoir.  A precision analysis would compare multiple
modern simulation-calibrated HMFs using consistent halo-mass definitions and
cosmological parameters and would propagate their calibration uncertainties
through the galaxy and AGN mappings.

The halo-mass definition is part of this consistency requirement.  A mass
inferred observationally or adopted in a halo-occupation model must be matched
to the mass convention used in the numerical HMF calibration.  Differences
among virial, friends-of-friends, and spherical-overdensity masses can become
important in the rare-halo regime, where even modest changes in mass can lead
to large abundance differences.

\subsection{Calibration Boundary at \texorpdfstring{$z=10$}{z=10}}
\label{subsec:discussion_calibration_boundary}

The quantitative KST calculations are restricted to the calibration interval
specified in Section~\ref{subsubsec:kst_scope},
\begin{equation}
0\leq z\leq10.
\label{eq:discussion_kst_range}
\end{equation}
The analytic correction can be evaluated mathematically at larger redshift,
but such values constitute extrapolations rather than simulation-calibrated
predictions.

This distinction is important because the high-redshift correction becomes
increasingly restrictive as $z$ increases.  Extending it beyond its calibrated
domain without additional numerical validation could therefore create an
artificially strong abundance constraint.  Observed systems at $z>10$ remain
important probes of early structure formation and of the halo-to-light
mapping, but they should not be used as the basis for a calibrated KST
exclusion.

Direct calibration at these redshifts will require simulations with sufficient
volume to sample rare massive halos while maintaining the mass resolution
needed to follow their progenitors and environments.  Consistent halo
definitions and controlled comparisons among numerical codes will be equally
important.

\subsection{Dominant Sources of Uncertainty}
\label{subsec:discussion_uncertainties}

The interpretation of high-redshift abundance tests depends on uncertainties
at several levels.  On the cosmological side, relevant ingredients include the
normalization and shape of the matter power spectrum, linear growth,
transfer-function modeling, halo definition, and numerical calibration of the
HMF.  For the framework considered here, HMF calibration is particularly
important because the observations probe the rapidly varying rare-halo tail.

The galaxy mapping introduces uncertainties in stellar-population synthesis,
the stellar initial mass function, star-formation history, stellar conversion
efficiency, metallicity, nebular emission, dust attenuation, feedback, and
scatter between halo mass and luminosity.  For AGN, additional uncertainties
arise from black-hole mass estimation, bolometric corrections, Eddington
ratios, obscuration, black-hole occupation, duty cycle, and the relation
between black-hole and halo mass.

Observational uncertainties include photometric and spectroscopic redshifts,
flux measurements, completeness corrections, source classification,
gravitational magnification, survey volume.  These
uncertainties are not independent.  For example, an AGN contribution can alter
an inferred stellar mass as well as a UV luminosity, while a redshift revision
changes both the inferred source properties and the predicted halo abundance.

The rare-halo tail amplifies the combined effect.  A relatively small shift in
an inferred stellar mass, Eddington ratio, or characteristic host mass may
correspond to a much larger change in the predicted number of halos.  Strong
cosmological conclusions therefore require a joint treatment of observational
errors, astrophysical scatter, and HMF calibration rather than variation of
only one ingredient at a time.

\subsection{Do the Current Data Require Physics Beyond
\texorpdfstring{$\Lambda$CDM}{Lambda-CDM}?}
\label{subsec:discussion_beyond_lcdm}

The present calculations do not require a departure from standard
$\Lambda$CDM.  Instead, the JWST populations probe regions of parameter space
in which the predicted halo supply can become small and strongly sensitive to
the adopted host mapping and HMF calibration.  The resulting observations are
therefore increasingly stringent stress tests of the combined cosmological and
astrophysical framework.

This interpretation is important because an abundance discrepancy does not
uniquely identify its physical origin.  A mismatch may arise from an
overestimated stellar or black-hole mass, an incorrect characteristic host
mass, an incomplete treatment of scatter, an inappropriate duty cycle or
occupation fraction, an HMF calibration error, or a genuine shortcoming of
the cosmological model.  These possibilities must be distinguished before a
departure from $\Lambda$CDM can be inferred.

If future observations, improved halo calibration, and more realistic
astrophysical modeling establish a statistically persistent discrepancy,
cosmological extensions could then be investigated.  Possibilities include
changes to the primordial power spectrum, non-Gaussian initial conditions,
early-dark-energy scenarios, or modified gravity.  Such extensions are not
required or fitted in the present analysis.  Any viable alternative would
also have to remain consistent with the cosmic microwave background,
large-scale structure, gravitational lensing, nucleosynthesis, and the other
successful tests of the standard cosmological model.

\subsection{Observational and Theoretical Tests Ahead}
\label{subsec:discussion_future_tests}

Progress on the observational side will come from larger survey volumes,
deeper spectroscopy, improved stellar-mass measurements, more secure AGN
classification, and better characterization of selection functions and
redshift distributions.  Clustering and gravitational lensing are especially
valuable because they can constrain characteristic host-halo masses more
directly than abundance measurements alone.

JWST will continue to provide deep measurements of the earliest galaxy and
AGN populations, while wide-field and multiwavelength observations from
facilities such as the Nancy Grace Roman Space Telescope, Euclid, Athena, the
Square Kilometre Array, and next-generation extremely large telescopes can
extend the available volumes and provide complementary constraints on stellar
populations, black-hole activity, environments, and clustering.

On the theoretical side, larger-volume and higher-resolution simulations at
$z\gtrsim10$ are needed to calibrate the rare-halo population directly.
Forward models should replace single-valued halo-to-light relations with
conditional distributions that include luminosity scatter, evolving
star-formation efficiency, bursty star formation, dust, feedback, and AGN
contamination.  AGN calculations should similarly incorporate distributions
of black-hole mass, Eddington ratio, obscuration, occupation, and duty cycle.

A particularly powerful approach will be to combine multiple observables.
Number densities constrain the available host population; luminosity functions
constrain the distribution of observable output; clustering constrains
characteristic host mass; lensing can test halo mass more directly; and
spectroscopy constrains redshift, stellar populations, ionization state, and
AGN activity.  Joint analysis of these observables can reduce degeneracies
that cannot be resolved by any single measurement.

\subsection{Cosmological Interpretation}
\label{subsec:discussion_summary}

The physical sequence connecting primordial fluctuations to the observed
high-redshift populations may be summarized as
\begin{equation}
\begin{aligned}
P(k)
&\longrightarrow
\sigma(M,z)
\longrightarrow
\nu(M,z)
\longrightarrow
\frac{dn_X}{dM}
\\
&\longrightarrow
\left\{
n_X(>M,z),\,
\rho_{h,X}(>M,z)
\right\}
\\
&\longrightarrow
\left\{
M_\star,\,
n_{\rm AGN},\,
M_{\rm UV}
\right\}
\longrightarrow
\Phi(M_{\rm UV},z).
\end{aligned}
\label{eq:discussion_complete_chain}
\end{equation}

Three conclusions follow from this framework.  First, the most massive
high-redshift galaxies and luminous AGN probe the rare-halo regime in which
small changes in host mass or redshift can produce large abundance changes.
Second, the calibration of the HMF is itself an important component of the
prediction; the ST--KST comparison demonstrates that uncertainty in the
high-redshift halo supply can materially affect massive-galaxy, AGN, and UV
luminosity-function tests.  Third, the cosmological halo supply and the
astrophysical observable-to-halo mapping must be treated as distinct but
coupled components of the inference.

The emerging JWST populations therefore provide quantitative stress tests of
$\Lambda$CDM structure formation and of the baryonic and black-hole physics
that connect halos to observable sources.  The most robust cosmological
conclusions will come from discrepancies that persist after observational
uncertainties, stochastic halo-to-light mappings, host-halo inference, and
high-redshift HMF calibration uncertainties are treated jointly.  In this
sense, the earliest galaxies and AGN provide a direct observational bridge
between primordial density fluctuations, nonlinear gravitational collapse,
baryonic conversion, and black-hole growth.

\section{Summary and Conclusions}
\label{sec:conclusions}

JWST observations of massive galaxies and luminous active galactic nuclei
during the first billion years have made the high-redshift halo population an
increasingly direct probe of cosmological structure formation.  The analysis
presented here connects the linear matter fluctuation spectrum to halo
formation and then to three complementary observational tests: the baryonic
reservoir available to massive galaxies, the abundance of halos capable of
hosting luminous AGN, and the rest-frame ultraviolet luminosity function of
high-redshift galaxies.

The calculation begins with a common Planck-like $\Lambda$CDM cosmology and
constructs the linear matter power spectrum, mass variance, peak height, and
halo abundance.  The Sheth--Tormen (ST) mass function is used as the principal
analytic baseline.  A Bolshoi/Klypin-corrected ST prescription (KST) is carried
as an illustrative simulation-motivated alternative in order to quantify the
sensitivity of the high-redshift conclusions to HMF calibration.  The two
branches share the same underlying power spectrum, growth history, mass
variance, peak-height definition, and astrophysical mappings; they differ only
in the adopted halo abundance.  This separation allows HMF uncertainty to be
distinguished from uncertainties in the conversion of halos into observable
galaxies and AGN.

The massive-galaxy test shows that an inferred stellar mass must satisfy both
an individual baryonic-reservoir condition and a population-level abundance
condition.  For a halo of mass $M_h$,
\begin{equation}
M_\star \leq \epsilon_\star f_b M_h ,
\qquad 0<\epsilon_\star\leq1 ,
\label{eq:conclusion_baryon_condition}
\end{equation}
where $\epsilon_\star=1$ represents an absolute baryonic ceiling rather than a
realistic star-formation efficiency.  As the required host mass increases,
the rapidly declining rare-halo tail makes the corresponding abundance
constraint progressively stronger.  The cumulative halo-mass-density
calculation provides an additional integral test by asking whether the total
matter and baryonic reservoir contained in the eligible halo population can
support the inferred stellar-mass density.

For AGN, halo abundance provides an upper envelope rather than a direct
prediction of the observable population.  The expected number density also
depends on black-hole occupation, duty cycle, and the relation between AGN
luminosity, black-hole mass, Eddington ratio, and host-halo mass.  In the
simplified threshold description,

\begin{equation}
n_{{\rm AGN},X}(z)
=
f_{\rm occ}f_{\rm duty}\,
n_X(>M_{\min},z),
\qquad
X\in\{\mathrm{ST},\mathrm{KST}\}.
\label{eq:conclusion_agn}
\end{equation}

Consequently, a shortage of sufficiently massive halos cannot be alleviated
by reducing the occupation fraction or duty cycle; those factors can only
lower the predicted active population.  Conversely, uncertainty in the
characteristic host mass can substantially alter the abundance constraint
because the calculation samples the steep high-mass tail.  Clustering and
lensing measurements are therefore valuable complementary probes of the
host-halo masses associated with high-redshift AGN.

The UV luminosity-function calculation extends the analysis from individual
extreme objects to the galaxy population as a whole.  A deterministic
halo-to-light relation specifies the UV luminosity associated with a halo of
given mass, redshift, and star-formation efficiency, while the HMF determines
how many such halos exist.  These two ingredients answer different physical
questions.  Increasing the stellar-conversion efficiency can lower the halo
mass required to produce a specified luminosity; changing the HMF calibration
changes the abundance of halos available at that mass.  The comparison with
representative measurements near $z\simeq6.4$, 8, and 10 therefore demonstrates
that high-redshift luminosity functions constrain both baryonic physics and
the cosmological halo supply.

Across all three tests, the same general result emerges: the interpretation of
the earliest massive and luminous systems depends sensitively on the mapping
between an observable and its host halo because these systems occupy the
rare, high-$\nu$ regime.  Relatively modest changes in inferred stellar mass,
star-formation efficiency, Eddington ratio, or characteristic host mass can
produce much larger changes in the predicted halo abundance.  HMF calibration
adds a separate source of uncertainty.  The ST--KST comparison demonstrates
that this uncertainty can materially affect the inferred high-redshift host
reservoir and should not be absorbed into an astrophysical efficiency
parameter.

The KST prescription is used quantitatively only within its adopted
simulation-calibrated interval, $0\leq z\leq10$.  Although its analytic form
can be evaluated at larger redshift, such use is extrapolative.  Objects at
$z>10$ remain important probes of early structure formation, but robust HMF
constraints in that regime require simulations that directly calibrate the
rare-halo population at those epochs.

The present results do not establish that current JWST observations require a
departure from standard $\Lambda$CDM.  Instead, they identify the combinations
of halo abundance, host mass, baryonic conversion, black-hole occupation,
duty cycle, and halo-to-light mapping that are most demanding within the
standard framework.  The appropriate interpretation of the most extreme
systems is therefore as quantitative stress tests of cosmological structure
formation and of the astrophysical assumptions used to connect halos with
observable sources.

Future progress will require these components to be constrained jointly.
Larger survey volumes and improved spectroscopy will refine galaxy and AGN
number densities, stellar masses, redshifts, and source classifications.
Clustering and lensing can constrain characteristic host masses.  On the
theoretical side, higher-redshift HMF calibrations and stochastic
halo-to-observable models are needed, including scatter, evolving
star-formation efficiency, bursty star formation, dust and feedback for
galaxies, and distributions of black-hole mass, Eddington ratio, occupation,
obscuration, and duty cycle for AGN.

The principal value of the framework is that it keeps these ingredients
physically distinct while placing them in a common abundance calculation.
The chain
\begin{equation}
\begin{aligned}
P(k)
&\longrightarrow
\sigma(M,z)
\longrightarrow
\nu(M,z)
\longrightarrow
\frac{dn_X}{dM}
\\
&\longrightarrow
\left\{
n_X(>M,z),\,
\rho_{h,X}(>M,z)
\right\}
\\
&\longrightarrow
\left\{
M_\star,\,
n_{\rm AGN},\,
M_{\rm UV},\,
\Phi(M_{\rm UV},z)
\right\}
\end{aligned}
\label{eq:conclusion_chain}
\end{equation}
provides a transparent connection from primordial density fluctuations to
nonlinear collapse and ultimately to the galaxies and AGN observed at cosmic
dawn.  As the observational samples and theoretical calibrations improve,
this combined approach can provide increasingly stringent tests of the
standard structure-formation picture while making clear whether any emerging
discrepancy originates primarily in the cosmological halo supply, the
astrophysical mapping, or the interaction between the two.

\section*{Acknowledgements}

The authors gratefully acknowledge the support of the Johns Hopkins
University Whiting School EP Applied Physics Program. R.S. also
acknowledges the support of the Johns Hopkins University Applied
Physics Laboratory.

\section*{Data Availability}

The numerical results presented in this work are derived from the analytic
models and observational datasets cited in the text. The numerical code used
to generate the theoretical calculations and figures is available from the
corresponding author upon reasonable request.

\bibliographystyle{apsrev4-1}
\bibliography{main}

%
%
%

%
%
%

\appendix

\section{Deriving Theoretical UV Luminosity Functions from the Standard and
Klypin-Corrected Sheth--Tormen Halo Mass Functions}
\label{app:uv_luminosity_function_ST_KST}

This appendix completes the bridge between the theoretical halo mass function
and the observable rest-frame ultraviolet (UV) luminosity function measured in
high-redshift galaxy surveys. The main text introduced the halo-to-light
mapping as an intermediate diagnostic. Here we formulate the luminosity
function in parallel for two halo-abundance prescriptions: the standard
Sheth--Tormen (ST) halo mass function and the Bolshoi-calibrated,
Klypin-corrected Sheth--Tormen (KST) halo mass function.

For a general halo mass function $dn_X/dM_h$, where
$X\in\{{\rm ST},{\rm KST}\}$, the UV luminosity function is
\begin{equation}
\Phi_X(M_{\rm UV},z)
=
\int
\frac{dn_X}{dM_h}(M_h,z)\,
P(M_{\rm UV}|M_h,z)\,
dM_h ,
\label{eq:app_lf_master_X}
\end{equation}
where $P(M_{\rm UV}|M_h,z)$ is the conditional probability that a halo
of mass $M_h$ at redshift $z$ is observed with UV absolute magnitude
$M_{\rm UV}$.

The purpose of this appendix is not to construct a complete semi-analytic
galaxy-formation model. Rather, it provides a transparent framework showing
how both the standard ST and KST halo mass functions propagate through the
same halo-to-light model. This formulation is designed to support direct
side-by-side numerical comparisons of
\begin{equation}
\Phi_{\rm ST}(M_{\rm UV},z)
\qquad \hbox{and} \qquad
\Phi_{\rm KST}(M_{\rm UV},z),
\end{equation}
while keeping all baryonic, stellar, duty-cycle, and scatter assumptions
identical.

\subsection{Standard and Klypin-corrected Sheth--Tormen halo mass functions}
\label{app:st_kst_hmf}

The halo-abundance calculation used throughout this work begins with the
standard Sheth--Tormen (ST) mass function
\citep{ShethTormen1999,ShethMoTormen2001},
\begin{equation}
\frac{dn_{\rm ST}}{dM_h}(M_h,z)
=
\frac{\bar{\rho}_m}{M_h}\,
f_{\rm ST}(\nu)
\left|
\frac{d\nu}{dM_h}
\right|,
\label{eq:app_st_hmf}
\end{equation}
where $\bar{\rho}_m$ is the mean comoving matter density and the peak-height
variable is defined by
\begin{equation}
\nu(M_h,z)
=
\frac{\delta_c}{\sigma(M_h,z)}
=
\frac{\delta_c}{D(z)\sigma(M_h,0)}.
\label{eq:app_peak_height}
\end{equation}
Here $\delta_c\simeq1.686$ is the spherical-collapse threshold and $D(z)$ is
the scale-independent linear growth factor, normalized such that $D(0)=1$.

For the spatially flat $\Lambda$CDM cosmology adopted in this work, the growth
factor is evaluated from
\begin{equation}
D(a)
=
\frac{
E(a)\displaystyle\int_0^a
\frac{da'}{a'^3E^3(a')}
}{
E(1)\displaystyle\int_0^1
\frac{da'}{a'^3E^3(a')}
},
\label{eq:app_growth_factor}
\end{equation}
where
\begin{equation}
E(a)
=
\frac{H(a)}{H_0}
=
\left(
\Omega_{m,0}a^{-3}+\Omega_{\Lambda,0}
\right)^{1/2},
\label{eq:app_Ea}
\end{equation}
with $a=(1+z)^{-1}$. The common multiplicative factor $5\Omega_{m,0}/2$ in
the integral solution cancels when the growth function is normalized to
$D(1)=D(z=0)=1$. Thus the calculation does not assume the
Einstein--de Sitter approximation $D(z)=(1+z)^{-1}$; the full flat-$\Lambda$CDM
growth history is used.

The Sheth--Tormen multiplicity function is
\begin{equation}
f_{\rm ST}(\nu)
=
A
\sqrt{\frac{2a_{\rm ST}}{\pi}}\,
\nu
\left[
1+\left(a_{\rm ST}\nu^2\right)^{-p}
\right]
\exp\left(
-\frac{a_{\rm ST}\nu^2}{2}
\right),
\label{eq:app_st_multiplicity}
\end{equation}
with the standard parameters
\begin{equation}
A=0.3222,
\qquad
a_{\rm ST}=0.707,
\qquad
p=0.3.
\label{eq:app_st_parameters}
\end{equation}
The notation $a_{\rm ST}$ is used here to distinguish the Sheth--Tormen
parameter from the cosmological scale factor $a$.

The variance entering equation~(\ref{eq:app_peak_height}) is calculated from
the linear matter power spectrum,
\begin{equation}
\sigma^2(M_h,z)
=
\frac{1}{2\pi^2}
\int_0^\infty
k^2P(k,z)W^2(kR)\,dk,
\label{eq:app_sigma}
\end{equation}
where $W(kR)$ is the Fourier-space spherical top-hat window function and
\begin{equation}
M_h
=
\frac{4\pi}{3}\bar{\rho}_mR^3.
\label{eq:app_mass_radius}
\end{equation}
Equivalently, with scale-independent linear growth,
$P(k,z)=D^2(z)P(k,0)$ and hence
$\sigma(M_h,z)=D(z)\sigma(M_h,0)$.

The Bolshoi simulation analysis of \citet{Klypin2011} found that the standard
ST expression increasingly overpredicts the abundance of distinct
spherical-overdensity halos toward high redshift. We therefore adopt the
redshift-dependent correction
\begin{equation}
F_{\rm K}(z)
=
\frac{
[5.501D(z)]^4
}{
1+[5.500D(z)]^4
},
\label{eq:app_klypin_factor}
\end{equation}
using the same normalized $\Lambda$CDM growth factor $D(z)$ defined above.

The Klypin-corrected Sheth--Tormen (KST) halo mass function used in this work
is therefore
\begin{equation}
\boxed{
\frac{dn_{\rm KST}}{dM_h}(M_h,z)
=
F_{\rm K}(z)
\frac{dn_{\rm ST}}{dM_h}(M_h,z)
}
\label{eq:app_kst_hmf}
\end{equation}
or, equivalently,
\begin{equation}
f_{\rm KST}(\nu,z)
=
F_{\rm K}(z)f_{\rm ST}(\nu).
\label{eq:app_kst_multiplicity}
\end{equation}

This distinction is important conceptually. Because $F_{\rm K}$ depends
explicitly on redshift through $D(z)$, $f_{\rm KST}$ is not a universal
multiplicity function of $\nu$ alone. Moreover, in the implementation adopted
here $F_{\rm K}$ is independent of halo mass at fixed redshift. The correction
therefore changes the normalization of the halo abundance but does not alter
the mass dependence of the underlying ST curve at that redshift. It also does
not modify the matter power spectrum, $\sigma(M,z)$, the peak-height
definition, or the baryonic halo-to-light mapping.

The representative numerical values of $F_{\rm K}(z)$ are given in
Equation~\eqref{eq:klypin_reference_values} of Section~\ref{subsec:kst}.

Thus the corresponding ST-to-KST abundance ratios are approximately
\begin{equation}
\frac{n_{\rm ST}}{n_{\rm KST}}
\simeq
1.27,\quad
2.02,\quad
3.77,\quad
7.18
\qquad
{\rm at}\quad
z=4,\ 6,\ 8,\ 10,
\label{eq:app_st_kst_ratios}
\end{equation}
respectively. These ratios apply at every halo mass at a given redshift for
the multiplicative implementation adopted here.

The corresponding cumulative abundances are
\begin{equation}
n_{\rm ST}(>M_{\min},z)
=
\int_{M_{\min}}^\infty
\frac{dn_{\rm ST}}{dM_h}(M_h,z)\,dM_h
\label{eq:app_st_cumulative}
\end{equation}
and
\begin{equation}
n_{\rm KST}(>M_{\min},z)
=
\int_{M_{\min}}^\infty
\frac{dn_{\rm KST}}{dM_h}(M_h,z)\,dM_h.
\label{eq:app_kst_cumulative}
\end{equation}
Since $F_{\rm K}(z)$ is independent of $M_h$ in this prescription, it may be
taken outside the mass integral, giving
\begin{equation}
\boxed{
n_{\rm KST}(>M_{\min},z)
=
F_{\rm K}(z)
n_{\rm ST}(>M_{\min},z)
}.
\label{eq:app_kst_cumulative_simple}
\end{equation}
The same multiplicative relation holds for any mass-weighted integral over
the halo mass function when the integration limits and all other assumptions
are held fixed.

Finally, the calibration boundary of the adopted correction must be kept
explicit. The Bolshoi-based fit is used quantitatively in this work only over
the approximate interval
\begin{equation}
0\leq z\leq10.
\label{eq:app_kst_calibration_range}
\end{equation}
Although equation~(\ref{eq:app_klypin_factor}) can be evaluated
mathematically at $z>10$, such values constitute an extrapolation of the fit
rather than a calibrated KST prediction. Observational systems beyond
$z=10$ may therefore be shown for context, but they are not used here as the
basis for calibrated KST abundance constraints.

\subsection{Deterministic halo-to-UV mapping}
\label{app:deterministic_halo_uv}

The halo mass function determines the abundance of halos, but an additional
baryonic prescription is required to associate a halo of mass $M_h$ with an
observable rest-frame ultraviolet luminosity. We adopt the same deliberately
simple deterministic mapping used in the main text so that the roles of halo
abundance and baryonic conversion remain clearly separated.

The total baryonic mass associated with a halo is taken to be
\begin{equation}
M_b
=
f_bM_h,
\label{eq:app_baryon_reservoir}
\end{equation}
where
\begin{equation}
f_b
=
\frac{\Omega_b}{\Omega_m}
\label{eq:app_fb}
\end{equation}
is the cosmic baryon fraction \citep{Planck2018,BoylanKolchin2023}. This is
the maximum cosmological baryonic reservoir associated with the halo; it
should not be interpreted as implying that all of the baryons are cold,
retained, or available for immediate star formation.

We parameterize the stellar mass as
\begin{equation}
M_\star
=
\epsilon_\star(M_h,z)\,f_bM_h,
\label{eq:app_stellar_mass}
\end{equation}
where $\epsilon_\star$ is an effective stellar-conversion efficiency. The
physical range satisfies
\begin{equation}
0<\epsilon_\star\leq1.
\label{eq:app_epsilon_range}
\end{equation}
The limiting value $\epsilon_\star=1$ represents an absolute baryonic ceiling
in which the entire cosmological baryon allotment is converted into stars; it
is not intended as a realistic galaxy-formation efficiency. In the baseline
calculations $\epsilon_\star$ is treated parametrically, while more complete
models may allow both halo-mass and redshift dependence
\citep{Behroozi2013,Moster2013,SomervilleDave2015}.

To obtain a characteristic star-formation rate, the stellar mass formed over
the adopted timescale $t_\star$ is written as
\begin{equation}
{\rm SFR}(M_h,z)
=
\frac{\epsilon_\star(M_h,z)f_bM_h}{t_\star(z)}.
\label{eq:app_sfr}
\end{equation}
For the first-order mapping used here, the timescale is parameterized as a
fixed fraction of the Hubble time,
\begin{equation}
t_\star(z)
=
f_{\rm dyn}H^{-1}(z),
\label{eq:app_tstar}
\end{equation}
where $f_{\rm dyn}$ is a dimensionless timescale parameter. Therefore
\begin{equation}
{\rm SFR}(M_h,z)
=
\frac{
\epsilon_\star(M_h,z)f_bM_hH(z)
}{
f_{\rm dyn}
}.
\label{eq:app_sfr_hubble}
\end{equation}
This prescription should be understood as an effective time-averaged mapping,
not as a detailed model of bursty star formation or gas accretion.

The corresponding intrinsic rest-frame UV spectral luminosity is taken to be
\begin{equation}
L_{\nu,{\rm UV}}
=
\kappa_{\rm UV}\,{\rm SFR},
\label{eq:app_uv_luminosity}
\end{equation}
where the baseline conversion adopted in the numerical calculations is
\begin{equation}
\kappa_{\rm UV}
\simeq
8\times10^{27}\,
{\rm erg\,s^{-1}\,Hz^{-1}}\,
(M_\odot\,{\rm yr}^{-1})^{-1}.
\label{eq:app_kappa}
\end{equation}
The precise value of $\kappa_{\rm UV}$ depends on assumptions concerning the
stellar initial mass function, metallicity, stellar age, star-formation
history, and population synthesis. It should therefore be regarded as part of
the astrophysical mapping rather than a cosmological prediction
\citep{Bouwens2021,FinkelsteinBagley2022,Napolitano2025}.

The AB absolute magnitude corresponding to a spectral luminosity
$L_{\nu,{\rm UV}}$ is
\begin{equation}
M_{\rm UV} = 51.60 - 2.5\log_{10}
\left(
\frac{L_{\nu,{\rm UV}}} {{\rm erg\,s^{-1}\,Hz^{-1}}}
\right).
\label{eq:app_MUV}
\end{equation}
Combining the preceding relations gives the deterministic mean
halo-to-magnitude mapping
\begin{equation}
\bar{M}_{\rm UV}(M_h,z)
=
51.60
-
2.5\log_{10}
\left[
\frac{
\kappa_{\rm UV}\,
\epsilon_\star(M_h,z)\,
f_bM_hH(z)
}{
f_{\rm dyn}
}
\right],
\label{eq:app_mean_MUV}
\end{equation}
where the quantities inside the logarithm are to be evaluated in mutually
consistent units so that the argument represents the numerical value of
$L_{\nu,{\rm UV}}$ in ${\rm erg\,s^{-1}\,Hz^{-1}}$.

Several features of equation~(\ref{eq:app_mean_MUV}) are useful for
interpreting the figures in the main text. At fixed redshift and fixed
$\epsilon_\star$, increasing $M_h$ increases the star-formation rate and UV
luminosity and therefore produces a more negative, brighter $M_{\rm UV}$.
Likewise, increasing $\epsilon_\star$ shifts a halo of fixed mass toward
brighter UV magnitude. The redshift dependence enters through $H(z)$ in the
adopted star-formation timescale as well as through any explicit redshift
dependence assigned to $\epsilon_\star$.

For the special baseline case in which $\epsilon_\star$ is independent of
halo mass at fixed redshift, equation~(\ref{eq:app_mean_MUV}) has the form
\begin{equation}
M_{\rm UV}
=
C(z,\epsilon_\star)
-
2.5\log_{10}M_h,
\label{eq:app_MUV_powerlaw}
\end{equation}
where $C(z,\epsilon_\star)$ contains the baryon fraction, UV conversion,
star-formation timescale, and the unit conversion associated with the chosen
mass unit. It follows that
\begin{equation}
\frac{dM_{\rm UV}}{dM_h}
=
-\frac{2.5}{\ln 10}\frac{1}{M_h}
\label{eq:app_dMUV_dMh}
\end{equation}
and hence
\begin{equation}
\boxed{
\left|
\frac{dM_h}{dM_{\rm UV}}
\right|
=
\frac{\ln 10}{2.5}M_h
}.
\label{eq:app_jacobian}
\end{equation}
This Jacobian is the factor required to transform the halo mass function into
the deterministic UV luminosity function in the following subsection.

Most importantly, the mapping in this subsection is common to the ST and KST
branches. The Klypin correction modifies the statistical abundance assigned
to halos but does not alter $M_b$, $\epsilon_\star$, $t_\star$,
$\kappa_{\rm UV}$, or the luminosity assigned to an individual halo. Thus, at
fixed $(M_h,z,\epsilon_\star)$,
\begin{equation}
M_{{\rm UV},{\rm ST}}
=
M_{{\rm UV},{\rm KST}}.
\label{eq:app_same_MUV_ST_KST}
\end{equation}
The distinction between ST and KST appears only when the halo-to-light mapping
is combined with their different halo abundances to construct a population
luminosity function.

The deterministic prescription is intentionally a controlled baseline. It
does not include scatter in luminosity at fixed halo mass, dust attenuation,
bursty star formation, stochastic gas accretion, feedback, or AGN
contamination. These effects can be incorporated through the conditional
probability $P(M_{\rm UV}\mid M_h,z)$ introduced at the beginning of this
Appendix and are considered in later subsections.

\subsection{Deterministic ST and KST luminosity functions}
\label{app:deterministic_lf}

The deterministic halo-to-UV relation derived in the preceding subsection
allows the halo mass function to be transformed directly into a rest-frame
UV luminosity function. Define
\begin{equation}
\Phi_X(M_{\rm UV},z)
\equiv
\frac{dn_X}{dM_{\rm UV}},
\qquad
X\in\{{\rm ST},{\rm KST}\},
\label{eq:app_lf_definition}
\end{equation}
so that $\Phi_X$ has units of comoving number density per unit magnitude.

For a monotonic, single-valued mapping between $M_h$ and $M_{\rm UV}$,
conservation of objects gives
\begin{equation}
\Phi_X(M_{\rm UV},z)\,
\left|dM_{\rm UV}\right|
=
\frac{dn_X}{dM_h}(M_h,z)\,
dM_h.
\label{eq:app_number_conservation_X}
\end{equation}
Therefore,
\begin{equation}
\boxed{
\Phi_X(M_{\rm UV},z)
=
\left.
\frac{dn_X}{dM_h}(M_h,z)
\left|
\frac{dM_h}{dM_{\rm UV}}
\right|
\right|_{M_h=M_h(M_{\rm UV},z)}
}
\label{eq:app_lf_deterministic_X}
\end{equation}
where the halo mass on the right-hand side is the mass corresponding to the
specified UV magnitude under the adopted halo-to-light mapping.

For the ST branch,
\begin{equation}
\Phi_{\rm ST}(M_{\rm UV},z)
=
\left.
\frac{dn_{\rm ST}}{dM_h}
\left|
\frac{dM_h}{dM_{\rm UV}}
\right|
\right|_{M_h=M_h(M_{\rm UV},z)},
\label{eq:app_lf_deterministic_ST}
\end{equation}
whereas for the KST branch,
\begin{equation}
\Phi_{\rm KST}(M_{\rm UV},z)
=
\left.
\frac{dn_{\rm KST}}{dM_h}
\left|
\frac{dM_h}{dM_{\rm UV}}
\right|
\right|_{M_h=M_h(M_{\rm UV},z)}.
\label{eq:app_lf_deterministic_KST}
\end{equation}

Using the multiplicative KST prescription from
equation~(\ref{eq:app_kst_hmf}),
\begin{equation}
\frac{dn_{\rm KST}}{dM_h}
=
F_{\rm K}(z)
\frac{dn_{\rm ST}}{dM_h},
\end{equation}
and using exactly the same halo-to-light mapping for both branches gives
\begin{equation}
\boxed{
\Phi_{\rm KST}(M_{\rm UV},z)
=
F_{\rm K}(z)\,
\Phi_{\rm ST}(M_{\rm UV},z)
}.
\label{eq:app_lf_deterministic_scaling}
\end{equation}
Thus, within the deterministic model adopted here, the KST correction changes
the normalization of the predicted UV luminosity function at a given redshift
but does not change its shape as a function of $M_{\rm UV}$. This result is a
direct consequence of the assumed mass independence of $F_{\rm K}(z)$ and
would not generally hold for a mass-dependent halo-mass-function correction.

For the baseline case used in the numerical luminosity-function calculation,
$\epsilon_\star$ is independent of halo mass at fixed redshift. The
halo-to-light relation then has the form
\begin{equation}
M_{\rm UV}
=
C(z,\epsilon_\star)
-
2.5\log_{10}
\left(
\frac{M_h}{M_\odot}
\right),
\label{eq:app_MUV_powerlaw_lf}
\end{equation}
where $C(z,\epsilon_\star)$ contains the adopted baryon fraction,
star-formation efficiency, star-formation timescale, UV conversion factor,
and the corresponding unit conversions. Differentiation gives
\begin{equation}
\frac{dM_{\rm UV}}{dM_h}
=
-\frac{2.5}{\ln 10}\frac{1}{M_h},
\label{eq:app_derivative_MUV_Mh}
\end{equation}
and therefore
\begin{equation}
\left|
\frac{dM_h}{dM_{\rm UV}}
\right|
=
\frac{\ln 10}{2.5}M_h.
\label{eq:app_jacobian_lf}
\end{equation}

The corresponding deterministic luminosity functions become
\begin{equation}
\boxed{
\Phi_{\rm ST}(M_{\rm UV},z)
=
\left.
\frac{\ln 10}{2.5}
M_h
\frac{dn_{\rm ST}}{dM_h}
\right|_{M_h=M_h(M_{\rm UV},z)}
}
\label{eq:app_lf_simple_ST}
\end{equation}
and
\begin{equation}
\boxed{
\Phi_{\rm KST}(M_{\rm UV},z)
=
\left.
\frac{\ln 10}{2.5}
M_h
\frac{dn_{\rm KST}}{dM_h}
\right|_{M_h=M_h(M_{\rm UV},z)}
}.
\label{eq:app_lf_simple_KST}
\end{equation}
Both expressions have units of comoving number density per magnitude.

An important distinction follows from these equations. Changing
$\epsilon_\star$ alters the relation between $M_h$ and $M_{\rm UV}$ and
therefore shifts the predicted luminosity function horizontally in magnitude
space as well as changing which part of the halo mass function contributes at
a specified luminosity. By contrast, at fixed redshift the adopted KST
correction multiplies the ST luminosity function vertically by
$F_{\rm K}(z)$. The two effects are therefore physically and mathematically
distinct: $\epsilon_\star$ controls the baryonic halo-to-light mapping, whereas
$F_{\rm K}$ controls the cosmological halo supply.

For the representative redshifts used in the population-level comparison,
the vertical KST suppression is substantial. The values
$F_{\rm K}(6)\simeq0.496$, $F_{\rm K}(8)\simeq0.265$, and
$F_{\rm K}(10)\simeq0.139$ imply that the deterministic KST luminosity
function is approximately $49.6$, $26.5$, and $13.9$ per cent of the
corresponding ST prediction at those redshifts. Equivalently, ST predicts
approximately $2.02$, $3.77$, and $7.18$ times the KST abundance,
respectively. These are the population-level differences illustrated in the
ST/KST UV-luminosity-function comparison in the main text.

The deterministic luminosity function should be interpreted as a controlled
baseline rather than a complete galaxy-formation prediction. It assumes a
one-to-one relation between halo mass and UV magnitude and therefore neglects
scatter, burstiness, dust attenuation, stochastic assembly histories, and
other sources of variation at fixed halo mass. A more general treatment
replaces the deterministic transformation with a conditional probability
distribution $P(M_{\rm UV}\mid M_h,z)$, as developed in the following
subsection.

\subsection{Including realistic scatter in the halo-to-light relation}
\label{app:scatter_halo_light}

The deterministic mapping developed above assigns a single UV magnitude to
each halo mass at fixed redshift. Real galaxies, however, need not obey a
one-to-one halo--luminosity relation. Variations in star-formation history,
gas accretion, merger history, metallicity, dust attenuation, feedback, halo
assembly, and possible AGN contamination can produce a distribution of UV
luminosities at fixed halo mass. A more general formulation therefore
introduces the conditional probability distribution
\begin{equation}
P(M_{\rm UV}\mid M_h,z),
\label{eq:app_conditional_probability}
\end{equation}
normalized according to
\begin{equation}
\int_{-\infty}^{+\infty}
P(M_{\rm UV}\mid M_h,z)\,dM_{\rm UV}
=
1.
\label{eq:app_conditional_normalization}
\end{equation}

A useful illustrative model is Gaussian scatter in UV magnitude about the
deterministic mean relation $\bar{M}_{\rm UV}(M_h,z)$,
\begin{equation}
P(M_{\rm UV}\mid M_h,z)
=
\frac{1}{\sqrt{2\pi}\sigma_{\rm UV}}
\exp\left[
-\frac{
\left[
M_{\rm UV}-\bar{M}_{\rm UV}(M_h,z)
\right]^2
}{
2\sigma_{\rm UV}^2
}
\right],
\label{eq:app_gaussian_scatter}
\end{equation}
where $\sigma_{\rm UV}$ denotes the intrinsic scatter in magnitude at fixed
halo mass. Values of order
\begin{equation}
\sigma_{\rm UV}\sim0.3\text{--}0.7~{\rm mag}
\label{eq:app_scatter_values}
\end{equation}
may be used illustratively to explore the sensitivity of the bright end to
stochasticity. These values are not fitted parameters in the present work and
are not used in the baseline luminosity-function curves shown in the main
text.

With scatter included, the luminosity function is no longer obtained from a
simple Jacobian transformation. Instead,
\begin{equation}
\Phi_X(M_{\rm UV},z)
=
\int_{M_{\rm min}}^{M_{\rm max}}
\frac{dn_X}{dM_h}(M_h,z)\,
P(M_{\rm UV}\mid M_h,z)\,dM_h,
\qquad
X\in\{{\rm ST},{\rm KST}\}.
\label{eq:app_lf_scatter_general}
\end{equation}
Thus,
\begin{equation}
\Phi_{\rm ST}(M_{\rm UV},z)
=
\int_{M_{\rm min}}^{M_{\rm max}}
\frac{dn_{\rm ST}}{dM_h}(M_h,z)\,
P(M_{\rm UV}\mid M_h,z)\,dM_h,
\label{eq:app_lf_scatter_ST}
\end{equation}
and
\begin{equation}
\Phi_{\rm KST}(M_{\rm UV},z)
=
\int_{M_{\rm min}}^{M_{\rm max}}
\frac{dn_{\rm KST}}{dM_h}(M_h,z)\,
P(M_{\rm UV}\mid M_h,z)\,dM_h.
\label{eq:app_lf_scatter_KST}
\end{equation}

For the particular KST prescription adopted in this work,
\begin{equation}
\frac{dn_{\rm KST}}{dM_h}
=
F_{\rm K}(z)
\frac{dn_{\rm ST}}{dM_h},
\end{equation}
and $F_{\rm K}(z)$ is independent of halo mass at fixed redshift. If the same
conditional halo-to-light distribution $P(M_{\rm UV}\mid M_h,z)$ is adopted
for ST and KST, the correction may therefore be taken outside the integral:
\begin{align}
\Phi_{\rm KST}(M_{\rm UV},z)
&=
F_{\rm K}(z)
\int_{M_{\rm min}}^{M_{\rm max}}
\frac{dn_{\rm ST}}{dM_h}(M_h,z)
P(M_{\rm UV}\mid M_h,z)\,dM_h
\nonumber\\
&=
\boxed{
F_{\rm K}(z)\,
\Phi_{\rm ST}(M_{\rm UV},z)
}.
\label{eq:app_lf_scatter_scaling}
\end{align}
Consequently, scatter can alter the shape of both luminosity functions while
preserving their multiplicative ST/KST ratio at a fixed redshift under the
assumptions used here.

The physical importance of scatter is greatest where the underlying halo mass
function is steep. At the bright end, very massive halos are intrinsically
rare, whereas somewhat lower-mass halos are much more numerous. A symmetric
scatter in magnitude can therefore produce an asymmetric change in number
counts: more objects can scatter from the numerous moderate-mass population
into a rare bright bin than in the reverse direction. This is the familiar
Eddington-like enhancement of a steeply declining distribution.

The effect can be represented schematically as
\begin{equation}
\Phi_{\rm bright}^{\rm scatter}
>
\Phi_{\rm bright}^{\rm deterministic}
\label{eq:app_bright_scatter}
\end{equation}
over portions of the bright tail, although the magnitude of the enhancement
depends on $\sigma_{\rm UV}$, the local slope and curvature of the halo mass
function, the mean halo-to-light relation, and the adopted mass integration
limits. Scatter therefore cannot in general be represented by a single
redshift-independent multiplicative factor.

This distinction is especially relevant at high redshift. A deterministic
model may assign a very bright source only to an exceptionally massive and
rare halo, whereas a stochastic model permits a more common lower-mass halo
to occupy the same observed magnitude temporarily. Scatter can consequently
reduce the severity of an apparent bright-end abundance discrepancy without
requiring a corresponding increase in the abundance of the most massive
halos.

However, this possibility should not be confused with the calculations
actually presented in the main figures. The ST and KST UV luminosity
functions used for the baseline population comparison in this work employ
the deterministic mapping of the preceding subsections; no luminosity scatter
is folded into those plotted curves. Equations
(\ref{eq:app_lf_scatter_general})--(\ref{eq:app_lf_scatter_scaling}) are
included to show how the framework can be generalized and to identify one of
the principal astrophysical uncertainties that should be incorporated in a
future statistical analysis.

A fully developed implementation could allow
$\sigma_{\rm UV}=\sigma_{\rm UV}(M_h,z)$ and could replace the Gaussian model
with a non-Gaussian or physically generated conditional distribution. It
could also include correlations among star-formation efficiency, dust
attenuation, halo assembly history, and AGN contribution. Such extensions
would modify the detailed luminosity-function shape, but as long as the same
conditional mapping is applied to both ST and KST and the adopted
$F_{\rm K}(z)$ remains mass-independent, their relative normalization at fixed
redshift continues to be set by the Klypin factor.

\subsection{Duty cycle and occupation fraction}
\label{app:duty_occupation}

The halo mass function specifies the abundance of dark matter halos, but not
every halo necessarily hosts the luminous population under consideration, nor
is every potential source observable at a given time. This distinction is
particularly important for active galactic nuclei (AGN), whose observable
number density depends on both the presence of a central black hole and the
fraction of time for which that black hole is in an active, sufficiently
luminous accretion state.

We describe these effects using a halo occupation fraction
$f_{\rm occ}(M_h,z)$ and a duty cycle $f_{\rm duty}(M_h,z)$. The occupation
fraction represents the probability that a halo of mass $M_h$ at redshift $z$
contains a black hole or other source belonging to the population of interest,
whereas the duty cycle represents the probability that the source is active
and observable during the relevant epoch. Both quantities satisfy
\begin{equation}
0\leq f_{\rm occ}(M_h,z)\leq1,
\qquad
0\leq f_{\rm duty}(M_h,z)\leq1.
\label{eq:app_occ_duty_bounds}
\end{equation}

For a threshold-based AGN host model in which only halos above
$M_{\min}$ are capable of hosting the selected AGN population, the active AGN
number density is
\begin{equation}
n_{{\rm AGN},X}(z)
=
\int_{M_{\min}}^\infty
\frac{dn_X}{dM_h}(M_h,z)\,
f_{\rm occ}(M_h,z)\,
f_{\rm duty}(M_h,z)\,
dM_h,
\qquad
X\in\{{\rm ST},{\rm KST}\}.
\label{eq:app_agn_general}
\end{equation}
This expression makes explicit that halo abundance, black-hole occupation, and
AGN activity are separate ingredients.

If $f_{\rm occ}$ and $f_{\rm duty}$ are approximated as mass-independent over
the relevant halo range, equation~(\ref{eq:app_agn_general}) reduces to
\begin{equation}
\boxed{
n_{{\rm AGN},X}(z)
=
f_{\rm occ}(z)\,
f_{\rm duty}(z)\,
n_X(>M_{\min},z)
}.
\label{eq:app_agn_simple}
\end{equation}
Because both fractions are bounded by unity,
\begin{equation}
n_{{\rm AGN},X}(z)
\leq
n_X(>M_{\min},z),
\label{eq:app_agn_ceiling}
\end{equation}
so the cumulative halo abundance provides an absolute host-supply ceiling for
a one-AGN-per-halo interpretation.

For the multiplicative KST prescription adopted in this work,
\begin{equation}
n_{\rm KST}(>M_{\min},z)
=
F_{\rm K}(z)\,
n_{\rm ST}(>M_{\min},z).
\label{eq:app_occ_kst_cumulative}
\end{equation}
If the same occupation and duty-cycle prescription is applied to both halo
mass functions, then
\begin{equation}
\boxed{
n_{{\rm AGN},{\rm KST}}(z)
=
F_{\rm K}(z)\,
n_{{\rm AGN},{\rm ST}}(z)
}.
\label{eq:app_agn_kst_scaling}
\end{equation}
Thus the KST correction reduces the available active-AGN population by the
same redshift-dependent factor as the underlying halo population under these
assumptions.

More generally, a luminosity-selected AGN population may be written using a
conditional luminosity distribution,
\begin{align}
\Phi_{{\rm AGN},X}(L,z)
&=
\int
\frac{dn_X}{dM_h}(M_h,z)\,
f_{\rm occ}(M_h,z)\,
f_{\rm duty}(M_h,z)
\nonumber\\
&\qquad\times
P(L\mid M_h,z,{\rm active})\,dM_h .
\label{eq:app_agn_lf_general}
\end{align}
where $P(L\mid M_h,z,{\rm active})$ describes the luminosity distribution of
an active source in a halo of mass $M_h$. In an AGN application this
distribution may depend on the black-hole--halo relation, black-hole mass,
Eddington-ratio distribution, radiative efficiency, obscuration, and
bolometric correction. These ingredients are conceptually distinct from the
cosmological halo abundance.

It is important not to confuse this AGN duty-cycle formalism with the galaxy
UV luminosity-function calculation developed in the preceding subsections.
The baseline galaxy UV luminosity functions in the main text are obtained by
applying the deterministic stellar halo-to-UV mapping directly to the ST and
KST halo mass functions. No separate AGN occupation fraction or AGN duty-cycle
factor is applied to those galaxy luminosity-function curves. The quantities
$f_{\rm occ}$ and $f_{\rm duty}$ instead enter the AGN host-supply constraints
and any future AGN-specific luminosity-function calculation.

The occupation fraction is primarily connected to black-hole seeding and
retention. At high redshift it may depend on whether black holes originate
from stellar remnants, direct-collapse channels, or other seed populations,
and on whether subsequent mergers or dynamical processes retain those black
holes within their hosts \citep{Volonteri2010,Inayoshi2020}. The duty cycle
instead describes the episodic nature of accretion and may depend on gas
supply, angular-momentum transport, mergers, feedback, and the luminosity
threshold used to define an ``active'' source.

Consequently, an observed AGN abundance does not uniquely determine a halo
mass. A low observed number density may reflect rare massive hosts, a low
occupation fraction, a short duty cycle, or some combination of these
effects. Conversely, explaining a large AGN abundance with very massive host
halos can require occupation and duty-cycle factors approaching their physical
upper limits. The abundance analysis in the main text uses this logic to
identify combinations of $M_{\min}$, $f_{\rm occ}$, and $f_{\rm duty}$ that
are compatible with the cosmological supply of halos.

The treatment adopted here is intentionally phenomenological. A more complete
halo-occupation model could allow $f_{\rm occ}$ and $f_{\rm duty}$ to vary
with halo mass and redshift and could distinguish central and satellite
populations. Such extensions would be necessary for a detailed forward model
of AGN luminosity functions and clustering. For the present purpose, however,
the simple formulation isolates the key constraint: the observable active
population cannot exceed the supply of suitable host halos once occupation
and activity fractions are taken into account.

\subsection{Bright-end scatter and Eddington-bias-like effects}
\label{app:bright_end_scatter}

The general role of scatter in the halo-to-light relation was introduced in
Appendix~\ref{app:scatter_halo_light}. Here we isolate its consequence for the
bright end of the UV luminosity function, where the underlying halo abundance
declines most rapidly. This subsection is intended as a physical interpretation
of the scatter formalism rather than as an additional ingredient in the
baseline luminosity-function curves presented in the main text.

At a fixed redshift, the number of moderate-mass halos is much larger than the
number of very massive halos. If the UV luminosity at fixed halo mass has a
finite dispersion, some of these more numerous moderate-mass systems can
scatter temporarily toward brighter magnitudes. Although scatter may operate
in both directions, the steep decline of the parent halo population means that
the net effect can enhance the abundance in the rare bright tail. This is
analogous to the familiar Eddington-bias effect that arises when a steep
underlying distribution is convolved with observational or intrinsic scatter.

For either halo-mass-function prescription, the contribution to a bright
magnitude interval is contained in the general convolution
\begin{equation}
\Phi_X(M_{\rm UV},z)
=
\int
\frac{dn_X}{dM_h}(M_h,z)\,
P(M_{\rm UV}\mid M_h,z)\,dM_h,
\qquad
X\in\{{\rm ST},{\rm KST}\}.
\label{eq:app_bright_scatter_general}
\end{equation}
The bright-end enhancement therefore does not require the observed luminous
object to occupy the halo mass assigned by a strictly deterministic
halo-to-light relation. A somewhat lower-mass and more abundant halo may
contribute to the same bright magnitude bin through the high-luminosity tail
of $P(M_{\rm UV}\mid M_h,z)$.

It is useful to define the enhancement relative to the deterministic
prediction as
\begin{equation}
{\cal E}_X(M_{\rm UV},z)
\equiv
\frac{
\Phi_X^{\rm scatter}(M_{\rm UV},z)
}{
\Phi_X^{\rm det}(M_{\rm UV},z)
}.
\label{eq:app_bright_enhancement}
\end{equation}
Over portions of the steep bright tail one generally expects
${\cal E}_X>1$, although neither its magnitude nor its detailed dependence on
$M_{\rm UV}$ is universal. The enhancement depends on the width and shape of
the conditional luminosity distribution, the local slope and curvature of the
halo mass function, the mean halo-to-light relation, dust and stellar
population assumptions, and the adopted mass integration range. It should
therefore not be represented by a single constant correction factor.

For the KST prescription used in this paper,
\begin{equation}
\frac{dn_{\rm KST}}{dM_h}
=
F_{\rm K}(z)
\frac{dn_{\rm ST}}{dM_h},
\label{eq:app_bright_kst_hmf}
\end{equation}
where $F_{\rm K}(z)$ is independent of halo mass at fixed redshift. If the
same conditional distribution $P(M_{\rm UV}\mid M_h,z)$ is used in both
branches, equation~(\ref{eq:app_bright_scatter_general}) immediately gives
\begin{equation}
\boxed{
\Phi_{\rm KST}^{\rm scatter}(M_{\rm UV},z)
=
F_{\rm K}(z)\,
\Phi_{\rm ST}^{\rm scatter}(M_{\rm UV},z)
}.
\label{eq:app_bright_scatter_scaling}
\end{equation}
Thus scatter may substantially reshape and enhance the bright tail, but under
the assumptions adopted here it does not erase the relative ST/KST abundance
suppression. In fact, because the same scatter kernel is applied to both
branches, their ratio at any fixed $M_{\rm UV}$ and $z$ remains
\begin{equation}
\frac{
\Phi_{\rm KST}^{\rm scatter}
}{
\Phi_{\rm ST}^{\rm scatter}
}
=
F_{\rm K}(z).
\label{eq:app_bright_ratio}
\end{equation}

This result also implies that, under the present assumptions, the fractional
bright-end enhancement produced by scatter is identical for ST and KST:
\begin{equation}
{\cal E}_{\rm KST}(M_{\rm UV},z)
=
{\cal E}_{\rm ST}(M_{\rm UV},z).
\label{eq:app_equal_scatter_enhancement}
\end{equation}
The absolute KST number density nevertheless remains lower because the
available halo reservoir has already been reduced by $F_{\rm K}(z)$.

The physical interpretation is therefore straightforward. Scatter can reduce
the halo mass that would be inferred for an individual very luminous object
under a deterministic mapping and can increase the predicted abundance of
bright objects by drawing from a much larger reservoir of lower-mass halos.
This may lessen an apparent bright-end abundance tension. It does not,
however, create additional dark matter halos or remove the KST suppression of
the underlying halo population.

No scatter correction or Eddington-bias adjustment is applied to the baseline
ST and KST luminosity-function curves shown in the main text. Those curves
retain the deterministic mapping described in
Appendix~\ref{app:deterministic_halo_uv} and
Appendix~\ref{app:deterministic_lf}. The discussion here instead identifies a
physically motivated extension that should be included in future
forward-modeling and statistical comparison with high-redshift luminosity
function measurements. In such an analysis, intrinsic luminosity scatter
should also be distinguished from observational uncertainties in measured
fluxes, photometric redshifts, completeness corrections, and inferred number
densities.

\subsection{Side-by-side numerical implementation of the ST and KST branches}
\label{app:numerical_implementation}

The numerical calculations evaluate the ST and KST predictions as parallel
branches of the same underlying cosmological calculation.
This is important because the KST prescription is applied only as a correction
to the halo abundance; it does not replace or modify the common power spectrum,
mass variance, peak-height calculation, or baryonic halo-to-light mapping.

The computational sequence may be summarized as follows.

\begin{enumerate}

\item Adopt the cosmological parameters
\begin{equation}
\left(
\Omega_m,\Omega_b,\Omega_\Lambda,h,\sigma_8,n_s
\right)
\label{eq:app_cosmological_parameter_set}
\end{equation}
used throughout the main calculation.

\item Construct the linear matter power spectrum at $z=0$,
\begin{equation}
P(k,0)
=
A_{\rm norm}\,
k^{n_s}T^2(k),
\label{eq:app_pk_numerical}
\end{equation}
using the adopted transfer function and determine $A_{\rm norm}$ by requiring
the calculated variance on the $8\,h^{-1}{\rm Mpc}$ scale to reproduce the
adopted value of $\sigma_8$. The same normalized power spectrum is used for
both ST and KST.

\item Construct the halo-mass grid and corresponding smoothing radii through
\begin{equation}
R(M_h)
=
\left(
\frac{3M_h}{4\pi\bar{\rho}_m}
\right)^{1/3}.
\label{eq:app_mass_radius_numerical}
\end{equation}
The numerical mass interval must be sufficiently broad to cover all halo
thresholds, cumulative integrals, and halo masses contributing to the plotted
luminosity range. The precise grid limits and resolution are numerical choices
rather than additional physical assumptions.

\item Compute the present-day mass variance,
\begin{equation}
\sigma^2(M_h,0)
=
\frac{1}{2\pi^2}
\int_0^\infty
k^2P(k,0)W^2[kR(M_h)]\,dk,
\label{eq:app_sigma0_numerical}
\end{equation}
using the same spherical top-hat window and numerical $k$ grid for both halo
mass-function branches.

\item At each requested redshift, compute the flat-$\Lambda$CDM growth factor
$D(z)$, normalized to
\begin{equation}
D(0)=1,
\end{equation}
and then evaluate
\begin{equation}
\sigma(M_h,z)
=
D(z)\sigma(M_h,0)
\label{eq:app_sigma_z_numerical}
\end{equation}
and
\begin{equation}
\nu(M_h,z)
=
\frac{\delta_c}{D(z)\sigma(M_h,0)}.
\label{eq:app_nu_numerical}
\end{equation}
The same $D(z)$, $\sigma(M_h,z)$, and $\nu(M_h,z)$ are supplied to the ST and
KST calculations.

\item Evaluate the standard ST differential halo mass function,
\begin{equation}
\frac{dn_{\rm ST}}{dM_h}(M_h,z),
\label{eq:app_st_numerical}
\end{equation}
and retain it as the unmodified baseline prediction.

\item Independently evaluate the Klypin factor
\begin{equation}
F_{\rm K}(z)
=
\frac{[5.501D(z)]^4}
{1+[5.500D(z)]^4},
\label{eq:app_fk_numerical}
\end{equation}
and construct the second halo-abundance branch,
\begin{equation}
\frac{dn_{\rm KST}}{dM_h}(M_h,z)
=
F_{\rm K}(z)
\frac{dn_{\rm ST}}{dM_h}(M_h,z).
\label{eq:app_kst_numerical}
\end{equation}
The ST array is stored independently from the KST array.

\item Compute the cumulative halo abundances independently,
\begin{equation}
n_X(>M_{\min},z)
=
\int_{M_{\min}}^\infty
\frac{dn_X}{dM_h}(M_h,z)\,dM_h,
\qquad
X\in\{{\rm ST},{\rm KST}\},
\label{eq:app_cumulative_numerical}
\end{equation}
together with any mass-weighted cumulative quantities required by the
baryonic-reservoir analysis. Because $F_{\rm K}$ is mass-independent in the
adopted implementation, these arrays provide a direct numerical check of
\begin{equation}
\frac{
n_{\rm KST}(>M_{\min},z)
}{
n_{\rm ST}(>M_{\min},z)
}
=
F_{\rm K}(z).
\label{eq:app_cumulative_ratio_check}
\end{equation}

\item Apply the common deterministic halo-to-UV prescription to obtain
\begin{equation}
M_{\rm UV}
=
\bar{M}_{\rm UV}(M_h,z;\epsilon_\star),
\label{eq:app_common_mapping_numerical}
\end{equation}
for each selected value of the stellar-conversion efficiency
$\epsilon_\star$. This mapping is calculated independently of whether the
halo abundance is labeled ST or KST.

\item For the baseline luminosity-function calculation presented in the main
text, transform each differential halo mass function using
\begin{equation}
\Phi_X(M_{\rm UV},z)
=
\left.
\frac{dn_X}{dM_h}
\left|
\frac{dM_h}{dM_{\rm UV}}
\right|
\right|_{M_h=M_h(M_{\rm UV},z)},
\qquad
X\in\{{\rm ST},{\rm KST}\}.
\label{eq:app_lf_numerical}
\end{equation}
For the baseline mass-independent $\epsilon_\star$ mapping,
\begin{equation}
\left|
\frac{dM_h}{dM_{\rm UV}}
\right|
=
\frac{\ln 10}{2.5}M_h.
\label{eq:app_jacobian_numerical}
\end{equation}
No luminosity scatter, AGN occupation fraction, or AGN duty-cycle factor is
included in these baseline galaxy UV luminosity-function curves.

\item Store the ST and KST luminosity functions separately and verify the
diagnostic ratio
\begin{equation}
R_\Phi(M_{\rm UV},z)
\equiv
\frac{
\Phi_{\rm KST}(M_{\rm UV},z)
}{
\Phi_{\rm ST}(M_{\rm UV},z)
}.
\label{eq:app_phi_ratio}
\end{equation}
For the mass-independent Klypin correction and identical halo-to-light
mapping adopted here,
\begin{equation}
\boxed{
R_\Phi(M_{\rm UV},z)
=
F_{\rm K}(z)
}
\label{eq:app_phi_ratio_fk}
\end{equation}
throughout the numerically sampled magnitude range, apart from interpolation,
finite-grid, and floating-point effects.

\item Repeat the calculation over the required redshift grid. Quantitative KST
interpretation is restricted to the approximate Bolshoi calibration interval
\begin{equation}
0\leq z\leq10.
\label{eq:app_numerical_calibration_range}
\end{equation}
Any evaluation at $z>10$ is treated explicitly as an extrapolation.

\end{enumerate}

Several internal consistency checks are useful for validating the numerical
implementation. First, the $z=0$ power spectrum should reproduce the adopted
$\sigma_8$ normalization. Second, the same $\sigma(M_h,z)$ and $\nu(M_h,z)$
arrays must enter both the ST and KST branches. Third, at every fixed redshift
the ratios
\begin{equation}
\frac{dn_{\rm KST}/dM_h}{dn_{\rm ST}/dM_h},
\qquad
\frac{n_{\rm KST}(>M_h)}{n_{\rm ST}(>M_h)},
\qquad
\frac{\Phi_{\rm KST}}{\Phi_{\rm ST}}
\label{eq:app_three_ratio_checks}
\end{equation}
should each reproduce $F_{\rm K}(z)$ wherever the corresponding numerical
quantities are well resolved. These provide particularly direct diagnostics
for detecting accidental changes in normalization, inconsistent interpolation,
or unintended modification of the ST baseline.

For graphical comparisons, the ST and KST predictions should be evaluated on
the same mass, magnitude, and redshift grids. Corresponding ST and KST curves
for the same physical parameter value---for example the same redshift,
$\epsilon_\star$, minimum halo mass, or duty cycle---should also be displayed
with a common color and distinguished by line style. This plotting convention
does not affect the numerical calculation, but it makes the physical
ST--KST comparison unambiguous.

The scatter formalism developed in Appendix~\ref{app:scatter_halo_light} may be
added as a separate extension by replacing the deterministic Jacobian
transformation with the convolution
\begin{equation}
\Phi_X(M_{\rm UV},z)
=
\int
\frac{dn_X}{dM_h}
P(M_{\rm UV}\mid M_h,z)\,dM_h.
\label{eq:app_optional_scatter_numerical}
\end{equation}
Likewise, AGN-specific predictions may subsequently include the occupation and
duty-cycle factors described in Appendix~\ref{app:duty_occupation}. These
extensions should be kept separate from the deterministic baseline so that
changes caused by the cosmological halo supply can be distinguished from
changes caused by baryonic or AGN population modeling.

The resulting numerical architecture therefore preserves the complete
calculation chain
\begin{align}
P(k)
&\rightarrow \sigma(M_h,z)
\rightarrow \nu(M_h,z)
\rightarrow \frac{dn_{\rm ST}}{dM_h}
\nonumber\\
&\rightarrow \frac{dn_{\rm KST}}{dM_h}
\rightarrow n_X(>M_h,z)
\nonumber\\
&\rightarrow M_{\rm UV}(M_h,z)
\rightarrow \Phi_X(M_{\rm UV},z).
\label{eq:app_numerical_chain}
\end{align}
while isolating the Klypin correction to the halo-abundance step. This
side-by-side construction is the numerical basis for the ST/KST comparisons
presented in the main text.

\subsection{Physical interpretation of the ST--KST comparison}
\label{app:physical_interpretation_st_kst}

The side-by-side ST and KST calculations are most useful when the different
physical ingredients in the prediction are kept conceptually separate. In the
framework adopted in this work, three principal levels enter the connection
between cosmology and the observed high-redshift population:
\begin{align}
\frac{dn_X}{dM_h}(M_h,z)
&\quad &&\text{cosmological halo supply}, \nonumber\\
\bar{M}_{\rm UV}(M_h,z;\epsilon_\star)
&\quad &&\text{mean stellar halo-to-light mapping}, \nonumber\\
P(M_{\rm UV}\mid M_h,z)
&\quad &&\text{optional luminosity scatter and baryonic variability},
\label{eq:app_three_physical_levels}
\end{align}
where $X\in\{{\rm ST},{\rm KST}\}$.

For the deterministic galaxy UV calculation used as the baseline in this
paper, the third ingredient is replaced by the single-valued relation
$M_{\rm UV}=\bar{M}_{\rm UV}(M_h,z;\epsilon_\star)$. Number conservation then
gives
\begin{equation}
\boxed{
\begin{aligned}
\Phi_X(M_{\rm UV},z)
&=
\left.
\frac{dn_X}{dM_h}(M_h,z)
\left|
\frac{dM_h}{dM_{\rm UV}}
\right|
\right|_{M_h=M_h(M_{\rm UV},z)},\\
&\hspace{1.0em}X\in\{{\rm ST},{\rm KST}\}.
\end{aligned}
}
\label{eq:app_physical_deterministic_lf}
\end{equation}
This expression is the appropriate physical chain for the baseline galaxy UV
luminosity functions presented in the main text. In particular, no AGN duty
cycle or black-hole occupation fraction is inserted into these galaxy
luminosity-function curves.

If intrinsic scatter in the stellar halo-to-light relation is introduced, the
corresponding expression becomes
\begin{equation}
\boxed{
\Phi_X(M_{\rm UV},z)
=
\int
\frac{dn_X}{dM_h}(M_h,z)\,
P(M_{\rm UV}\mid M_h,z)\,dM_h
}.
\label{eq:app_physical_scatter_lf}
\end{equation}
The conditional probability distribution then describes baryonic variability
at fixed halo mass, while the halo mass function continues to determine the
cosmological supply of potential hosts.

The AGN abundance problem is related but physically distinct. For an
AGN-selected population, the abundance additionally depends on whether a halo
contains a black hole and whether that black hole is active at the epoch of
observation. The corresponding host-supply expression is
\begin{equation}
n_{{\rm AGN},X}(z)
=
\int_{M_{\min}}^\infty
\frac{dn_X}{dM_h}(M_h,z)\,
f_{\rm occ}(M_h,z)\,
f_{\rm duty}(M_h,z)\,dM_h.
\label{eq:app_physical_agn_abundance}
\end{equation}
Thus $f_{\rm occ}$ and $f_{\rm duty}$ belong to the AGN population model and
should not be conflated with the stellar halo-to-UV mapping used for the
galaxy luminosity function.

This separation is important for interpreting the ST--KST comparison. Both
branches begin with the same cosmological parameters, linear matter power
spectrum, growth factor, mass variance, peak height, and collapse prescription
entering the ST baseline. They also use the same baryonic mapping when a
galaxy or AGN observable is assigned to a halo. The distinction between the
two branches is isolated to the adopted simulation-calibrated abundance
correction,
\begin{equation}
\boxed{
\frac{dn_{\rm KST}}{dM_h}(M_h,z)
=
F_{\rm K}(z)
\frac{dn_{\rm ST}}{dM_h}(M_h,z)
}.
\label{eq:app_physical_kst_relation}
\end{equation}

Because $F_{\rm K}(z)$ is independent of halo mass in the prescription used
here, the correction changes the normalization of the halo population at a
fixed redshift without changing the mass dependence of the ST mass function.
Consequently,
\begin{equation}
\frac{
dn_{\rm KST}/dM_h
}{
dn_{\rm ST}/dM_h
}
=
F_{\rm K}(z)
\label{eq:app_physical_hmf_ratio}
\end{equation}
at every resolved halo mass. The same factor propagates into any quantity that
is a linear integral or deterministic transformation of the halo mass
function, provided all other assumptions are held fixed.

For example,
\begin{equation}
n_{\rm KST}(>M_{\min},z)
=
F_{\rm K}(z)
n_{\rm ST}(>M_{\min},z),
\label{eq:app_physical_cumulative_ratio}
\end{equation}
and for the deterministic stellar UV mapping,
\begin{equation}
\Phi_{\rm KST}(M_{\rm UV},z)
=
F_{\rm K}(z)
\Phi_{\rm ST}(M_{\rm UV},z).
\label{eq:app_physical_lf_ratio}
\end{equation}
Likewise, if identical black-hole occupation and duty-cycle functions are used
for the two branches,
\begin{equation}
n_{{\rm AGN},{\rm KST}}(z)
=
F_{\rm K}(z)
n_{{\rm AGN},{\rm ST}}(z).
\label{eq:app_physical_agn_ratio}
\end{equation}

These relations make the physical interpretation particularly transparent.
Changing $\epsilon_\star$ modifies the mapping between halo mass and stellar UV
luminosity. Changing $M_{\min}$ modifies which portion of the halo population
is available to host a selected object. Changing $f_{\rm occ}$ or
$f_{\rm duty}$ modifies the fraction of suitable halos that contribute to an
AGN sample. Introducing $\sigma_{\rm UV}$ or a more general conditional
distribution changes the spread of luminosity at fixed halo mass. By contrast,
the ST--KST distinction modifies the cosmological abundance assigned to the
same underlying halo population.

This separation also clarifies the meaning of an apparent discrepancy with
high-redshift observations. A mismatch in luminosity at a given halo mass
primarily probes the baryonic or black-hole mapping, whereas a mismatch in the
number of available hosts probes the halo abundance and the population
occupation assumptions. These effects can partially compensate one another in
an observational comparison, but they represent physically different
mechanisms and should therefore not be combined into a single phenomenological
correction.

The interpretation is especially important toward high redshift, where
$F_{\rm K}(z)$ decreases rapidly. Relative to the ST baseline, the KST branch
therefore supplies fewer halos before any assumptions about stellar efficiency,
black-hole occupation, duty cycle, or luminosity scatter are applied. Higher
star-formation efficiency or favorable AGN activity parameters can make an
individual halo more luminous or increase the fraction of halos appearing in
an observed sample, but they cannot increase the underlying cosmological
number of halos predicted by the adopted KST mass function.

The physical structure of the calculation can therefore be summarized
schematically as
\begin{equation}
\boxed{
\begin{aligned}
P(k)
&\rightarrow \sigma(M_h,z)
\rightarrow \nu(M_h,z)
\rightarrow \frac{dn_X}{dM_h}\\
&\rightarrow
\text{halo-to-observable mapping}\\
&\rightarrow
\text{observable population}.
\end{aligned}
}
\label{eq:app_physical_separation}
\end{equation}
Here the first line represents the cosmological halo supply, while the
halo-to-observable mapping contains the baryonic and/or black-hole physics.
For the present analysis, the principal purpose of carrying ST and KST
side-by-side is therefore not to introduce two different galaxy-formation
models, but to quantify how a change in the calibrated high-redshift halo
supply propagates through otherwise identical astrophysical assumptions.

\subsection{Connection to JWST observations}
\label{app:jwst_connection}

The purpose of the halo-to-light construction developed in this Appendix is to
place the theoretical halo-abundance calculation in the same observable plane
as high-redshift galaxy measurements. Rest-frame ultraviolet luminosity
functions are particularly useful for this purpose because they constrain not
only whether halos can produce galaxies of a given luminosity, but also whether
the cosmological halo population is sufficiently abundant to reproduce the
observed number density of such systems.

The baseline comparison in this work uses representative measurements of the
high-redshift galaxy UV luminosity function from \citet{Franco2025},
\citet{Donnan2023}, and \citet{Donnan2024}. These data provide population-level
constraints spanning the cosmic-dawn regime and complement comparisons based
on individual unusually massive or luminous JWST sources. The theoretical
curves and observational measurements should nevertheless be regarded as a
controlled first-order comparison rather than a formal statistical fit.

For each theoretical branch,
\begin{equation}
X\in\{{\rm ST},{\rm KST}\},
\end{equation}
the deterministic prediction is
\begin{equation}
\begin{aligned}
\Phi_X(M_{\rm UV},z)
&=
\left.
\frac{dn_X}{dM_h}(M_h,z)
\left|
\frac{dM_h}{dM_{\rm UV}}
\right|
\right|_{M_h=M_h(M_{\rm UV},z)} .
\end{aligned}
\label{eq:app_jwst_lf}
\end{equation}
The observational comparison therefore probes two logically distinct
requirements. First, the adopted baryonic mapping must assign sufficiently
bright UV luminosities to halos of plausible mass. Second, the halo mass
function must provide enough such halos to reproduce the measured number
density.

This distinction is central to the interpretation of the JWST comparison. A
model can reproduce the luminosity of an individual source by increasing the
stellar-conversion efficiency or modifying the star-formation timescale, yet
still fail to provide enough host halos. Conversely, a cosmology may provide
an adequate number of halos while an adopted baryonic prescription produces
galaxies that are systematically too faint. The luminosity-function
comparison therefore helps separate a halo-supply problem from a
halo-to-light-conversion problem.

In the present deterministic model, changing $\epsilon_\star$ modifies the
mapping between $M_h$ and $M_{\rm UV}$. Larger $\epsilon_\star$ produces a
larger stellar mass and star-formation rate at fixed halo mass and therefore a
brighter, more negative $M_{\rm UV}$. The resulting theoretical luminosity
function is shifted relative to the observational points because a specified
UV magnitude is then associated with a different part of the halo mass
function.

By contrast, the KST correction acts directly on the halo abundance. Since
\begin{equation}
\frac{dn_{\rm KST}}{dM_h}
=
F_{\rm K}(z)
\frac{dn_{\rm ST}}{dM_h}
\label{eq:app_jwst_kst_hmf}
\end{equation}
and the same deterministic halo-to-light relation is used for both branches,
\begin{equation}
\boxed{
\Phi_{\rm KST}(M_{\rm UV},z)
=
F_{\rm K}(z)
\Phi_{\rm ST}(M_{\rm UV},z)
}.
\label{eq:app_jwst_kst_lf}
\end{equation}
Thus the ST and KST curves at a common redshift and common
$\epsilon_\star$ have the same deterministic shape in this implementation but
different normalizations.

The population-level comparisons used in the main text are chosen to sample
several redshifts at which the theoretical ST--KST difference becomes
progressively more important. The COSMOS-Web analysis of
\citet{Franco2025} provides a large-area JWST determination of the UV
luminosity function over a broad high-redshift interval and is used for the
lower-redshift comparison. The measurements of \citet{Donnan2023} provide
UV-luminosity-function constraints extending from approximately $z\simeq8$
into the very high-redshift regime, while the later PRIMER analysis of
\citet{Donnan2024} provides an independent multifield determination extending
through approximately $z\simeq9$--15. These datasets therefore provide a
useful observational benchmark for the redshift evolution predicted by the
halo-based model.

The representative numerical values of $F_{\rm K}(z)$ are given in
Equation~\eqref{eq:klypin_reference_values} of Section~\ref{subsec:kst}.

Consequently, for identical baryonic assumptions the KST luminosity function
is approximately $49.6$, $26.5$, and $13.9$ per cent of the corresponding ST
prediction at $z=6$, $8$, and $10$, respectively. Equivalently, the ST
prediction is approximately $2.02$, $3.77$, and $7.18$ times larger. The
increasing separation toward high redshift illustrates why the calibration of
the halo mass function becomes increasingly important when interpreting
cosmic-dawn number densities.

The comparison should not be interpreted as establishing a unique value of
$\epsilon_\star$ or as a precision goodness-of-fit test of either halo mass
function. The baseline theoretical curves omit several effects that can alter
the observed UV luminosity function, including intrinsic luminosity scatter,
bursty star formation, dust attenuation, mass- and redshift-dependent
star-formation efficiency, feedback, halo assembly history, and possible AGN
contamination. Observational measurements also carry uncertainties associated
with sample selection, completeness, photometric and spectroscopic redshifts,
survey volume, cosmic variance, lensing where relevant, and contamination.

Scatter is especially relevant at the bright end. As discussed in
Appendices~\ref{app:scatter_halo_light} and
\ref{app:bright_end_scatter}, a distribution
$P(M_{\rm UV}\mid M_h,z)$ can allow more numerous moderate-mass halos to
contribute to bright magnitude bins. Such scatter can increase the predicted
bright-end abundance relative to a deterministic model. It is not included in
the baseline curves used in the present observational comparison and should
therefore be treated as an important extension rather than as an implicit
correction to the plotted results.

Similarly, the galaxy UV luminosity-function comparison should be distinguished
from the AGN abundance constraints developed elsewhere in this work. The
baseline UV luminosity functions use a stellar halo-to-light mapping and do
not multiply the galaxy counts by an AGN occupation fraction or duty cycle.
AGN-specific predictions require additional assumptions concerning black-hole
occupation, accretion duty cycle, Eddington-ratio distributions, radiative
efficiency, obscuration, and the relation between black-hole and halo mass.

The Bolshoi calibration boundary also remains important in the observational
interpretation. Quantitative KST comparisons in this work are restricted to
approximately
\begin{equation}
z\leq10.
\label{eq:app_jwst_calibration_limit}
\end{equation}
JWST measurements at $z>10$ remain highly informative for early structure
formation and galaxy evolution, but evaluation of the same Klypin fitting
expression beyond this range constitutes an extrapolation rather than a
simulation-calibrated KST prediction. Such measurements should therefore be
used to motivate new high-redshift numerical calibrations rather than to claim
a calibrated KST discrepancy.

The observational logic of the framework can finally be summarized as
\begin{equation}
\boxed{
\begin{aligned}
P(k)
&\rightarrow \sigma(M_h,z)
\rightarrow \nu(M_h,z)
\rightarrow \frac{dn_X}{dM_h}\\
&\rightarrow M_{\rm UV}(M_h,z;\epsilon_\star)\\
&\rightarrow \Phi_X(M_{\rm UV},z)
\rightarrow \text{JWST comparison}.
\end{aligned}
}
\label{eq:app_jwst_chain}
\end{equation}
The first line describes the cosmological halo supply, the second the stellar
halo-to-light mapping, and the third the observable population comparison.

The main value of this construction is therefore diagnostic. It provides a
transparent way to ask whether an apparent high-redshift discrepancy is driven
primarily by the abundance of dark matter halos, by the efficiency with which
their baryons produce UV light, or by astrophysical and observational effects
not included in the deterministic baseline. As JWST surveys continue to
improve the precision and dynamic range of high-redshift luminosity functions,
this separation will allow increasingly stringent tests of both early galaxy
formation and the underlying halo-abundance model.
\end{document}